\documentclass[11pt, reqno,preprint]{article}
\usepackage{jheppub}
\usepackage{epsfig}
\usepackage{amssymb}
\usepackage{amsmath}
\usepackage{mathrsfs}
\usepackage{hyperref}
\usepackage{multirow}
\usepackage[normalem]{ulem}
\usepackage{graphicx}
\usepackage{textcomp}
\usepackage{subcaption}

\usepackage{soul,xcolor}
\setstcolor{red} 

\def\be{\begin{equation}}
\def\ee{\end{equation}}
\def\ba{\begin{eqnarray}}
\def\ea{\end{eqnarray}}
\def\nl{\nonumber \\}
\def\a{\alpha}
\def\ab{\overline{\alpha}}
\def\zb{\overline{z}}
\def\wb{\overline{w}}
\def\r{\rho}
\def\rb{\overline{\rho}}
\def\j{J}

\def\Res{\operatorname*{Res}}
\def\Df{\Delta_\phi}
\def\OPE{\lambda^2}  
\def\OPEfree{\lambda^{2,\rm free}}
\def\Konishi{K}

\def\Qmat{\left[Q_{\Delta,\j}\right]}

\def\Gt{\Upsilon(\mS,\mT)}
\def\RAdS{R_{\rm AdS}}
\def\etaAdS{\eta_{\rm AdS}}

\def\cM{\mathcal{M}}
\def\<{\langle}
\def\>{\rangle}
\def\Nfour{\mathcal{N}=4}

\newcommand{\beq}{\begin{equation}}
\newcommand{\eeq}{\end{equation}}
\newcommand{\bba}{\begin{align}}
\newcommand{\eea}{\end{align}}

\newcommand{\mS}{\mathrm{s}}
\newcommand{\mT}{\mathrm{t}}
\newcommand{\mU}{\mathrm{u}}

\def\OO{\mathcal{O}}
\def\cH{\mathcal{H}}

\def\cP{\mathcal{P}}
\def\dDisc{{\rm dDisc}}
\def\D{\Delta}

\title{\centering 
Bootstrapping $\mathcal{N}=4$ sYM correlators using Integrability and Localization}
\author{Simon Caron-Huot$^1$, Frank Coronado$^{2}$, Zahra Zahraee$^{3}$}
\affiliation{
${}^1$Department of Physics, McGill University, 3600 Rue University, Montr\'eal, H3A 2T8, QC Canada
}
\affiliation{${}^2$ Institut f\"ur Theoretische Physik, ETH Z\"urich, CH-8093 Z\"urich, Switzerland}
\affiliation{
${}^3$ CERN, 
Theoretical Physics Department, 
CH-1211 Geneva 23, 
Switzerland}
\emailAdd{schuot@physics.mcgill.ca}
\emailAdd{fcidrogo@gmail.com}
\emailAdd{zahra.zahraee@cern.ch}

\abstract{We study four-point correlation functions of the stress-tensor multiplet in $\mathcal{N}=4$ super Yang-Mills (sYM) theory by leveraging integrability and localization techniques.  We combine dispersive sum rules and spectral information from integrability, used previously, with integrated constraints from supersymmetric localization.
We obtain two-sided bounds on the OPE coefficient of the so-called Konishi operator in the planar limit at any value of the 't Hooft coupling ranging from weak to strong coupling. In addition to individual OPE coefficients, we discuss how to bound the correlation function itself and obtain two-sided bounds at various values of the cross-ratios and coupling.  Lastly, considering the limit of large 't Hooft coupling, we connect the analysis with that of an analogous flat space problem involving the Virasoro-Shapiro amplitude.
}

\begin{document}

\maketitle
\section{Introduction}

It is sometimes said that a conformal field theory can be deemed ``solved'' once we know the scaling dimensions and operator-product-expansion (OPE) coefficients of all local operators.
Enormous progress towards this goal has been achieved in maximal supersymmetric Yang-Mills (sYM) theory in its planar `t Hooft limit. Thanks to the AdS/CFT duality, this also addresses the problem of finding the spectrum of masses and couplings in type IIB string theory in $AdS_5\times S^5$.  In practice, since there are infinitely many local operators,
it is also natural to ask about various observables that repackage this data, such as vacuum correlation functions of simple operators.

In the planar limit, $\Nfour$ sYM enjoys a symmetry enhancement known as integrability \cite{Beisert:2010jr}. Integrability has brought us a solution of the spectral problem, ultimately known as the Quantum Spectral Curve of sYM \cite{Gromov:2013pga,Gromov:2014caa}. In practice, this leads to analytic results to high order in weak and strong coupling expansions, as well as accurate numerical results at finite coupling. This was first achieved for the lightest non-protected scalar operator, the so-called Konishi operator, in \cite{Gromov:2009zb}, whose anomalous dimension was computed for a wide range of values interpolating between weak and strong `t Hooft coupling. This has been extended to arbitrary non-protected operators, notably for the first hundreds of lightest operators \cite{Gromov:2023hzc,Ekhammar:2024rfj}.

Progress has been more moderate for three-point functions and more generally higher-point functions, despite the existence of non-perturbative integrability-based methods to tackle them \cite{Basso:2015zoa,Fleury:2016ykk,Basso:2022nny,Bercini:2022jxo}.  Although non-pertubative on the `t Hooft coupling, these methods still tend to be most readily applicable in certain limits such as when some operators carry large R-charge \cite{Basso:2019diw,Jiang:2019zig,Coronado:2018cxj,Kostov:2019stn,Belitsky:2020qrm}. 
More recently, thanks to the combination of CFT dispersive sum rules and a special basis of functions, a strong coupling expansion of the four-point correlator of stress-tensors has been obtained \cite{Alday:2022uxp,Alday:2023mvu}. Through AdS/CFT, this novel result goes beyond the supergravity limit and provides the second curvature correction to the AdS Virasoro-Shapiro amplitude around its flat-space limit. Nevertheless, there is still no handle on the finite coupling regime comparable to what is available for the Konishi scaling dimension.  

Another independent and powerful non-perturbative method is the numerical conformal bootstrap \cite{Rattazzi:2008pe, El-Showk:2012cjh, Poland:2018epd}, which implements general consistency constraints such as crossing and unitarity, in order to bound OPE data in conformal field theory given possibly additional input or assumptions. See \cite{Poland:2022qrs} for a recent review. This method has been applied to the sYM theory at finite number of colors in \cite{Beem:2016wfs} to bootstrap the stress-tensor correlator at finite gauge coupling along the conformal manifold. In \cite{Chester:2021aun}, by including constraints from supersymmetric localization on this correlation function, numerical bounds were obtained for arbitrary values of Yang-Mills coupling for $SU(2)$ and $SU(3)$. Recently, in \cite{Chester:2023ehi}, this was extended to higher-rank gauge groups, providing results up to $SU(11)$. Numerical bootstrap has also been successfully applied to one-dimensional defect CFT that lives on 1/2-BPS Wilson line in planar $\Nfour$ sYM \cite{Cavaglia:2021bnz,Cavaglia:2022qpg, Cavaglia:2023mmu}.  Exceptional mileage was gained for these observables by combining the bootstrap method with spectral data from integrability.

In a previous paper \cite{Caron-Huot:2022sdy}, we combined known single-trace spectral information with conformal bootstrap techniques to bound the OPE coefficient of the lightest non-protected single-trace  in the planar limit.
We did so by studying the four-point function of stress tensors, where both single and double traces are exchanged as intermediate operators in the planar limit, and using dispersive transforms to reconstruct the full correlator from single-traces alone.  These dispersive sum rules resulted in interesting upper bounds on OPE coefficients (seemingly saturated in known limits), although lower bounds remained elusive.

In this paper we continue the study of the four-point stress-tensor correlator in planar sYM using a numerical bootstrap method based on dispersive sum rules.  The main novelty will be the addition of
the supersymmetric localization constraints from \cite{Chester:2020dja,Chester:2021aun}. As mentioned above, this approach has been successful for both finite-$N_c$ correlators and defects; here, however, we will continue to specialize to the planar limit in order to benefit from the integrability spectra.

Our main new results will be nontrivial lower bounds at finite coupling, which remained elusive in our preceding work, as well as a sharpening of upper bounds. We will thus provide two-sided bounds which can be compared with analytic results at weak and strong coupling, and give a small allowed region for the OPE coefficient at intermediate couplings, where no other analytic nor numerical techniques exist.
Similar two-sided bounds will be obtained for other observables such as the four-point correlator itself at various values of the cross-ratio.
In the limit of strong `t Hooft coupling, our results will be compared with a holographically dual S-matrix problem.

This paper is organized as follows. In section~\ref{sec:setup} we review general properties of the four-point stress-tensor correlator, the CFT dispersive sum rules of \cite{Caron-Huot:2022sdy} and introduce the integrated constraints as additional sum rules. We also review known analytic results for the OPE coefficient and its connection with flat-space physics. Section~\ref{sec:numericalbootstrap} contains our main results: two-sided bounds for the OPE coefficient of the Konishi operator and for the correlator at finite coupling. In section~\ref{sec:flat-space} we show how our numerical optimization problem at strong coupling approximates an analogous S-matrix problem for the four-graviton amplitude in the flat space limit of ${\rm AdS}_5$. Finally in section~\ref{sec:discussion} we summarize our findings and discuss future directions. The first three  appendices discuss: the  Regge limit of the dispersive functionals, the Mellin representation of the integrated correlators, and the known strong coupling OPE data in the leading Regge trajectory. The final appendix contains table \ref{tab:bounds} recording the numerical bounds plotted in figure \ref{fig:ope-results}.

\section{Setup and ingredients}
\label{sec:setup}
\subsection{Stress-tensor multiplet correlators and Mellin representation}

We consider the correlation function of four
operators which generate the stress-tensor supermultiplet in $\Nfour$ super Yang-Mills.
These are scalar operators transforming in the $[0,2,0]$ representation of the SU(4) global symmetry (e.g. a traceless symmetric two-index tensor of so(6)).
In index-free notation this operator can be viewed as
\begin{align} \label{020 operator}
    \OO(x,y) \propto {\rm Tr }[(y{\cdot}\phi(x))^2],
\end{align}
which is a function of $x\in \mathbb{R}^{3,1}$, a spacetime point, and $y\in \mathbb{C}^6$ a null 6-vector.
We use the canonical normalization
$\< \OO(x_1,y_1)\OO(x_2,y_2)\>=(y_{12}^2/x_{12}^2)^2$. Due to conformal symmetry the four-point correlation function depend on $x_i$ and $y_i$ through spacetime and R-charge cross-ratios,
\begin{align}
u= \frac{x_{12}^{2}x_{34}^2}{x_{13}^2x_{24}^2} = z \bar{z}\,, \qquad &v = \frac{x_{23}^2x_{14}^2}{x_{13}^2x_{24}^2} = (1-z)(1-\bar{z})\,, \label{cross-ratios u,v} \\
\sigma= \frac{y_{12}^2 y_{34}^2}{y_{13}^2 y_{24}^2} = \alpha \bar{\alpha}\,, \qquad& \tau = \frac{y_{23}^2 y_{14}^2}{y_{13}^2 y_{24}^2} = (1-\alpha)(1-\bar{\alpha}) \label{cross-ratios sigma tau}\,.
\end{align}

In addition, thanks to
the superconformal Ward identities \cite{Eden:2000bk,Dolan:2004mu, Nirschl:2004pa}, we can constrain the dependence on R-charge vectors and write the correlator as a free part ($g\rightarrow 0$) plus an interacting part:
\begin{align}
\frac{x_{13}^4x_{24}^4}{y_{13}^4y_{24}^4} \langle \OO(x_1,y_1)\cdots \OO(x_4,y_4)\rangle
&=1+\frac{\sigma^2}{u^2}+\frac{\tau^2}{v^2} +\frac{1}{c}\left(\frac{\sigma}{u}+\frac{\tau}{v}+\frac{\sigma\tau}{u v}\right)
\nl &\quad + \frac{1}{c}(z-\a) (z - \ab) (\zb- \a) (\zb - \ab)\cH(z,\zb)\,, \label{G ansatz}
\end{align}
where $c$ is the central charge ($c= \frac{N_c^2-1}{4}$ in SU($N_c$) gauge theory). The dynamical part $\cH$  has no $\alpha, \bar{\alpha}$ dependence and it satisfies the same crossing properties as a correlator of four identical scalars of effective dimension $\Delta_{\rm \phi}^{\rm eff}=4$:
\begin{align}
    \cH(u,v)=\cH(v,u)=u^{-4}\cH(\tfrac{1}{u},\tfrac{v}{u}).
\end{align}
Furthermore, $\cH(u,v)$ admits a conventional OPE decomposition in terms of four-dimensional conformal blocks, which include in the planar limit single-trace and double-trace operators. In \cite{Caron-Huot:2022sdy}, we explained another expansion in terms of the so-called Polyakov-Regge blocks where only single traces contribute to the sum. Here we write the sum as given by the protected and non-protected single-trace contributions:\footnote{The coefficients here differ from standard three-point couplings by a factor of $c$: $f_{{\cal O}{\cal O},[\Delta,J]}^2=\frac{1}{c}\lambda^2_{\Delta,J}$.}
\be \cH(u,v)= \cH^{\rm sugra}(u,v) + \sum_{(\Delta,J)\,\rm long} \lambda^2_{\Delta,J} \cP^{\Nfour}_{u,v}(\Delta,J)\,,
 \label{PR expansion uv}
\ee
where $\cP^{\Nfour}_{u,v}(\Delta,J)$ are Polyakov-Regge blocks defined in \cite{Caron-Huot:2022sdy}  as a dispersive transform of the standard conformal block (see eq. 2.17 therein); they represent conventional blocks supplemented by infinite sums of double-trace blocks.
The part outside the sum, accounting for protected operators, is precisely the answer in the supergravity limit and is given by:
\begin{equation}
    \cH^{\rm sugra}(u,v)=-\bar{D}_{2,4,2,2} = \partial_{u}\partial_{v}(1+u\partial_{u}+v\partial_{v})F_{1}(u,v)
\end{equation}
where the $\bar{D}$-function can be obtained as a differential operator acting on the box integral:
\be \label{F1}
\begin{aligned}
    F_1(u,v)\equiv \frac{2\text{Li}_{2}(z)-2\text{Li}_{2}(\bar{z})+\log(z\bar{z})(\log(1-z)-\log(1-\bar{z}))}{z-\bar{z}}\,.
\end{aligned}
\ee

For many calculations, as discussed in \cite{Caron-Huot:2022sdy}, the Mellin space representation of the correlator and Polyakov-Regge blocks are also useful:
\be
\cH(u,v) = \iint\!\!\frac{d\mS\,d\mT}{(4\pi i)^2}\,u^{\tfrac{\mS}{2}-4}v^{\tfrac{\mT}{2}-4}
\Gamma\!\left(4-\tfrac{\mS}{2}\right)^2\Gamma\!\left(4-\tfrac{\mT}{2}\right)^2\Gamma\!\left(4-\tfrac{\mU}{2}\right)^2
\widehat{\cH}(\mS,\mT)\,,
\label{eq:mellinRep}
\ee
where $\mS+\mT+\mU=4\Delta_{\phi}^{\rm eff}=16$,
and $\widehat{\cH}(\mS,\mT)$ is invariant under any permutation
of $\{\mS,\mT,\mU\}$.
For future reference, we record
the Mellin amplitude in the limits of weak and strong coupling (with $g^2=\frac{\lambda}{16\pi^2}$) in our conventions:
\begin{subequations} \label{H weak strong}
\begin{align}
 \lim_{g\to 0}\widehat{\cH} &=\frac{-2g^2}{(\tfrac{\mS}{2}-3)^2(\tfrac{\mT}{2}-3)^2(\tfrac{\mU}{2}-3)^2} + O(g^4), \\
 \lim_{g\to\infty}\widehat{\cH} &= \frac{1}{(\tfrac{\mS}{2}-3)(\tfrac{\mT}{2}-3)(\tfrac{\mU}{2}-3)}
\equiv \widehat{\cH}^{\rm sugra}\,. 
\end{align}
\end{subequations}
The \emph{Polyakov-Regge} expansion in Mellin space admits a similar form to the position space
one in \eqref{PR expansion uv}
(Regge boundedness of the reduced correlator ensures that the Mellin transform converges term by term \cite{Caron-Huot:2020adz,Caron-Huot:2022sdy}):
\be
\widehat{\cH}(\mS,\mT) = \widehat{\cH}^{\rm sugra}(\mS,\mT)\,+\! \sum_{(\Delta,J)\,\rm long} \lambda^2_{\Delta,J}{\cP}^{\Nfour}_{\Delta,J}(\mS,\mT)\,,
\label{PR expansion st}
\ee
where the Polyakov-Regge blocks are given explicitly as sums over poles:
\be \widehat{\cP}^{\Nfour}_{\Delta,\j}(\mS,\mT)
= \sum\limits_{n=0}^\infty
\mathcal{Q}^{n}_{\Delta+4,\j}(16 - \mS- \mT)
\left[\frac{1}{\mS-(\Delta-\j+2n+4)}+\frac{1}{\mT-(\Delta-\j+2n+4)}\right] \,.
\label{PR Mellin explicit}
\ee
In Mellin space, the Polyakov-Regge expansion
is a conventional dispersion relation which reconstructs a meromorphic function of $\mS$ from its poles (at fixed $\mU$).
The residues ${\cal Q}$ are \emph{Mack polynomials} $Q$, which have degree $\j$ in $n$ and $\mU$,
times specific Gamma functions:
\begin{align}\label{eq:Qcal_rep}
 {\cal Q}^{n}_{\Delta,\j}(\mU) &=
 K_{\Delta,\j}^{n}
\times \left(Q^{n}_{\Delta,\j}(\mU)\equiv \sum_{k,q=0}^{\j} (-n)_q  \Qmat_{q,k} \big(\tfrac{8-\mU}{2}\big)_k\right),\\
 K^n_{\Delta,\j}&= \frac{2\Gamma(\Delta+\j)\Gamma(\Delta+\j-1)}{n!\, \Gamma(\Delta-1+n)
 \Gamma\big(\tfrac{8-\Delta+\j}{2}-n\big)^2\Gamma\big(\tfrac{\Delta+\j}{2}\big)^4}\,.
\end{align}
The coefficients $\Qmat_{q,k}$, which are each rational functions of $\Delta$,
can be computed efficiently using the recursion in (C.8) of \cite{Caron-Huot:2022sdy} (see also \cite{Costa:2012cb}).
They are only nonvanishing for $q+k\leq \j$ and we normalize them by the leading power of $u$: $\Qmat_{0,\j}=(-1)^\j$.
Note that here we have specialized formulas to $\Delta_\phi^{\rm eff}=4$ and $d=4$ and we refer to that reference for more general formulas.

For our application, the key property of the
Polyakov-Regge blocks is that they enjoy double zeros for $\Delta-J=8+2m$ with $m=0,1,2\ldots$, explicit from the $1/\Gamma\big(\tfrac{8-\Delta+\j}{2}-n\big)^2$ factor. These ensure the decoupling of all double traces in the planar limit
(notice the shift $\Delta+4$ in \eqref{PR Mellin explicit} related to supersymmetry).
The fact that poles in the Mellin amplitude are saturated by single traces is a longstanding observation \cite{Fitzpatrick:2011ia}.

\subsection{Dispersive Constraints}
\label{sec:Dispersive Constraints}

The difference between Polyakov-Regge expansions in different channels must vanish, which give rises to crossing relations known as \emph{dispersive sum rules} which constrain the OPE data. 
The Polyakov-Regge blocks manifest $u\leftrightarrow v$ symmetry (corresponding to $\mS\leftrightarrow \mT$ in Mellin space), however $\mS\leftrightarrow \mU$ crossing is non-trivial and amounts to infinite numbers of constraints,
\begin{equation}
  0 = \sum_{(\Delta,\j)\ \rm long} \OPE_{\Delta,\j} X_{u,v}[\Delta,\j]
 \quad\mbox{with}\quad X_{u,v}\equiv \cP^{\Nfour}_{u,v} -u^{-4}\,\cP^{\Nfour}_{1/u,v/u}\,,
  \quad\mbox{for $(u,v)$ Euclidean}.
\end{equation}
The equivalent constraints in Mellin space read
\begin{equation}\label{eq:XstMellin}
   0 = \sum_{(\Delta,\j)\ \rm long} \OPE_{\Delta,\j} \widehat{X}_{s,t}[\Delta,\j]
 \quad\mbox{with}\quad  \widehat{X}_{\mS,\mT}\equiv \widehat{\cP}_{\mS,\mT}^{\Nfour}-\widehat{\cP}_{16-\mS-\mT,\mT}^{\Nfour}\,.
\end{equation}

In addition we also impose the so-called antisubtracted sum rules, which
implement the extra constraints from the good Regge behavior of the reduced correlator:
\begin{equation} \label{Bt}
    0=\widehat{B}_\mT^{\rm protected}+\sum_{\Delta,J}\OPE_{\Delta,\j}\widehat{B}_\mT[\D,\j]
\end{equation}
where we have
\begin{equation}\label{eq:BtMellin}
    \begin{split}
        \widehat{B}_\mT^{\rm protected}=&\frac{2}{(\tfrac{\mT}{2}-3)(\tfrac{\mT}{2}-2)},\\ \widehat{B}_\mT[\D,\j]=&\sum_{n=0}^{\infty}\frac{2(\D-\j+2n)+2-\mT}{\mT-6}\mathcal{Q}^{n}_{\Delta+4,\j}(10-\mT)
    \end{split}
\end{equation}
We also use the position space version of this sum rule, $B_v$:
\begin{equation}
    0=B_v^{\rm protected}+\sum_{\Delta,J}\OPE_{\Delta,\j}B_v[\D,\j] \qquad (v>0 \,, \rm real)
\end{equation}
where
\begin{equation}
    B_v^{\rm protected}=\frac{v^2-1-2v\log v}{v(1-v)^3}\,.
\end{equation}
Physically, the $B$ sum rules relate
stress-tensor exchanges at low energy (graviton exchange in the bulk) to the
high-energy spectral density. In addition, we use specific infinite linear combinations of the $B$ functionals, called
$\Psi_{\ell}$ and $\Phi_{\ell,\ell+2}$,
designed to diagonalize the action on operators of specific spin and for twist near two. We refer to \cite{Caron-Huot:2022sdy} for full details.

Each of the above functionals can be written in the form:
\begin{equation} \label{W generic}
    0=W^{\rm protected}+\sum_{\Delta,J}\OPE_{\Delta,\j}W[\D,\j],
\end{equation}
where for $W=\Phi_{\ell,\ell+2}$ or $\Psi_{\ell}$ the protected parts read:
\begin{equation}
    \Phi^{\rm protected}_{\ell,\ell+2}=0, \qquad \Psi^{\rm protected}_{\ell}=-\frac{2\Gamma(\ell+3)^2}{\Gamma(2\ell+5)}\,.
\end{equation}
This concludes the list of dispersive functionals from \cite{Caron-Huot:2022sdy}, which we used in this paper. To produce a finite list of functionals to use in numerical bootstrap, we sample a range of values. We start from the crossing-symmetric points: $z=\bar{z}=-1$ for position-space crossing $X_{u,v}$, $\mS=\mT=\mU=16/3$ for Mellin-space crossing $\widehat{X}_{\mS,\mT}$, and $v=1$ or $\mT=5$ for $B_v$ and $\widehat{B}_\mT$ respectively.
In addition, in this paper we enrich this menu by adding two integrated constraints.

\subsection{Integrated constraints from supersymmetric localization}
\label{sec:Integrated constraints}

Supersymmetric localization enables to calculate exactly the partition function of sYM theory on $S^4$ with certain deformation turned on.
Taking derivatives with respect to deformation parameters then predicts certain correlation functions integrated over the sphere.

Two such constraints have been obtained for the stress tensor four-point correlator that we study. We quote them from eqs.~(2.15)-(2.16) of \cite{Chester:2020dja}, after adapting to our conventions
($\cH(u,v)^{\rm here}=u^{-2}f(u,v)^{\rm there}$) and flipping the sign of $I_4$ to be consistent with later publications from the same authors \cite{Chester:2021aun}:
\begin{subequations}
\label{I24}
\begin{align} \label{Ie}
 I_2(g) &= -\frac{1}{2\pi^2} \int d^4x\ \mathcal{H}(u,v)\big|_{u=x^2,v=(e-x)^2}\,, \\
 I_4(g) &= -\frac{16}{2\pi^2} \int d^4x\ (1+u+v)F_1(u,v)\mathcal{H}(u,v)\big|_{u=x^2,v=(e-x)^2} 
\label{I4}
\end{align}
\end{subequations}
where $e^\mu$ is an arbitrary (constant) unit vector and $F_1$ is the special function in \eqref{F1}.
To implement these constraints,
we insert the Mellin representation \eqref{eq:mellinRep} into \eqref{I24} and integrate over $d^4x$ analytically.
This was carried out for $I_2$ in \cite{Alday:2021vfb} and we described it for $I_4$ in appendix \ref{app:Integrated Mellin}. Using crossing symmetry of $\widehat{\cH}$ to simplify the latter result, this allows to rewrite \eqref{I24} equivalently as:
\begin{subequations}
\label{cI24}
\begin{align}
\label{cI2}
     I_2(g) &= -\frac12\iint\!\!\frac{d\mS\,d\mT}{(4\pi i)^2}
\widehat{\cH}(\mS,\mT) \Gt, \\
\label{cI4}
I_4(g) &=
{-}48\iint\!\!\frac{d\mS\,d\mT}{(4\pi i)^2}\,\widehat{\cH}(\mS,\mT)\Gt\left[
\frac{2(\mU- 5)}{(\mS- 6) (\mT - 6)}+
\frac{\mT-\mS}{\mU-6}\left( H_{\frac{\mS}{2}-3}+H_{3-\frac{\mS}{2}}\right)\right]  ,
\end{align}
\end{subequations}
where $\Gt=\prod_{x=\mS,\mT,\mU} \Gamma\big(\tfrac{x}{2}-2\big)\Gamma\!\left(4-\tfrac{x}{2}\right)$ is a product of six Gamma functions and $H_a=\Gamma'(a+1)/\Gamma(a+1)-\Gamma'(1)$ are analytically continued hamornic sums.  The integration contours must satisfy $4\leq {\rm Re}\, \mS,\mT,\mU\leq 6$.

Localization predicts the left-hand-sides of these integrated correlators for any $N_c$ and $g$ in terms of integrals over the eigenvalues of SU($N_c$) matrices. In the 't Hooft limit the eigenvalues condense and the formulas simplify to involve integrals over the (Fourier transform of) eigenvalue distributions.  Combining (A.8), (A.17) and (A.24) of \cite{Chester:2020dja} we find\footnote{We used $\omega^{\rm there}=\frac12\,t^{\rm here}$ to make the integrals more similar to formulas from the integrability context \cite{Beisert:2006ez}.}:
\begin{subequations}
\label{eq:I-prot}
\begin{align}
I_2(g) &= \int_0^\infty \frac{t dt e^{-t}}{(1-e^{-t})^2}\left( J_1(2gt)^2-J_2(2gt)^2\right) \\
I_4(g) &= 48\zeta_3-\frac{8}{g^2} \int_0^\infty \frac{t dt e^{-t}}{(1-e^{-t})^2}J_1(2gt)^2 \\\nonumber
&\phantom{=}-\frac{192}{g}\int_0^\infty  \frac{t dt e^{-t}J_1(2gt)}{(1-e^{-t})^2} \int_0^\infty\frac{t' dt' e^{-t'}J_1(2gt')}{(1-e^{-t'})^2} 
\left( \frac{t J_0(2gt)J_1(2gt')-(t{\leftrightarrow}t')}{t'^2-t^2}\right)
\label{I24(g)}
\end{align}\end{subequations}
where as before $g^2=\frac{\lambda}{16\pi^2}$ and $J$ are Bessel functions.

Analytic results for these quantities at weak and strong coupling can be found in \cite{Chester:2020dja},
which we also verified numerically. 
By inserting the Polyakov-Regge expansion \eqref{PR expansion st} into \eqref{cI24} and equating it to  eq.~\eqref{eq:I-prot} we obtain sum rules on single-trace OPE data:
\be\label{integrated constraints}
 0= I_p^{\rm protected}(g) + \sum_{(\Delta,\j)\ \rm long} \OPE_{\Delta,\j} I_p[\D,\j] \qquad (p=2,4),
\ee
where the ``protected''\footnote{We use ``$I^{\text{protected}}_p$'' to keep the notation similar to other sum rules in \eqref{W generic}, even though it is coupling-dependent.} part includes the integral over the supergravity term in \eqref{PR expansion st}, recorded in \eqref{Mellin model}, minus the integrals \eqref{eq:I-prot}:
\be
 I_2^{\rm protected}(g) \equiv \frac14-I_2(g), \qquad I_4^{\rm protected}(g) \equiv 24(2\zeta_3-1) - I_4(g)\,.
\ee
The limits of the $I_p^{\rm protected}(g)$ are then:
\begin{subequations}\label{limits I24}
\begin{align}
\label{limits I2}
 I_2^{\rm protected}(g) &\to \left\{\begin{array}{p{55mm}l}
  $\frac14-6g^2\zeta_3 + O(g^4)$, & g\to 0, \\
  $\frac{3\zeta_3}{(4\pi g)^3}- \frac{45\zeta_5}{4(4\pi g)^5}+O(g^{-7})$ , & g\to \infty, \end{array}\right. \\
 I_4^{\rm protected}(g) &\to \left\{\begin{array}{p{55mm}ll}
  $24(2\zeta_3{-}1)-960g^2\zeta_5 + O(g^4)$, & g\to 0, \\
  $\frac{384\zeta_3}{(4\pi g)^3}-\frac{1152\zeta_5}{(4\pi g)^5}+O(g^{-7})$, & g\to \infty. \end{array}\right.
\label{limits I4}
\end{align}
\end{subequations}
Note that the $I^{\rm protected}$ vanish in the supergravity limit $g\to\infty$, as required by \eqref{integrated constraints} and the fact that all long single-trace operators become heavy and decouple. The two integrated constraints thus effectively determine the two leading corrections (contact interactions) to supergravity.
(More precisely, we can anticipate $I_2$ and $(I_4-128I_2)\propto \zeta_5$ to give two nondegenerate constraints at strong coupling.)

To summarize, the two sum rules \eqref{integrated constraints} define the integrated constraints for our purposes, 
in terms of unprotected single-traces to leading order in the large-$N_c$ limit.  A method to rapidly evaluate the contribution $I_p[\D,\j]$ of individual Polyakov-Regge blocks is detailed below \eqref{Ip app}.

\subsection{Flat space limit and OPE coefficients at strong coupling}\label{sec:OPEandFlat}

In section~\ref{sec:flat-space} we will explain how the CFT bootstrap in the holographic limit (large `t Hooft coupling) reduces to an analogous flat-space S-matrix problem.    Here we review well-understood kinematical aspects of the relation between CFT and S-matrix data \cite{Gary:2009ae,Paulos:2016fap,Li:2021snj}, using simple heuristic arguments that produce the correct normalization factors.

The common starting point is radial quantization of the CFT$_d$, or equivalently, the theory placed on the Lorentzian cylinder $R{\times}S^{d-1}$.
We create a scattering state in AdS${}_{d+1}$ by inserting two local operators at the same time on the boundary cylinder. In the CFT, the resulting (primary) states can be classified by their scaling dimension and spin $\Delta,\j$.  The basic idea is that these quantum numbers map simply to the center-of-mass energy and angular momentum of the bulk scattering process: $\sqrt{s}=\Delta/\RAdS$, while $\j$ is the same.  Flat space physics is probed by intermediate operators with $\Delta\gg 1$.

We first consider the situation with no scattering, by pairing the above state with a copy of itself at a nearby time.  By applying the Euclidean OPE to this four-point function, is shown that the density of OPE coefficients squared of any CFT must match, at large $\Delta$ and fixed $\j$, that of an uninteracting theory \cite{Mukhametzhanov:2018zja}:
\be\label{Euclidean OPE smeared}
  \sum_{\Delta}
  \frac{\OPE_{\Delta,\j}}{\OPEfree_{\Delta,\j}} (\cdots)\simeq
  \int \frac{d\Delta}{2} (\cdots) ,
\ee
where $(\cdots)$ represents any sufficiently smooth function of $\Delta$.
The $\frac12$ ensures the correct match in generalized free field theory, where the spacing between double-twist operators is $2$, however we stress that the sum on the left includes all operators.
This result is proved by comparing with identity exchange in a cross-channel; corrections by inverse powers of $\Delta$ can be predicted systematically from exchanges of light operators \cite{Mukhametzhanov:2018zja}.

To discuss scattering, we now evolve the
initial state by a time $\pi$ and rotate it by $\pi$, which adds a phase to left-hand-side. The idea is that the above must give the correctly normalized ``no scattering'' part of the S-matrix, $S_\j^{\rm flat}(s)=1+ia_\j^{\rm flat}(s)$, and therefore:\footnote{Our normalizations are such that unitarity implies $|S^{\rm flat}_\j|\leq 1$}
\be\label{flat OPE smeared}
  \sum_{\Delta}
  \frac{\OPE_{\Delta,\j}}{\OPEfree_{\Delta,\j}} e^{-i\pi(\Delta-\j-2\Delta_\phi)}(\cdots)\simeq
\int \frac{d\Delta}{2}\ \left(1+i a_\j^{\rm flat}(s)\right)_{s=\Delta^2/\RAdS^2}(\cdots)\,.
\ee
The ``amplitude'' defined by the above relations enjoys interesting properties, such as analyticity and dispersion relations \cite{vanRees:2023fcf}, which are inherited from corresponding exact CFT statements.
For example, the CFT version of the dispersion relation which reconstructs the amplitude from its imaginary part, for example, reconstructs the correlator from its ``double-discontinuity'' \cite{Carmi:2019cub}.
The double discontinuity admits an OPE with coefficients multiplied by trigonometric factor \cite{Caron-Huot:2017vep,vanRees:2023fcf}:
\be\label{dDisc OPE smeared}
  \sum_{\Delta}
  \frac{\OPE_{\Delta,\j}}{\OPEfree_{\Delta,\j}} 2\sin^2\big(\tfrac{\pi(\Delta{-}\j{-}2\Df)}{2}\big)(\cdots)\simeq
    \int \frac{d\Delta}{2}
    \,{\rm Im}\,a_\j^{\rm flat}(s)_{s=\Delta^2/\RAdS^2} (\cdots)\,.
\ee
In the large-$N_c$ limit, this implies the following simple relation between the
residue of an S-matrix pole (or narrow resonance) at $s=m^2$ and finite spin, 
and the OPE coefficient of single-trace operator with large dimension $\Delta=m\,\RAdS$:
\be
 a_\j^{\rm flat}(s) \approx \frac{C^2_{m,\j}}{m^2-s-i0} \quad\Longleftrightarrow\quad
 2\sin^2\big(\tfrac{\pi(\Delta{-}\j{-}2\Df)}{2}\big)\frac{\OPE_{\Delta,\j}}{\OPEfree_{\Delta,\j}}
  = \frac{\pi \RAdS}{4\,m}C^2_{m,\j}.
\label{flat OPE}
\ee
This relation will enable us to simply compare the CFT and S-matrix problems.
In this context, we will treat the intermediate state as heavy, and generally expect the approximations to work up to $1/(m\RAdS)^2$ curvature corrections.

Let us specialize to $\mathcal{N}=4$ sYM theory. The free OPE coefficients $\OPEfree_{\Delta,\j}$ for long double-trace operators are given by:
\be\label{OPEfree N4}
 \OPEfree_{\Delta,\j} = 2(\Delta+2)(\j+1)
 \frac{\Gamma\big(\tfrac{\Delta-\j}{2}+1\big)^2\Gamma\big(\tfrac{\Delta+\j}{2}+2\big)^2}{\Gamma(\Delta-\j+1)\Gamma(\Delta+\j+3)},
\ee
which is a priori defined only for integers $\Delta=4+\j+2n$ but viewed here as a smooth function of $\Delta$.
In the CFT, we consider the correlator of identical complex scalars given by $u^2{\cal H}$ in \eqref{G ansatz} and satisfying a conventional OPE. In the bulk, this represents the scattering of gravitons with polarizations along the $S^5$ whose amplitude is given as $s^4\cM(s,t)$ where
$\cM$ is a crossing symmetric super-amplitude. In the flat space limit this is given by the Virasoro-Shapiro amplitude restricted to five dimensions.

\paragraph{Virasoro-Shapiro amplitude}
As a cross-check, let us apply the above to the Virasoro-Shapiro (super)-amplitude:
\be\label{eq:MVSflat}
 \cM(s,t) = \frac{8\pi G_5}{stu}
 \frac{\Gamma\big(1-\tfrac{\alpha's}{4}\big)\Gamma\big(1-\tfrac{\alpha't}{4}\big)\Gamma\big(1-\tfrac{\alpha'u}{4}\big)}
 {\Gamma\big(1+\tfrac{\alpha's}{4}\big)\Gamma\big(1+\tfrac{\alpha't}{4}\big)\Gamma\big(1+\tfrac{\alpha'u}{4}\big)}\,.
\ee
Taking the residue of the first pole (at $s=4/\alpha'$) and applying \eqref{C2 integral flat} with $d=4$ ($D=5$) gives the residue defined in \eqref{flat OPE}:
\be
 C^2_{2/\sqrt{\alpha'},0} =\frac{2G_5}{(\alpha')^{\frac52}},
\ee
where $G_5$ is the five-dimensional Newton's constant.
Using the AdS/CFT dictionary,
\be
 8\pi G_5=\frac{\pi^2 \RAdS^3}{c}, \qquad \alpha'=\RAdS^2\lambda^{-\frac12},
\ee
eq.~\eqref{flat OPE} then predicts the leading behavior of the Konishi OPE coefficient at strong coupling: 
\be\label{Konishi strong prediction}
 \frac{2\sin^2(\tfrac{\pi\Delta_\Konishi}{2}\big) \OPE_\Konishi}{\OPEfree_{\Delta_\Konishi,0}} =
 \frac{\pi^2\lambda}{32c}\left(1+O(\lambda^{-\frac12})\right)\,. \  
\ee
Thus is in perfect agreement with \cite{Minahan:2014usa,Goncalves:2014ffa,Alday:2022uxp,Alday:2023mvu}. The latter references also provided sub-leading corrections which we review below. There we also highlight the benefits of normalizing by $\OPEfree_{\Delta_\Konishi,0}$ to reorganize the strong coupling series.

\subsection{Konishi normalization and analytic results at weak and strong coupling}
\label{sec:analyticKonishi}

The scaling dimension of the lightest unprotected scalar at strong coupling is given by\footnote{See \cite{Ekhammar:2024rfj} for the latest analytic results at strong coupling.}:
\beq \label{strong Konishi dim}
\left(\Delta_K+2\right)^2 = 4\lambda^{1/2} + 8 +\frac{6- 12\zeta_3}{\lambda^{1/2}} + \frac{4+12 \zeta_3 + 30 \zeta_5}{\lambda} + O(\lambda^{-3/2})
\,,
\eeq
with $\lambda \equiv 16\pi^2 g^2$, while its OPE coefficient has been expressed as \cite{Alday:2022uxp,Alday:2023mvu}:
\bba \label{strong Konishi OPE}
\lambda_{K}^2 &= \frac{\pi^3 \Delta_K^{6} 4^{-\Delta_K-6} }{\sin^2(\pi \Delta_K/2)}\,\left(1+f_1 \lambda^{-1/4}+f_2 \lambda^{-1/2} + f_3 \lambda^{-3/4} + f_4 \lambda^{-1}+O(\lambda^{-5/4})\right)
\end{align}
with the coefficients $f_1 = \frac{23}{4}$, $f_2 = \frac{405}{32}+2\zeta_3$, $\ldots$ given in this reference.

The appearance of $(\Delta_K+2)^2$ in \eqref{strong Konishi dim} is natural since the superconformal Casimir is invariant under the shadow symmetry $\Delta_K\mapsto -4-\Delta_K$.\footnote{The usual shadow symmetry in $d$ spacetime dimensions is $\Delta\mapsto d-\Delta$ but we recall the shift by 4 in the superconformal blocks \eqref{PR Mellin explicit}.}  
Note that the series for $(\Delta_K+2)^2$ itself contains only integer powers of $\lambda^{1/2}$, which is to be contrasted with the series for $\Delta_K$, which includes both odd and even powers of $\lambda^{1/4}$.

In a similar fashion, we will now confirm that the $\lambda^{1/4}$ powers in \eqref{strong Konishi OPE} are automatically removed by simply leveraging the shadow symmetry. 

At the same time, let us introduce a convenient normalization for the Konishi OPE coefficient which will allow us to continuously plot it from weak to strong coupling.
According to the preceding discussion, we should divide by the free theory coefficient in \eqref{OPEfree N4}, times a function symmetrical under $\Delta_K\mapsto -4-\Delta_K$.  There is a rather unique choice which accounts for all the double zeros at
even dimensions $\Delta_K\geq 4$ but remains nonsingular in the weak-coupling limit $\Delta_K\to 2$:\footnote{
Recall from eqs.~\eqref{G ansatz} and \eqref{PR expansion uv} that $\lambda^2_{\Delta,J}$ is the standard OPE coefficient squared multipled by $c$.
}
\begin{equation}\label{eq:Ktilde}
 \tilde{\lambda}_K^2 \equiv
 \frac{2^8\sin^2\left(\frac{\pi}{2}\Delta_K\right)}
 {\big[\frac{\pi}{2}(\Delta_K-2)(\Delta_K+6)\big]^2}
 \frac{\lambda_{K}^2}
 {\lambda^{2,\text{free}}_{\Delta_{K},J=0}}\,.
\end{equation}
At weak coupling we have $\tilde{\lambda}_K^2\to 2+O(\lambda)$, more explicitly:
\beq\label{eq:weaktilde}
\tilde{\lambda}^{2}_{K} = 2 - 10 g^2 +\frac{288\zeta_3 -97}{2}g^{4} + O(g^6) 
\eeq
while at strong coupling we find
that \eqref{strong Konishi OPE} simplifies to:
\begin{align}
\tilde{\lambda}_K^2 &= 1+ \frac{5+8\zeta_3}{4} \lambda^{-1/2} + \frac{13-136 \zeta_3 +32 \zeta_3^2 -48 \zeta_5}{16} \lambda^{-1} + O(\lambda^{-3/2})
\label{eq:KtildeStrongSeries}
\\
&= \exp\left(\frac{5+8\zeta_3}{\left(\Delta_K+2\right)^2} + \frac{\frac{81}{2}-112\zeta_3 -48 \zeta_5}{\left(\Delta_{K}+2\right)^4} +O(\Delta_K^{-6})\right)\,. \label{eq:expK}
\end{align}
In the last line we highlight the exponentiation of the OPE coefficient, manifesting the absence of $\zeta_3^2$. See \eqref{eq:logOPEhigherSpin} for a similar exponentiation on the higher-spin OPE data in the first Regge trajectory. It would be interesting to test if our normalization \eqref{eq:Ktilde} leads to a strong coupling series with simpler transcendental numbers. Numerically, we can anticipate
$\tilde{\lambda}_K^2$ to smoothly interpolate between its limits 2 and 1 as function of the `t Hooft coupling.

\section{Numerical bootstrap}
\label{sec:numericalbootstrap}
In this section, we combine dispersive sum rules, integrated constraints from supersymmetric localization, and spectral information from integrability to bound the OPE coefficient of the Konishi operator and the four-point correlator itself. We focus on a range of the t'Hooft coupling from weak ($g=0.1,\,\Delta_K\approx 2.1$) to strong regime ($g=3.7,\,\Delta_K\approx 12$). 

\subsection{Generalities}\label{sec:OptimizationSetUp}

Let us start by briefly reviewing the numerical bootstrap problem previously discussed in \cite{Caron-Huot:2022sdy}. The idea is to use the menu of dispersive functionals discussed in table.~3 of \cite{Caron-Huot:2022sdy} as well as functionals which implement the integrated constraints discussed in section~\ref{sec:Integrated constraints}. This is in contrast with the traditional functionals which come from the derivative expansion of crossing equations used in the conformal bootstrap. 

\def\Ofunc{O}

Employing these, we set up a linear optimization problem with a menu of functionals each having different behaviour in various limits of spin and twist. Following \cite{Caron-Huot:2022sdy}, if we label each of these functionals $W_k$, we have
\be
 0=W_{k}^{\rm protected} + \sum_{(\Delta,\j)\ \rm long} \OPE_{\Delta,\j} W_{k}[\D,\j].
 \label{many sum rules}
\ee
Now, in this setup one can bound any OPE coefficient, $\OPE_{\Delta,\j}$ as discussed in \cite{Caron-Huot:2022sdy}. Alternatively one can bound any linear combination of OPE coefficients with arbitrary weight $\Ofunc$,
\begin{equation}
    \Ofunc=\sum_{(\Delta,\j)\ \rm long}\OPE_{\Delta,\j}\Ofunc[\D,\j].
    \label{eq:mathcalO}
\end{equation} 

To formulate a bootstrap problem which bounds the observable $O$, conceptually, we simply need
to add the equation defining $O$ as a new functional to the previous problem \cite{Collier:2017shs, Antunes:2021abs, Paulos:2020zxx}.
Following \cite{Cavaglia:2023mmu}, we implement this in a way that simultaneously cancels the Konishi contribution from the OPE sum, by taking the combination $\eqref{many sum rules}\times \Ofunc[K]-\eqref{eq:mathcalO}\times W_k[K]$ as:
\begin{equation}
-W_{k}^{\rm protected}\,\Ofunc[K] -O\,W_k[K]=\sum_{(\Delta',\j')\ \rm long} \OPE_{\Delta',\j'} \left(W_{k}[\D',\j']\,\Ofunc[K]-W_k[K]\Ofunc[\D',\j']\right),
\end{equation}
where we use the notation for the action on the Konishi operator:
$\Ofunc[K]\equiv \Ofunc[\D_K,0]$ and $W_k[K]\equiv W_k[\D_K,0]$. 
The sum no longer contains the Konishi operator and the prime indicates that.
We then proceed by taking finite linear combinations and looking for coefficients $\alpha_k$ such that
\be
\hspace{0mm}\sum_{k} \alpha_{k}\left(W_{k}[\D',\j']\,\Ofunc[K]-W_k[K]\Ofunc[\D',\j']\right)\geq 0\qquad \forall \quad (\Delta',\j') \mbox{ single-traces except Konishi}.
\label{pos}
\ee
To find the optimal bound (for a particular finite list of functionals $\{W_k\}$) we then solve a standard linear optimization problem subject to the conditions:
\begin{equation}
\begin{split}
&\text{maximize} \qquad \sum_{k} \alpha_{k}\,W_{k}^{\rm protected}\times \Ofunc[K] \\
&\text{such that}\qquad \sum_{k} \alpha_{k}W_k[K]=\pm 1.
\end{split}
\end{equation}
The plus (minus) sign yields an upper (lower) bound, respectively.
We use the SDPB solver to efficiently solve this type of problem \cite{Simmons-Duffin:2015qma}. 

We will apply the procedure just discussed for different choices of the observable $O$.
In subsection~\ref{sec:numericsKonishi}, we bound the OPE coefficient of the Konishi operator itself using $\Ofunc[\D_K,0]=1$ and all other $\Ofunc[\D',J']=0$.
In subsection~\ref{sec:corr},  we consider the correlator, for which $\Ofunc[\D,J]=\mathcal{P}_{u,v}^{\Nfour}[\Delta,J]$ according to \eqref{PR expansion uv}.
However, this setup is general and could be used to bound other observables such as the Mellin amplitude \eqref{PR expansion st} with $\Ofunc[\D,J]=\mathcal{P}_{\mS,\mT}^{\Nfour}[\Delta,J]$.  One could also consider sYM analogs of the $\alpha$ and $\beta$ functionals from \cite{Mazac:2019shk,Caron-Huot:2020adz} 
which isolate the anomalous dimension and OPE coefficient of individual double-trace operators.
All of these observables can be written in the form of eq.~\eqref{eq:mathcalO}. We will not discuss all these examples in this manuscript and will focus on position space correlation function.

When imposing the inequality \eqref{pos}, we use information about the spectrum of single-trace operators of the theory from integrability.  We elucidate which data is used and the sensitivity of the bootstrap problem to the spectrum in the next subsection.

\subsection{Spectral input from Integrability}
\label{sec:Spectral assumptions}

The non-protected operators that can appear in the OPE of two stress tensor multiplets consist
of R-charge singlets with even Lorentz spin $J$. Thanks to our focus on the planar limit and the use of dispersive sum rules, we only need spectral information about the single traces. For our bootstrap implementation,  the most important input data are the scaling dimensions of the lightest operators in the leading and second Regge trajectories. Our results will most strongly depend on the scaling dimension of the lightest operator: Konishi, and the gap to the second lightest scalar operator.

We will use numerical data from the quantum spectral curve (QSC) at low spins, together with large spin asymptotics derived from the Asymptotic Bethe Ansatz and BES equation. At larger couplings, we will use a physically motivated interpolation based on the so-called semi-classical GKP string.

\begin{figure}[t]
\centering
\includegraphics[width=.9\textwidth]{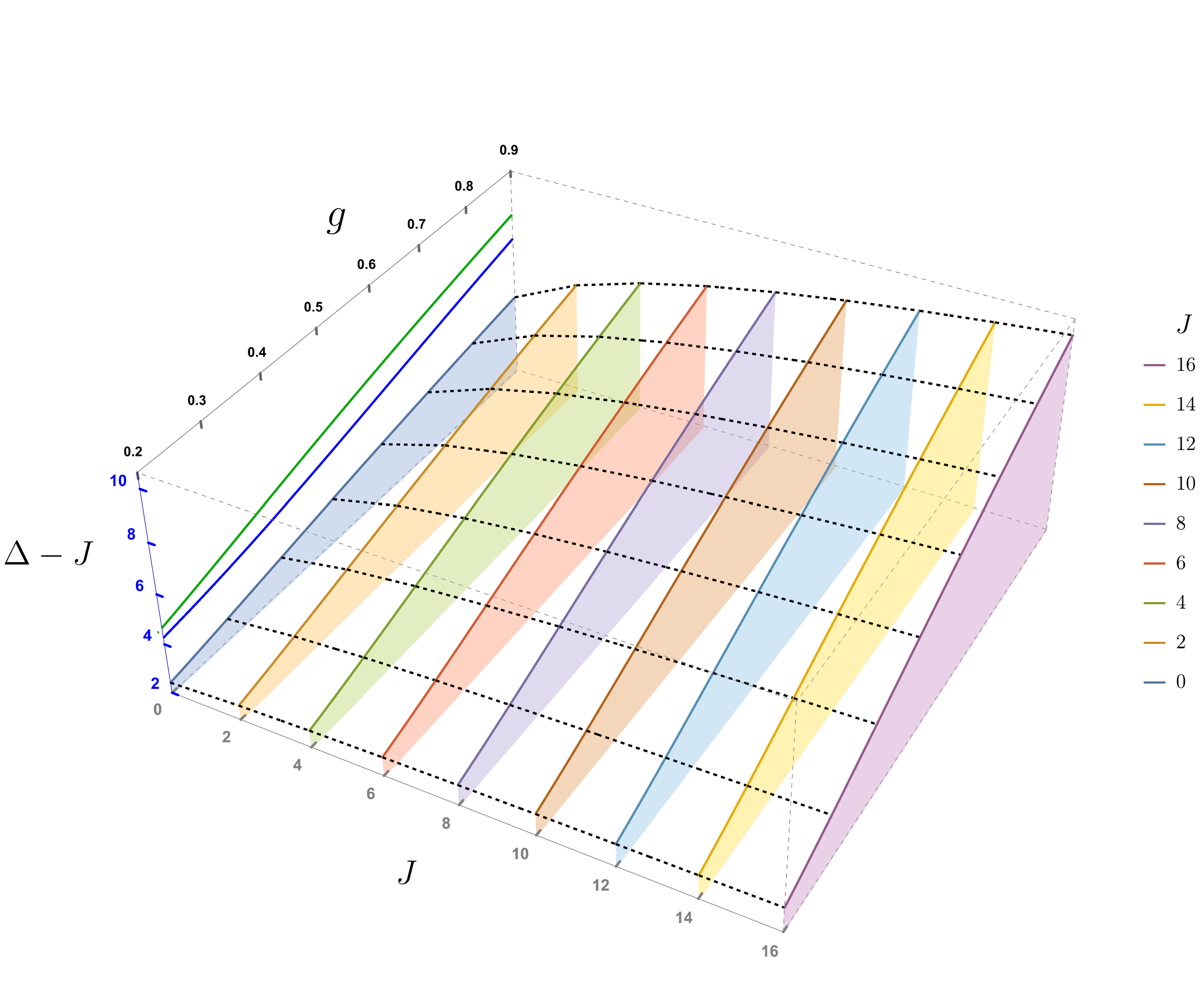}
\caption{Spectral data generated using the quantum spectral curve in \cite{Gromov:2023hzc}. We show the operators in the leading Regge trajectory connected by dashed lines at couplings $g\in\{0.2,\,0.3\,\cdots,0.9\}$. We also show the scaling dimension of the second and third lightest operators with spin $J=0$.
}
\label{fig:QSCspectrum}
\end{figure}

The QSC gives us high-precision numerical data for the scaling dimension of the Konishi operator ($\Delta_K$) and the second lightest scalar operator ($\Delta_{\rm gap}$) in a wide range of the `t Hooft coupling $g\in[0.1,\,3.7]$, corresponding to the range $\Delta_{K}\in [2.1,\,12]$. Moreover, in the range $g\in[0.1,\,1]$, we also use QSC data from \cite{Gromov:2023hzc}\footnote{We are grateful to Julius Julius and Nika Sokolova for producing this data for us using the code from \cite{Gromov:2023hzc}} for the next eight operators in the leading trajectory $J=2,4,\cdots, 16$. While in the range $g\in[1,\,3.7]$ we use QSC data from \cite{Hegedus:2016eop} for the next three spinning operators $J=2,4,6$. See figure \ref{fig:QSCspectrum} for a representation of the QSC data we use in our numerical bootstrap.

To complement this, we use the large spin asymptotics for the leading Regge trajectory and the gap to the second trajectory, see eqs.~2.38 and 2.40 of \cite{Caron-Huot:2022sdy} and references therein. Let us repeat both equations here for completeness $(\tau\equiv \Delta- J)$:
\begin{equation} 
\begin{split}
&\tau(\j)_{\text{twist-2},\, J\to \infty} \,=
 2+ 2\Gamma_{\text{cusp}}(g)\,\log \big(J e^{\gamma_E}\big) \,+\, 2\Gamma_{\text{virtual}}(g) + O(\log^\#\!(J)/J)\\
 &\tau(\j)_{\text{twist-4},\, J\to \infty}=\tau(\j)_{\text{twist-2},\, J\to \infty}+\Delta\tau_{J\rightarrow \infty}
 \end{split}
 \label{large-spin}
\end{equation}
These anomalous dimensions: $\Gamma_{\text{cusp}}(g)$, $\Gamma_{\text{virtual}}(g)$ and $\Delta\tau_{J\rightarrow \infty}$ can be computed at arbitrary coupling as shown in references \cite{Basso:2010in,Basso:2013aha}, see also appendix~A of \cite{Caron-Huot:2022sdy}. 
\begin{figure}[t]
\centering
\includegraphics[width=0.8\textwidth]{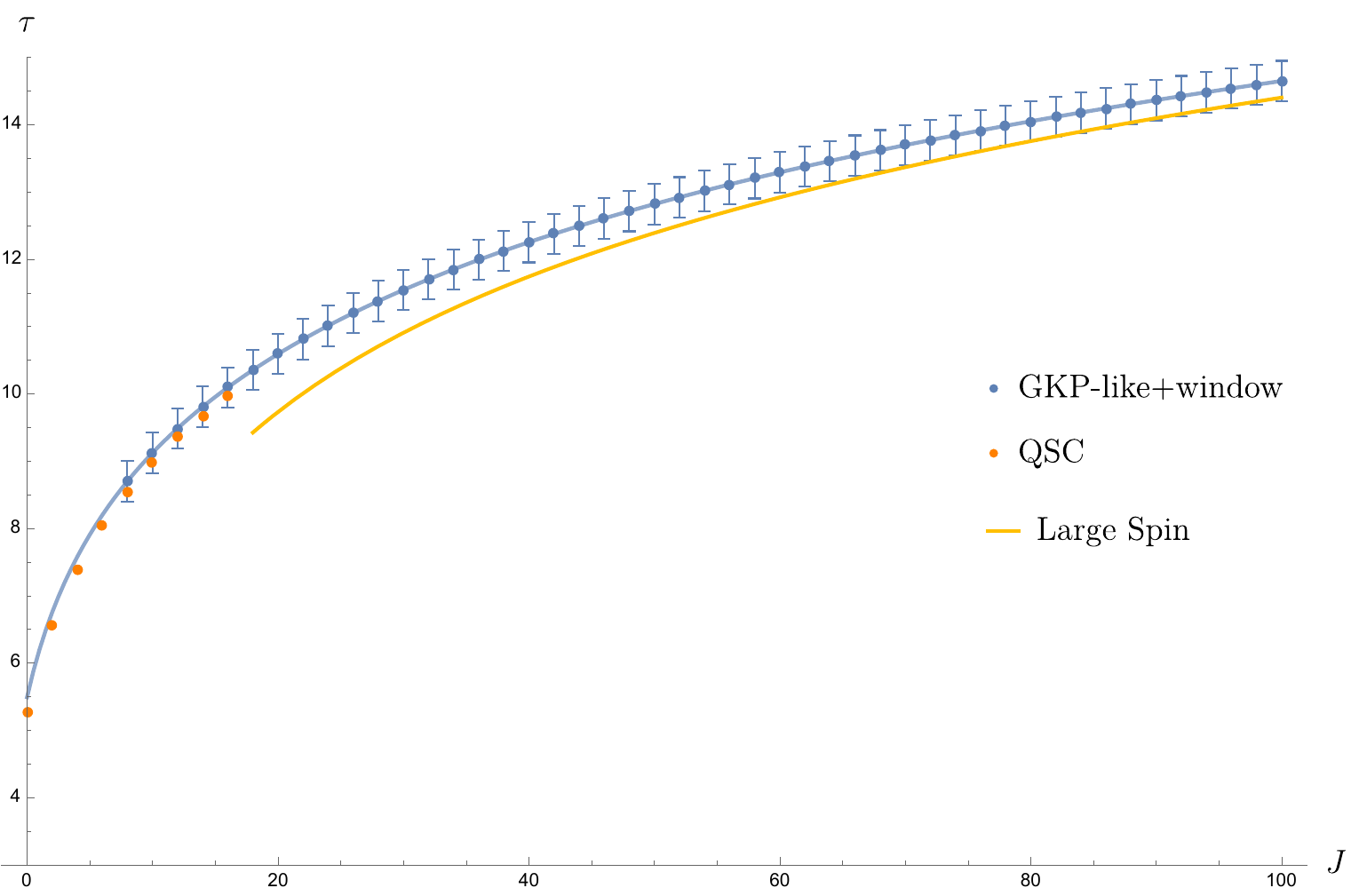}
\caption{The spectrum obtained using the GKP-like model in eq.~\eqref{eq:imp-gkp} and windows around it (blue dots and error bars) is plotted for $g=0.9$. For reference, QSC data (orange points) and the large spin asymptotic curve are shown as well. The GKP-like model predicts the low-spin data surprisingly well. We include conservative windows to account for the difference for $J\geq8$ where we typically rely on the GKP-like curve.}
\label{fig:QSCvsGKP}
\end{figure}

In the range $g\geq1$, where we work with low spin data from QSC ($J=2,4,6$), in order to interpolate between small and large spin on the leading trajectory, we use the GKP approximation at intermediate spin. This GKP regime \cite{Gubser:2002tv} corresponds to operators dual to classical strings in $AdS_5\times S^5$, whose quantum numbers lie in the regime $\Delta\sim \sqrt{\lambda},\,J\sim \sqrt{\lambda}$ (with $\lambda\equiv16\pi^2 g^2$). These strings lie at the equator of $S_3$ with the azimuthal angular velocity being $\omega=\phi/t$, and their energy and spin are given by:
\begin{equation}
    \begin{split}
       \Delta=&8g\int^{\rho_0}_0 d\rho\frac{\cosh^2\rho}{\sqrt{\cosh^2\rho-\omega^2\sinh^2\rho}}\,,\\
        J=&8g\int^{\rho_0}_0 d\rho\frac{\omega\sinh^2\rho}{\sqrt{\cosh^2\rho-\omega^2\sinh^2\rho}}\,.
    \end{split}
\end{equation}
Here $\rho_0$ is the maximal extension of the string in the radial coordinate of global\footnote{with metric $ds^2 = R^2 (-dt^2\cosh^2\rho+d\rho^2 +\sinh^2\rho\,d\Omega_3^2)$} $AdS_5$ and is determined by Virasoro constraints as $\coth^2\rho_0=\omega^2$.
We perform this integral for $\omega\geq 1$ to find $\Delta$ and $J$ for various points. This formula gives us the leading trajectory data $\Delta_{\rm GKP}(J)$
in the limit of strong coupling: $g\to\infty$.
In order to get a reasonable model at finite $g$,
we use $\Gamma_{\text{cusp}}$ as a proxy for the string tension and replace $g$ by $\tfrac12 \Gamma_{\text{cusp}}$
in the above:
\begin{equation}
\label{eq:imp-gkp}
    \Delta_{\rm GKP-like}(J)=\Delta^{g\rightarrow \frac12\Gamma_{\text{cusp}}(g)}_{\rm GKP}(J)+C(g)
\end{equation}
where the constant $C(g)$ is adjusted to match
the large-spin asymptotics given in \eqref{large-spin}.
For $g\gtrsim 0.4$ we find that this parameter-free ansatz agrees quite well with the QSC data even for small spins, as exemplified in fig.~\ref{fig:QSCvsGKP} for $g=0.9$ (there we stop the GKP-like curve at $J=8$).

Despite this evidence, that GKP-like is a good finite-$g$ approximation for the leading trajectory, we still include conservative windows (error bands) around the prediction for the position of the operators in $J$-$\Delta$ plane when QSC data is not available (see blue bars in fig.~\ref{fig:QSCvsGKP}). When using the optimization method of sec.~\ref{sec:OptimizationSetUp}, we demand positivity of the extremal functional throughout these windows, rather than only at the center points. This gives us confidence that the resulting bounds, while possibly not optimal, are rigorous.
A similar method has been used to incorporate analytic bootstrap results in \cite{Su:2022xnj}.

\begin{figure}
    \centering
    \includegraphics[width=0.45\linewidth]{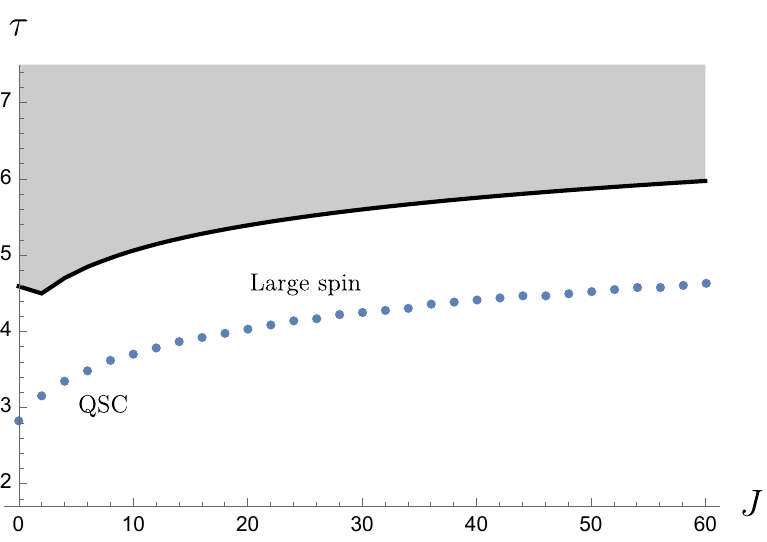}
    \includegraphics[width=0.45\linewidth]{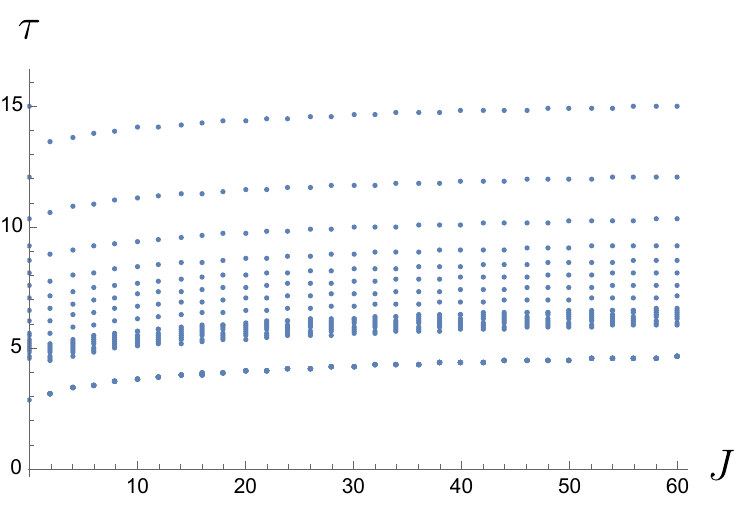}
    \caption{The spectrum used in the bootstrap $g=0.3$. The left hand side shows schematically the regions we would like to cover and the right-hand side shows the states we cover in practice.}
    \label{fig:spectrum}
\end{figure}

In addition to the leading trajectory, we impose positivity on the scalar sector ($J=0$) above the gap set by the second lightest long operator. For spinning operators, we impose positivity above the leading trajectory starting at a gap set by the large-spin gap in eq. \eqref{large-spin}.
Figure~\ref{fig:spectrum}, taken from \cite{Caron-Huot:2022sdy},  shows the regions where we impose positivity. In practice, we only cover a discrete subset of this region, with finite cutoffs in spin and twist. For instance, for $g=0.3$, the region and individual states on which we impose positivity are shown on the right panel of the same figure.  

As is visible in the grid of figure~\ref{fig:spectrum}, our discrete sampling does not reach very large values of the twist, $\tau_{\rm max}\sim 20$ in this example.  
This is sufficient for this study because we added analytic control over the $\tau\to\infty$ limit, which we find greatly stabilizes the numerical problem convergence.

The large-twist limit of each of the functionals introduced in the preceding section are derived in appendix~\ref{app:regge-limit}, using Regge moment technology from \cite{Caron-Huot:2021enk}.
We keep only the leading Regge moment, which
describes the limit of the functionals action on states with large twist and spin with fixed ratio, more precisely with the following quantity fixed:
\be\label{eq:etaAdS}
 \eta_{\rm AdS} \equiv  1+ \frac{2(\j+1)^2}{(\Delta-\j+1)(\Delta+\j+3)}.
\ee
The variable $\eta_{\rm AdS}-1$ can be interpreted as impact parameter of the scattering process in the bulk of AdS spacetime.

Numerically, we thus impose positivity of our functionals for a grid in $\eta_{\text{Ads}}\geq 1$, which automatically ensures positivity at large twist. This in turn is essential for stabilizing the numerical optimization problem once the antisubstracted functionals (the $\widehat{B}_{\mT}$, $B_v$ and derived functionals from section~\ref{sec:Dispersive Constraints}) are included, as anticipated in the discussion section of \cite{Caron-Huot:2022sdy}. This will also be discussed in fig.~\ref{fig:regge-limit} below.
This analytic control alleviates the need to sample the large-$\tau$ region as was done
in our previous paper. On the other hand, analytic control over the limit of large spin with fixed twist has not been derived and could be worth further investigating.

\subsection{Bounds on the Konishi OPE coefficient}\label{sec:numericsKonishi}
\begin{figure}[t]
\centering
\includegraphics[width=1\textwidth]{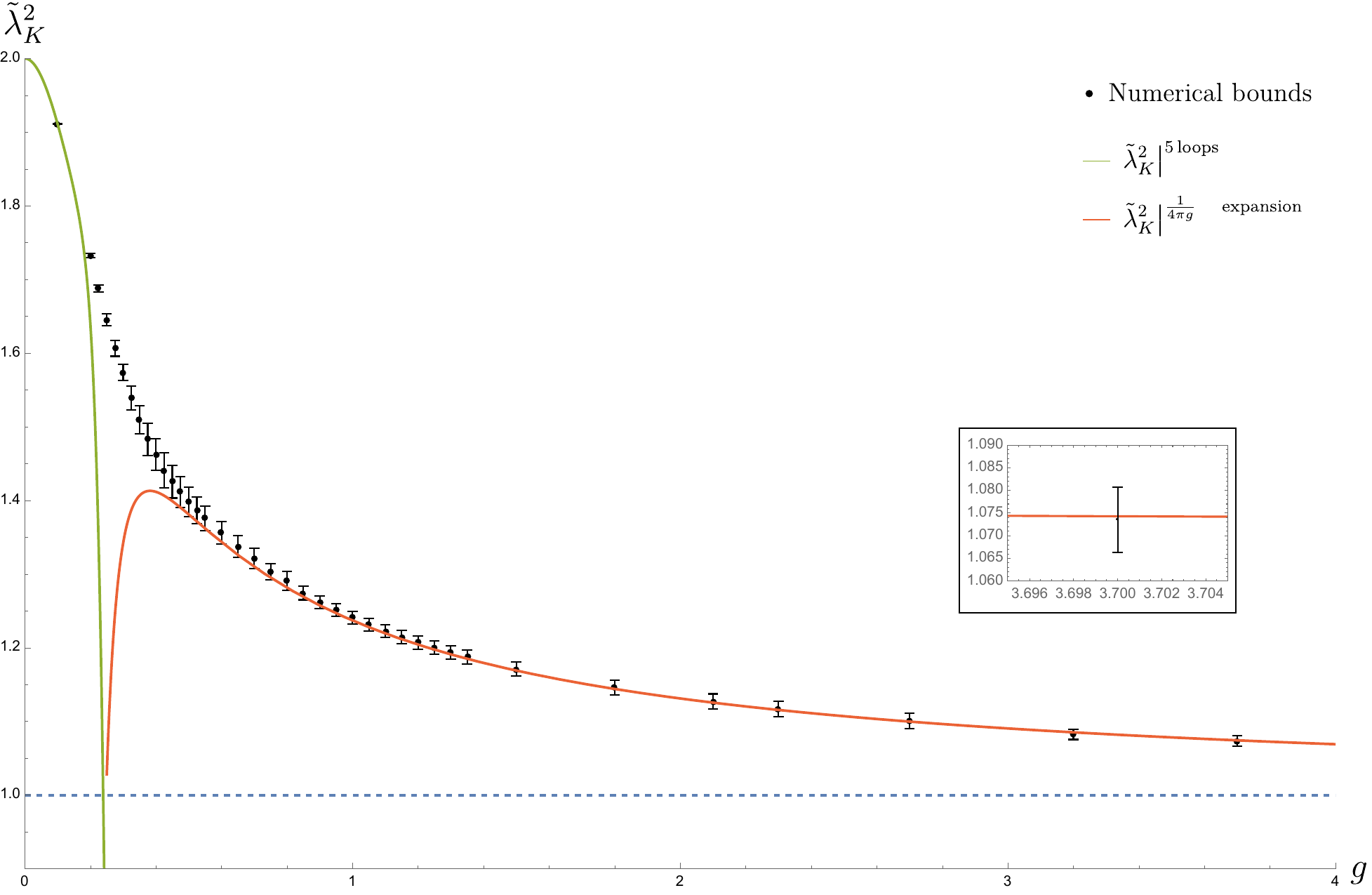}
\caption{OPE coefficient as a function of the coupling, normalized as in \eqref{eq:Ktilde} and recorded in table~\ref{tab:bounds}.  The error bars showing our rigorous lower and upper bounds are compared with the state of the art strong coupling expansion from \cite{Alday:2023mvu} and weak coupling expansion from \cite{Georgoudis:2017meq}. 
}
\label{fig:ope-results}
\end{figure}
In this section, we give upper and lower bounds on the OPE coefficient of the Konishi operator. This corresponds to choosing $\Ofunc[\D_K,0]=1$ and all other $\Ofunc[\D',J']=0$ in the setup of section~\ref{sec:OptimizationSetUp}. In \cite{Caron-Huot:2022sdy}, upper bounds including 40 dispersive functionals were obtained (in eq.~4.9 and eq.~4.12 therein), but no nontrivial lower bounds were found. We will now see that the integrated constraints that we add in this paper make a significant difference and makes it possible to obtain sandwiching two-sided bounds.

The main result of this section is summarised in fig.~\ref{fig:ope-results} which illustrates our bounds for a wide range of the `t Hooft coupling. 
In this plot, we use the tilde normalization introduced in eq.~\eqref{eq:Ktilde}. We also report these bounds in table~\ref{tab:bounds} in appendix \ref{app:tablebounds}. One characteristic of these numerical bounds is the change in the size of the windows between upper and lower bounds as a function of coupling, which represents (rigorous) error estimates. At weak coupling, the windows between the upper and the lower bounds are small. However, they grow as we increase the couplings beyond perturbative regime. As we reach strong enough coupling, they once again shrink. This emphasizes that the region with the least numerical control is the intermediate coupling $g\sim 0.4$.  The tight numerical bounds at weak coupling can be associated with the use of dispersive functionals: $\Psi_0$ and $\Phi_{0,2}$, which were tailored to solve the one-loop problem, as explained in \cite{Caron-Huot:2022sdy}. The simplicity at strong coupling has a different nature and it is related to the fact that in this regime the dual AdS problem is approximated by flat space physics as further explained in section~\ref{sec:flat-space}. 

\begin{figure}[t]
\centering
\includegraphics[width=.75\textwidth]{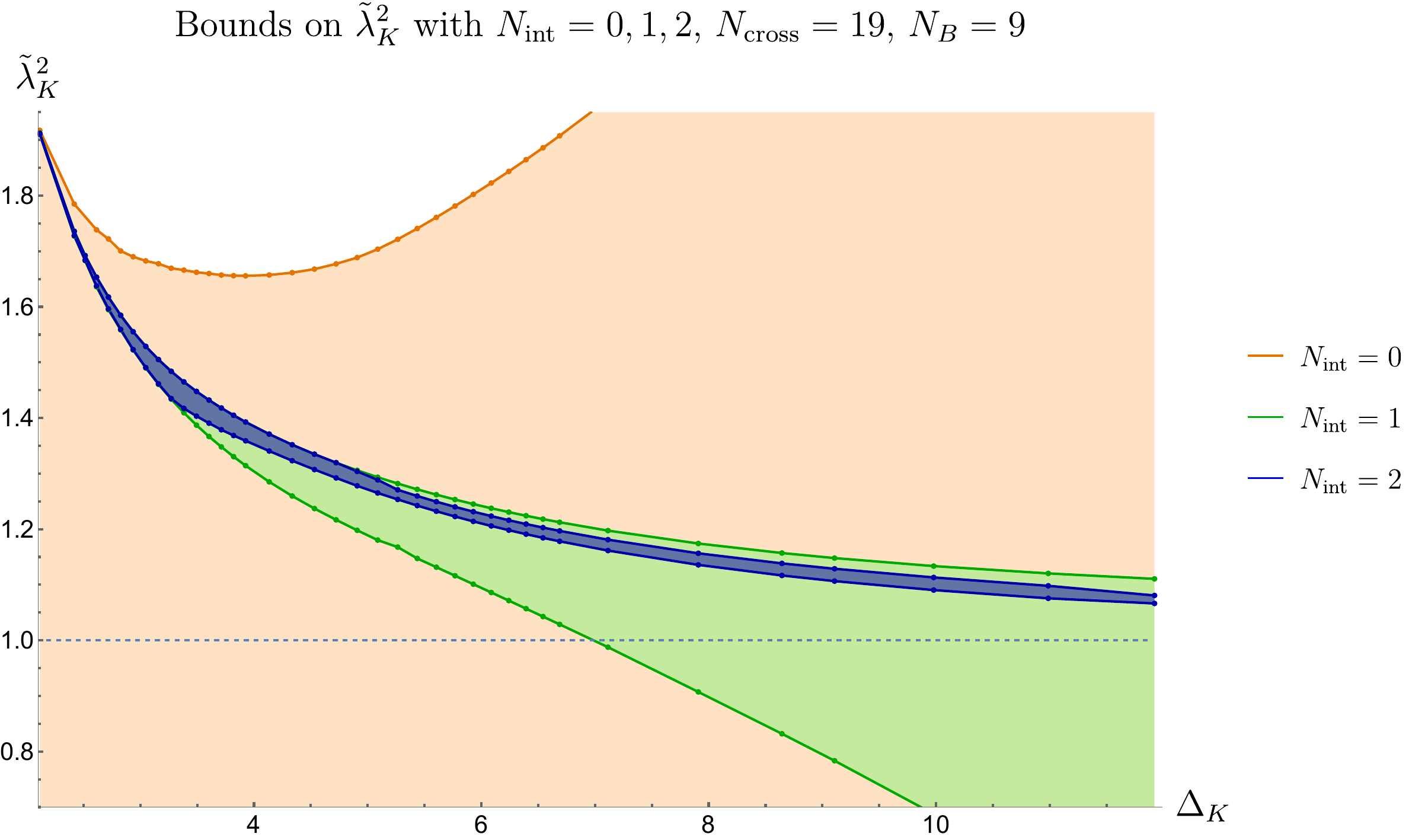}
\caption{The effect of including integrated constraints in the optimization of the Konishi OPE coefficient $\tilde{\lambda}^{2}_{K}$ in relation to 
its scaling dimension $\Delta_K$. Sample points range from $\Delta_{K}\approx 2.1$ to $\Delta_{K}\approx12$ ($g$ from 0.1 to 3.7). For all bounds in the plot we used 19 crossing functionals, plus 9 from the $B$-family: $\Psi_{0}$, four of $\Phi_{l,l+2}$, four of $B_v$, and with $0$, $1$ or $2$ integrated constraints. The bounds for $N_{\text{int}}=2$ shrink to the narrow allowed band shown in blue.}
\label{fig:Nint012}
\end{figure}
\begin{figure}[t]
\centering
\includegraphics[width=0.75\textwidth]{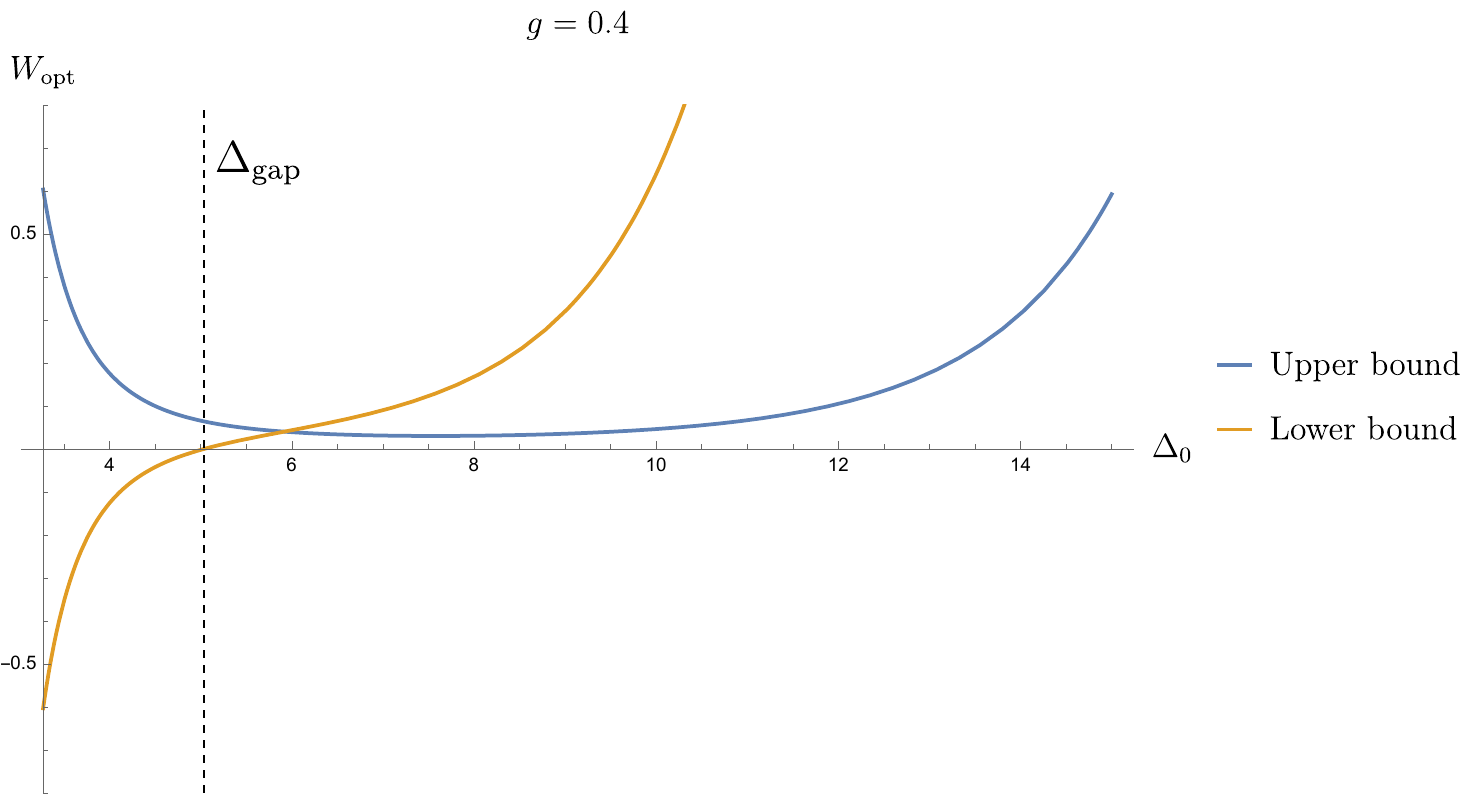}
\caption{Optimized functionals for the upper-bound and lower-bound problem acting on spin-0 states. Since for the lower-bound problem, the action of the functional must be negative on Konishi (at the leftmost end of the plot), the functional needs to change sign at some $\Delta$. This happens right at $\Delta_{\rm gap}.$ For the upper-bound problem acting on the spin 0 states, the optimized functional does not need to change sign and turns out to remain positive within the gap.}
\label{fig:opt-spin0}
\end{figure}
The effect of including integrated constraints yields tight lower bounds at all values of the couplings. It also makes the upper bounds reported in eq.~4.9 and ~4.11 of \cite{Caron-Huot:2022sdy} sharper, see figure \ref{fig:Nint012} for visual demonstration of including one or both integrated constraints on the optimization problem. For instance,  in our tilde normalization,  we have the update: 
\begin{align}
&g=0.1:\qquad\tilde{\lambda}_{K}^2\leq 1.9163 &\Rightarrow& & &\tilde{\lambda}_{K}^2\leq 1.9116 \nonumber\\
&g=0.2:\qquad\tilde{\lambda}_{K}^2\leq 1.747&\Rightarrow& & &\tilde{\lambda}_{K}^2\leq 1.735 \nonumber\\
&g=0.3:\qquad \tilde{\lambda}_{K}^2\leq 1.69 \mbox{ (our previous work)}&\Rightarrow& & &\tilde{\lambda}_{K}^2\leq 1.58 \mbox{ (this paper)}, 
\end{align}


\begin{figure}[t]
\centering
\includegraphics[width=0.65\textwidth]{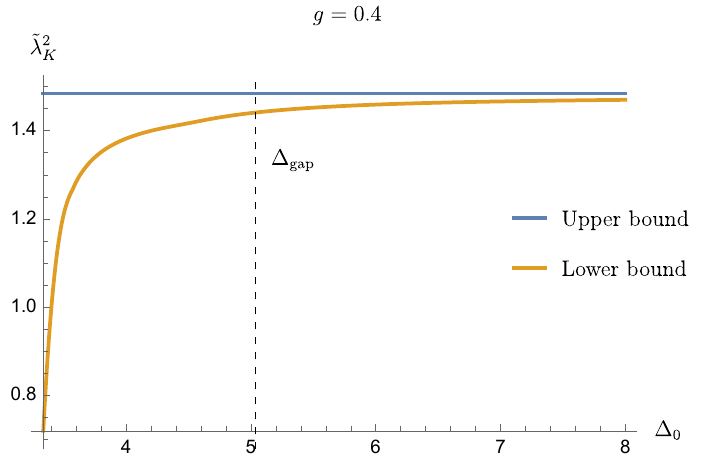}
\caption{Upper and lower bounds resulting from numerical bootstrap with changing spin-0 gap at fix $\Delta_K$ with $g=0.4$. The starting point on the horizontal axes is at  $\Delta_K+0.1\approx 3.37$. The vertical dashed line marks the QSC value: $\Delta_{\rm gap}\approx 5.04$. We see that if we impose positivity below $\Delta_{\rm gap}$, we get a weaker lower bound. In addition, we see that the accuracy of the numerical problem is not enough to rule out theories with gaps much larger than $\Delta_{\rm gap}$. For the upper bound the situation is more trivial and its value seems to be insensitive to the gap at this precision level.}
\label{fig:boundspin0gap}
\end{figure}

\begin{figure}[t]
\centering
\includegraphics[width=1\textwidth]{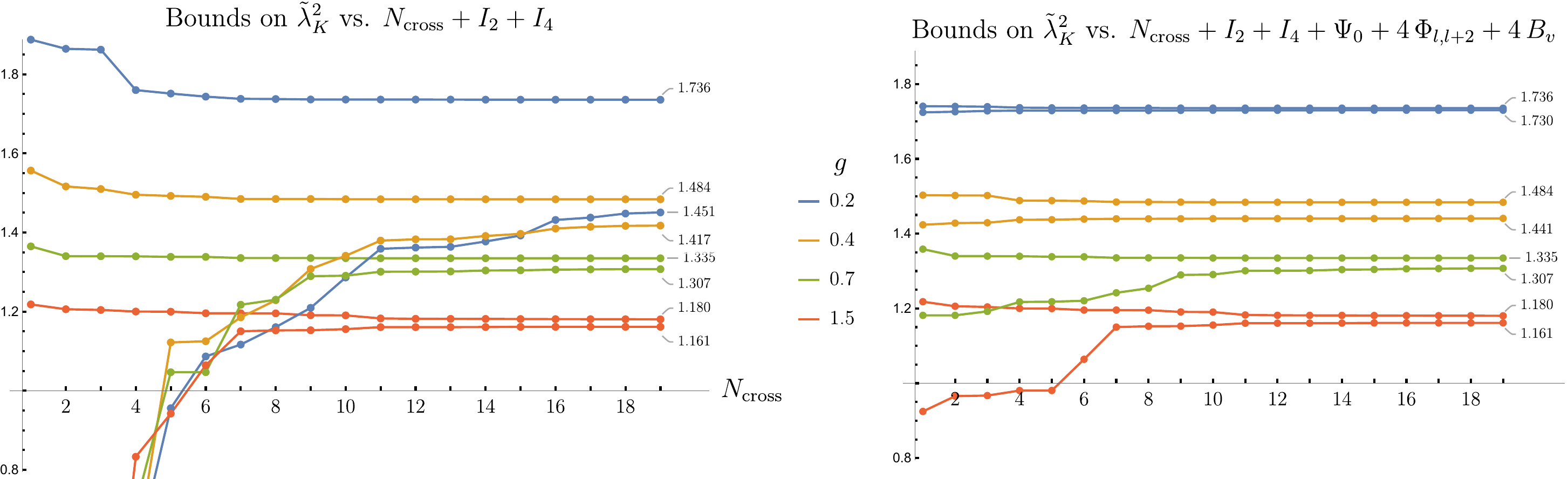}
\caption{Numerical upper and lower bounds on $\tilde{\lambda}^{2}_{K}$ and their dependence on the number and type of functionals we use in the optimization problem. In all plots, we use two integrated constraints: $I_{2}$ and $I_4$. On the left-hand plot, we add only the $X$-family of crossing functionals counted by $N_{\rm cross}$. For large couplings, the plateau is achieved with a low value of $N_{\rm cross}$, while for weaker coupling we require higher numbers to obtain the optimal bound. On the right-hand plot we include the B-family of functionals and the difference is remarkable on the weak and intermediate coupling regime $g\leq 0.4$, where the final plateau is already obtained by adding a few number of crossing functionals. The bounds indicated on the right plot correspond to $N_{\text{cross}}=19$ and are equal to the optimal bounds reported in table \ref{tab:bounds}.} 
\label{fig:boundsNcross}
\end{figure}

\begin{figure}[t]
\centering
\includegraphics[width=0.8\textwidth]{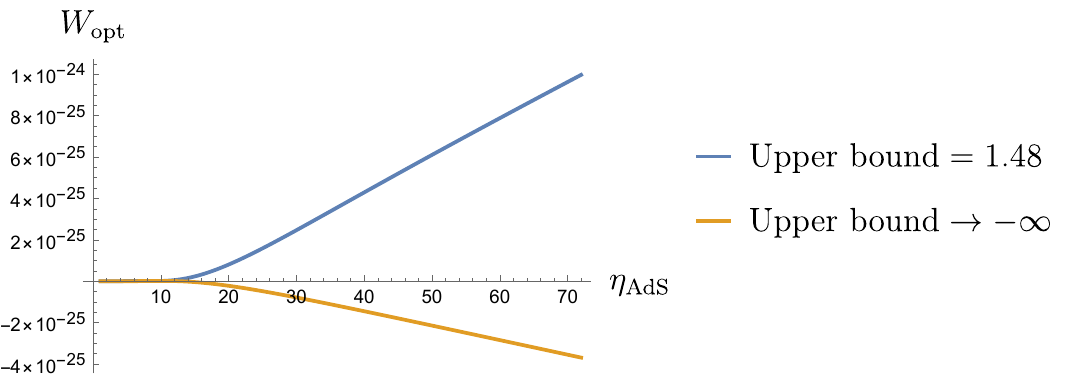}
\caption{Two optimized functionals obtained from linear optimization for finding upper bounds at $g=0.4$. The blue one corresponds to the problem where no positivity is imposed in the Regge limit and does not yield a finite upper bound and the yellow one giving finite upper bound is obtained once positivity in the AdS impact parameter space is imposed.}
\label{fig:regge-limit}
\end{figure}

It is worthwhile to illustrate the action of optimized functionals for lower bounds and upper bounds on the spin-0 states as they behave very differently. This manifests the gap dependence of the numerical bootstrap problem. We plot this for our benchmark case: $g=0.4$ in fig.~\ref{fig:opt-spin0}. The difference in the behavior for these two different optimized functionals is that for the lower-bound problem, the functional on Konishi must be negative, and the functional needs to change sign at some $\Delta$ which happens right at $\Delta_{\rm gap}$. Whereas for the upper-bound case, the functional is positive on Konishi and all other spin-0 states and it never needs to change sign. So the value of $\Delta_{\rm gap}$ is less important in this setup. We can also see this in the dependence of the bounds on $\Delta_{\rm gap}$. In fig.~\ref{fig:boundspin0gap}, we see that if we change $\Delta_{\rm gap}$, the lower bound changes whereas the upper bound stays the same within our accuracy. Note that, here the accuracy of the numerical problem is not enough to rule out theories with gaps much larger than $\Delta_{\rm gap}$ as the lower bound never becomes larger than the upper bound. We have also explored the dependence of our bounds on the position of the higher-spin spectrum and concluded this is subdominant, specially at stronger coupling.

\paragraph{On bounds convergence}
The results presented in tab.~\ref{tab:bounds} are obtained with only 30 functionals. However, it is important to note that the stability of the bounds against adding more $B$ functionals (anti-subtracted) and crossing functionals (unsubtracted) has been checked extensively. This is illustrated in fig.~\ref{fig:boundsNcross}. On the left plot, we only include the integrated constraints and follow how the bounds change as we increase the number of crossing functionals $N_{\text{cross}}$ from 1 to 19. 
We show that for strongish coupling, $g=0.7$ and $g=1.5$, both upper and lower bounds start converging after adding a few number of crossing functionals ($N_{\text{cross}}\approx 10$ for lower-bound convergence). However, at weaker couplings, we only have a convergent upper bound. The lower bound at $g=0.2$ has not yet reached the optimal value after using 19 crossing constraints and at $g=0.4$, it only approaches it after including at least 17 of them. On the right plot, we add the B-family of functionals including: $\Psi_0$, $\Phi_{l,l+2}$, and $B_v$. We see that the behaviour at strong coupling for $g=0.7$ and $g=1.5$ does not change much, still reaching the same final plateau as when using only a few crossing functionals. But at weak and intermediate coupling, for $g=0.2$ and $g=0.4$, we see that with this B-family the lower bound now fully converges and raises to a plateau narrowly close to the upper bound.  Clearly the numerical bound exhibits stability with changing the number of functionals.  For this reason, we believe that the recorded bounds have already converged.

\paragraph{On positivity in the Regge limit}
It is also instructive to study how our optimal functional behaves in the Regge limit. This is as discussed in appendix~\ref{app:regge-limit}. Due to non-decaying nature of $B$-functionals in the Regge limit, special limits of these functionals specifically engineered for this region, are essential to stabilize the numerical problem. In fig.~\ref{fig:regge-limit}, we explicitly see that if $B$ functionals are used, not imposing the Regge limit constraints causes the numerical problem to diverge and not yield a bound due to the fact that the optimized functional can become negative at large AdS impact parameter $\eta_{\text{AdS}}\geq 20$, see eq.~\eqref{eq:etaAdS} for its definition. 

We stopped our numerical analysis at $g=3.7$ ($\lambda\simeq 2160$ and $\Delta_K\approx 12$) where we already approach the strong coupling asymptotics (straight line at 1 in fig.~\ref{fig:ope-results}) predicted by the flat-space limit of the AdS dual, see sec.~\ref{sec:OPEandFlat}. However, there is no obstruction in pushing our numerics to stronger coupling. In fact in sec.~\ref{sec:numericalCFTvsFlat} we study this regime at $g=10$ and $g=100$, in order to compare with the numerical flat-space bootstrap.  There we explain this comparison in detail and provide the explicit correspondence between flat space partial waves and Polyakov-Regge blocks, the basis of functionals and other components of the numerical problem are also compared.

\subsection{Stress-tensor Correlator}
\label{sec:corr}
We now discuss bounds on the reduced correlator $\mathcal{H}(u,v)$ for different values of cross ratios $u$ and $v$ at finite coupling. This correlator has been studied in the weak coupling regime using a basis of conformal loop-integrals, see for instance \cite{Drummond:2013nda}. At strong coupling, from AdS/CFT, it is dual to a four-point closed string amplitude also known as the AdS Virasoro-Shapiro amplitude. This has been recently computed in a small curvature expansion around the flat space limit of AdS \cite{Alday:2022uxp,Alday:2023mvu}, providing novel results for the strong coupling series of the CFT correlator. At finite coupling, we know the integrated correlators of \cite{Chester:2020dja} thanks to localization, however, the correlator in cross-ratio space is still widely unexplored. We address this problem by placing two-sided bounds on the correlator in cross-ratio space. We mainly focus on $g=0.4$, which lies outside the perturbative regime, to demonstrate the applicability of our method. In addition, we also derive numerical bounds for a wide range of couplings for a single pair of cross ratios. We leave a more thorough study in a wider range of couplings and cross ratios for the future.
We also test our method at weak coupling by comparing against the two-loop perturbative series.

 We start by writing the decomposition of the correlator coming from dispersion relation in terms of single traces in eq.~\eqref{PR expansion uv},
\begin{equation}
    \mathcal{H}(u,v)=\mathcal{H}^{\rm sugra}(u,v)+\sum_{(\Delta,J) \hspace{1mm} \rm long}\lambda^2_{\Delta, J}\mathcal{P}_{u,v}^{\mathcal{N}=4}[\Delta,J].
\end{equation} 
In this section, we define $O$ to be the unprotected part of the above sum:
\beq
\label{eq:O-corr}
O_{u_0,v_0}\equiv\mathcal{H}^{\rm long}_{u_0,v_0}= \sum_{(\Delta,J) \hspace{1mm} \rm long}\lambda^2_{\Delta, J}\mathcal{P}_{u_0,v_0}^{\mathcal{N}=4}[\Delta,J]
\eeq
for some fixed value of $u_0$ and $v_0$. Then using the formalism discussed in section~\ref{sec:OptimizationSetUp} we look for two-sided bounds on $O_{u_0,v_0}$. After that, we add the value for $\mathcal{H}^{\rm sugra}(u_0,v_0)$ to the result to obtain upper bounds and lower bounds for the reduced correlator. This can be done at any value of the coupling. One subtlety that needs to be noticed is the dependence of the bounds on $J_{\rm max}$ and $\Delta_{\rm max}$ since in practice we are ignoring the contributions of operators with $J\geq _{\rm max}$ or $\Delta\geq \Delta_{\rm max}$.
Thus in practice, it is essential to check the stability of the bounds against changing $J_{\rm max}$ and $\Delta_{\rm max}$, especially as this can depend on the value of cross-ratios.

\begin{table}[t]
\centering
\begin{tabular}{c|c|c|c}
$(\text{Re}\,z,\,\text{Im}\,z )$ & Two-loop & Lower bound  & Upper bound \\ \hline
$(0.1, 0.1)$ & $-5.86$ &$-6.38$& $-6.22$  \\
$(0.1, 0.3)$ & $-0.788$ &$-0.852$ & $-0.826$ \\
$(0.1, 0.5)$ & $-0.196$ & $-0.212$ &$-0.205$ \\
$(0.1, 0.7)$ & $-0.065$ & $-0.070$& $-0.068$ \\
$(0.1, 0.9)$ & $-0.025$ & $-0.027$& $-0.026$ \\
$(0.1, 1)$& $-0.0166$ & $-0.0179$ & $-0.0173$\\
$(0.1, 1.25)$&$-0.00064$&$-0.0069$& $-0.0066$\\
$(0.2, 0.1)$&$-2.81$&$-3.05$& $-2.96$\\
$(0.2, 0.3)$&$-0.767$&$-0.829$& $-0.802$\\
$(0.2, 0.5)$&$-0.217$&$-0.235$& $-0.227$\\
\end{tabular}
\caption{List of two-sided bounds for the reduced correlator $\mathcal{H}(z,\bar{z})$ in the perturbative regime with $g=0.1$. We sample a list of pairs of cross ratios in the Euclidean kinematics $\bar{z}=z^{*}$. We also include the numerical evaluation of the 2-loop correlator in \eqref{eq:H2loop} for reference. This comparison is also presented in figure \ref{fig:2loop-conv}.}
\label{tab:Hbounds}
\end{table}

Let us start with the weak-coupling regime. Here we use $J_{\rm max}=260$ and $\Delta_{\rm max}=275$. At  $g=0.1$, we compare our two-sided bounds to a numerical evaluation of the 2-loop correlator for various cross ratios, see table~\ref{tab:Hbounds}. Here we record the 2-loop correlator for comparison:
\begin{align}\label{eq:H2loop}
\mathcal{H}(z,\bar{z})^{\text{2-loop}} &=\,-2g^2\frac{F_{1}(z)}{uv}+4g^{4}\frac{\frac{1+u+v}{4}\,F_{1}(z)^2+ F_{2}(z)+F_{2}(1-z)+\frac{F_{2}\left(\frac{z}{z-1}\right)}{v}}{uv} +O(g)^6 \nonumber \\
&\overset{z,\bar{z}\to 0}{=} \, \frac{-\frac{1}{3}+\lambda_K^2\times (z\bar{z})^{\frac{\Delta_K-2}{2}}+\cdots}{z\bar{z}} .
\end{align}
We use the notation for the ladder integrals: $F_L(z)\equiv \sum_{l=0}^L\frac{(-1)^l(2L-l)!\,\left[\log(z\bar{z})\right]^l}{L!(L-l)!l!}\frac{\text{Li}_{2L-l}(z)-\text{Li}_{2L-l}(\bar{z})}{z-\bar{z}}$, and recall that $u\equiv z\bar{z},\,v\equiv (1-z)(1-\bar{z})$. In the bottom line, we show the OPE limit which contains the data of Konishi operator at two loops: $\Delta_{K}=2+12g^2-48g^4+O(g)^6$ and $\lambda_K^2 = \frac{1}{3}-4 g^2 + g^4(56+24\zeta_3)+O(g)^6$, also recorded in our tilde normalization in \eqref{eq:weaktilde}. 

\begin{figure}[t]
\centering
\includegraphics[width=0.9\textwidth]{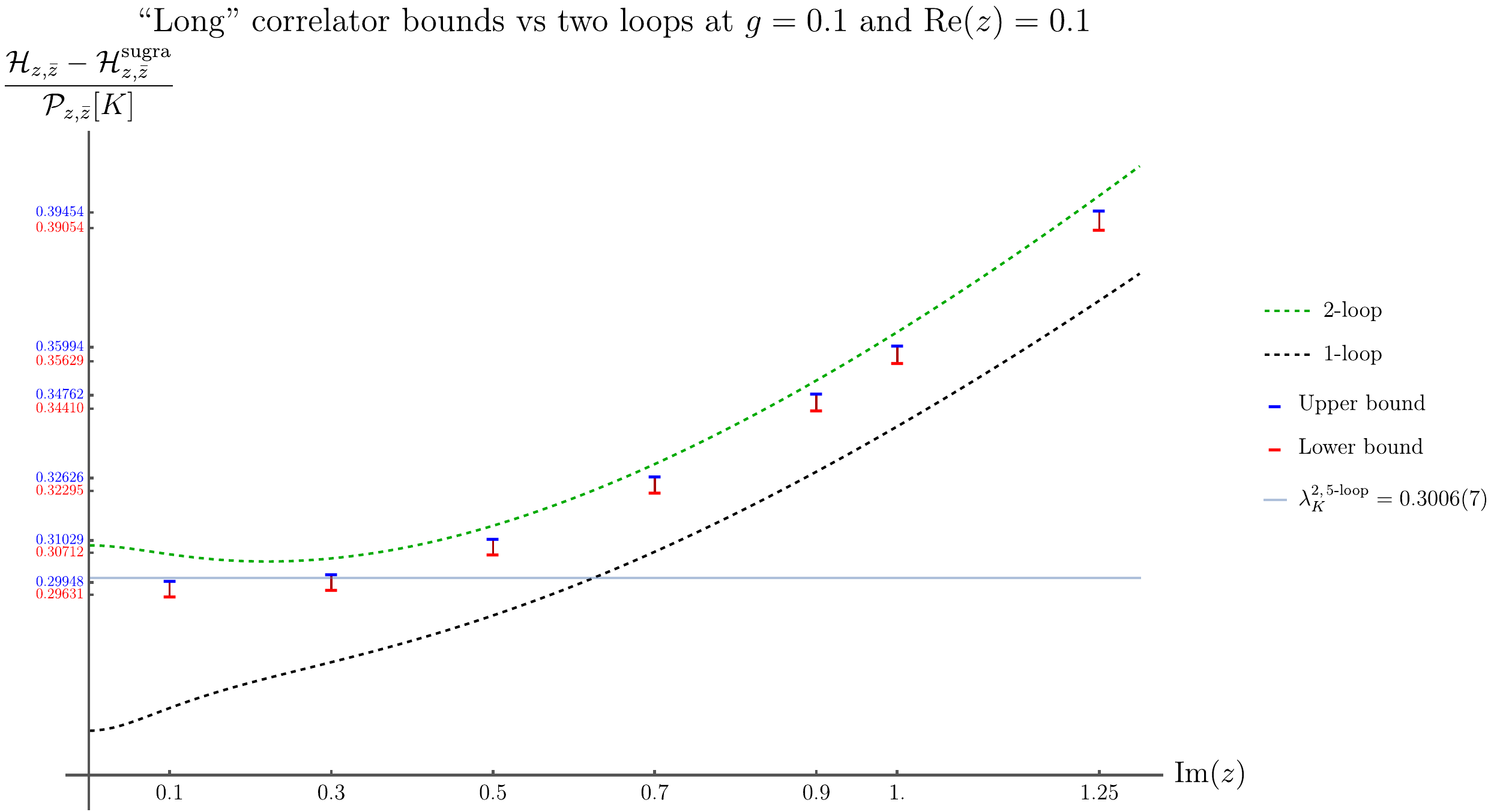}
\caption{Comparison between analytic perturbative results and numerical bounds on the reduced correlator $\mathcal{H}(z,\bar{z})$ at $g=0.1$, complex cross-ratio with $\bar{z} =z^{*}$ and fix $\text{Re}\,z=0.1$. We isolate the contribution of non-protected operators by subtracting the sugra-correlator $\mathcal{H}^{\text{sugra}}$, see decomposition in eq.~\eqref{PR expansion uv}, and normalize by  Polyakov-Regge block on Konishi operator: $\mathcal{P}_{z,\bar{z}}[K]\equiv \mathcal{P}_{z,\bar{z}}^{\mathcal{N}=4}[\Delta_K,0]$. In the limit  $|z|\to0$, this is dominated by the Konishi OPE coefficient and we show its 5-loop approximation for reference.}
\label{fig:2loop-conv}
\end{figure}

In figure \ref{fig:2loop-conv}, we plot our correlator bounds at coupling $g=0.1$ and contrast them with the numerical evaluation of the  1-loop and 2-loop correlators. We see that including the 2-loop correction brings perturbation theory closer to our numerical bounds. Besides, we check that in the OPE limit $|z|\to 0$, our  bounds on the ratio: $(\mathcal{H}-\mathcal{H}^{\rm sugra})/\mathcal{P}[\Delta_K,0]$, are close to those on the Konishi OPE coefficient. Away from this limit, increasing $\text{Im}\,z$, our bounds get farther from Konishi and remain close to the two-loop analytic curve. We expect that analytic results containing higher-loop corrections will improve the match with our numerical bounds \cite{Eden:2011we,Drummond:2013nda}.

\begin{figure}[t]
\centering
\includegraphics[width=.9\textwidth]{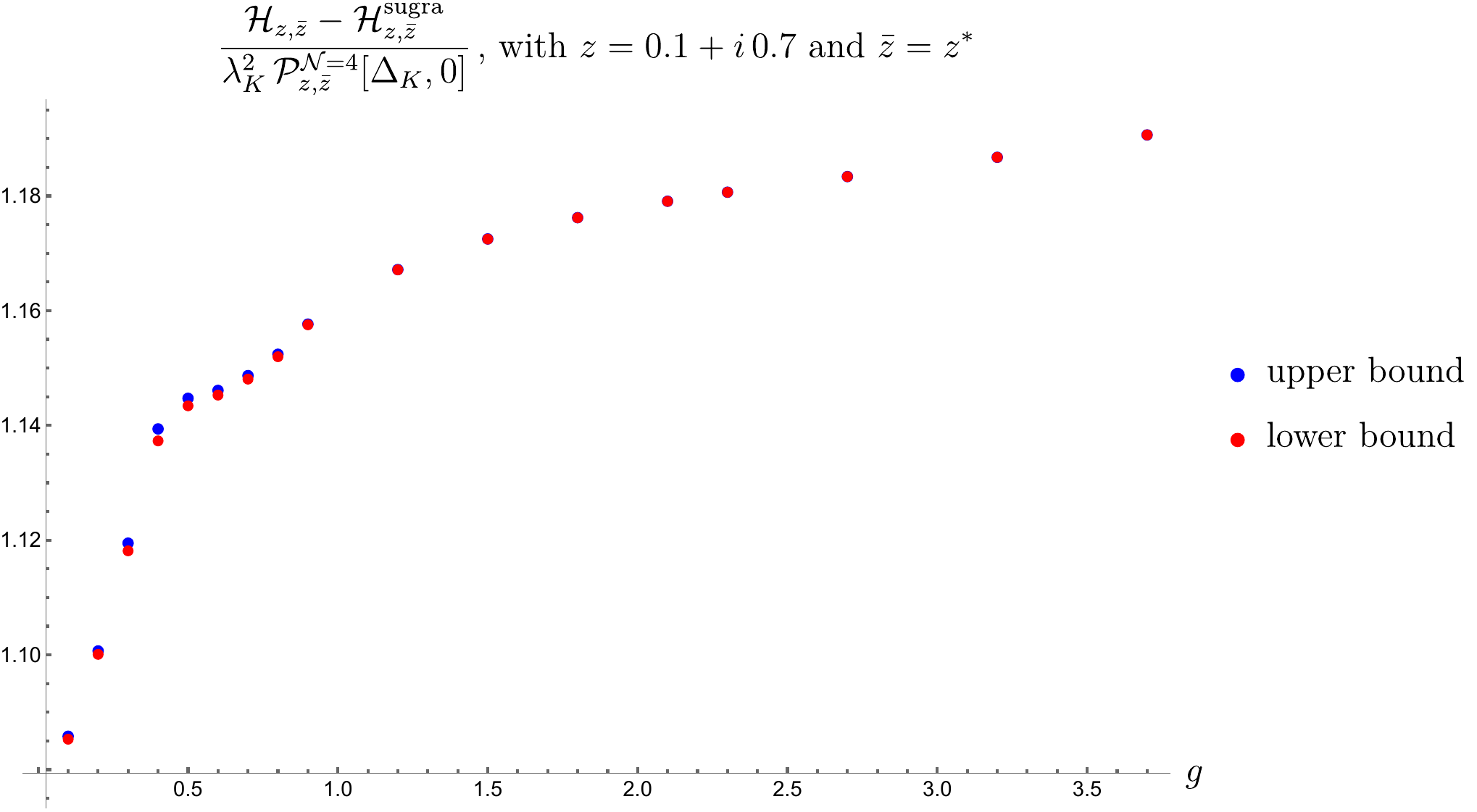}
\caption{Two-sided bounds on the non-protected part of the reduced correlator $\mathcal{H}_{z,\bar{z}}$ in Euclidean kinematics, with complex cross ratio $z=0.1+i\,0.7$  ($u=0.5,\,v=1.3$) and changing coupling $g\in[0.1,\,3.7]$, normalized by the Konishi  contribution to \eqref{PR expansion uv}.
In the denominator we used the midpoint of the two-sided bounds in figure \ref{fig:ope-results}, which may create some artificial features. 
}
\label{fig:HboundsManyCouplings}
\end{figure}

At intermediate couplings, no other results currently exist to compare with.
We present two ways of exploring this regime. First, we place bounds on the correlator for a wide range of couplings but with a fix pair of cross ratios. Second, we focus on a specific value of the coupling ($g=0.4$) and place bounds on the correlator for various pairs of cross ratios in Euclidean and Lorentzian kinematics. 

In figure \ref{fig:HboundsManyCouplings}, we present two-sided bounds on the non-protected part of the correlator, in Euclidean kinematics with cross ratios $u=0.5,\, v=1.3$ (corresponding to complex conjugates $z^{*}=\bar{z}$), away from any OPE limit. For comparison, we consider the ratio to the Konishi contribution to the OPE sum in \eqref{PR expansion uv}. This ratio grows as we increase the coupling, from $\approx 1$ at weak coupling to seemingly a constant at strong coupling, whose value depends on our cross-ratio choice. Furthermore, the gap between our bounds is very narrow and only visible at weaker coupling, while it closes in the strong-coupling region. Finally, as explained below for the case $g=0.4$, the sugra-correlator $\mathcal{H}^{\rm sugra}$ is still the dominant contribution to the reduced correlator $\mathcal{H}$ at intermediate values of the coupling, and becomes more dominant as we go to stronger coupling, as expected. 

\begin{figure}[t]
\centering
\includegraphics[width=0.45\textwidth]{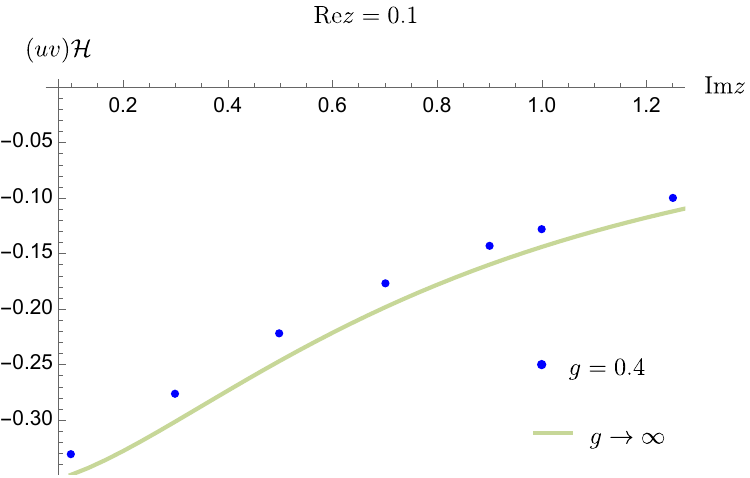}
\includegraphics[width=0.45\textwidth]{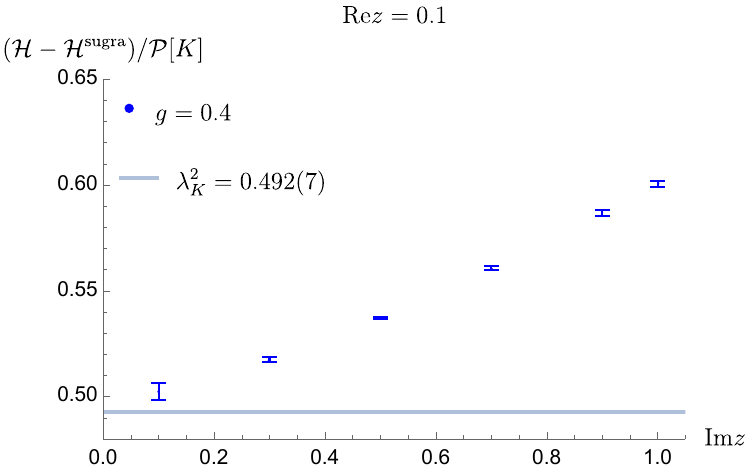}
\caption{Plots at fixed value of $\text{Re}\,z=0.1$. Left: $uv \mathcal{H}(z,\bar{z})$ is shown with the blue dots. The green curve illustrates the contribution of $(uv)\mathcal{H}^{\text{sugra}}$ for comparison. Right: contribution of non-protected operators, $(\mathcal{H}-\mathcal{H}^{\text{sugra}})/\mathcal{P}[K]$ is plotted with the blue bars. The gap between lower bounds and upper bounds in our numerical result is visible in this figure. For reference, we represent the contribution of Konishi operator to the sum by the bottom solid line.}
\label{fig:result-GzzbEx12-0.4}
\end{figure}

For our second way of exploring this finite-coupling regime, we focus on coupling $g=0.4$. Similar to weak coupling analysis, we focus on the bounds for fixed $\text{Re}\,z=0.1$ in fig.~\ref{fig:result-GzzbEx12-0.4} as a function of $\text{Im}\,z$. In the left plot, we compare the correlator $(uv)\mathcal{H}(z,\bar{z})$ (blue dots) with the contributions of the protected operators, $(uv)\mathcal{H}^{\text{sugra}}$ (green curve). In the right figure, the contribution of non-protected operators are isolated and shown as blue bars. This is then compared with the Konishi OPE coefficient (also numerical bound), indicated by the bottom line in the plot, which gives the dominant contribution in the limit $|z|\rightarrow 0$. 

\begin{figure}[t]
\centering
\includegraphics[width=0.45\textwidth]{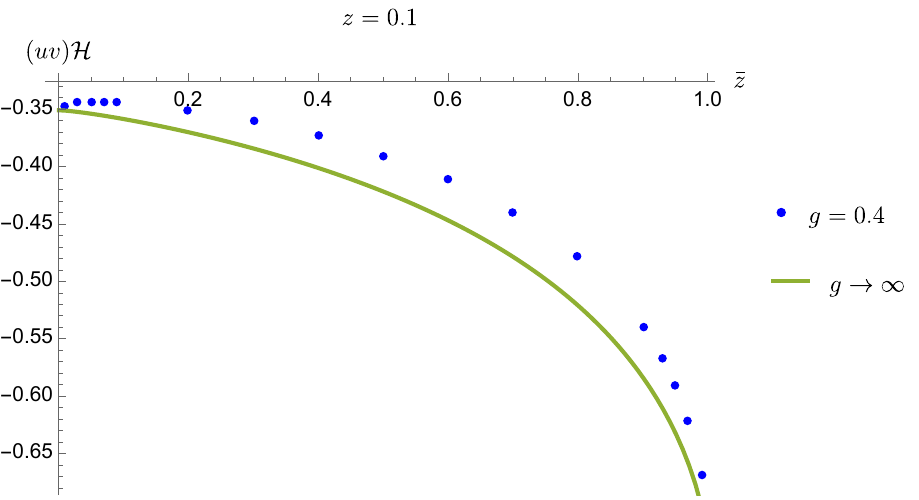}\qquad
\includegraphics[width=0.45\textwidth]{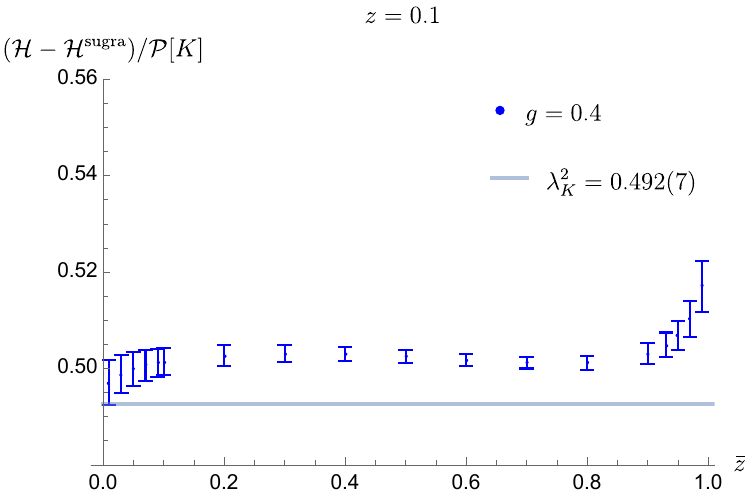}\qquad
\includegraphics[width=0.45\textwidth]{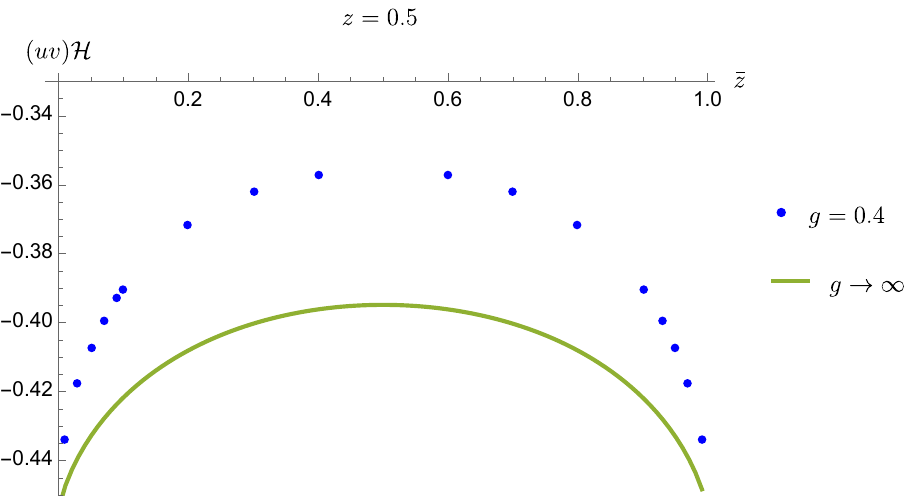}\qquad
\includegraphics[width=0.45\textwidth]{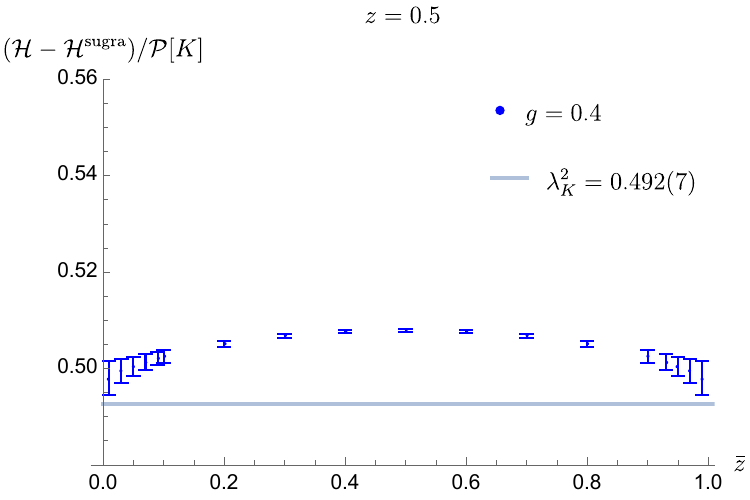}\qquad
\caption{Plot of correlator at $g=0.4$ at fixed $z=0.1$ (top) and $z=0.5$ (bottom). Left: bounds on $uv \mathcal{H}$ for different independent real $\bar{z}$. Here we see how our numerical results (blue dots) compare with supergravity contribution (green curve). Right: contribution of non-protected operators, $(\mathcal{H}-\mathcal{H}^{\text{sugra}})/\mathcal{P}[K]$ are illustrated by blue bars to show the gap between upper and lower bounds. The contribution of Konishi is also shown for reference.}
\label{fig:result-Gzzb-0.4}
\end{figure}
\begin{figure}[t]
\centering
\includegraphics[width=0.45\textwidth]{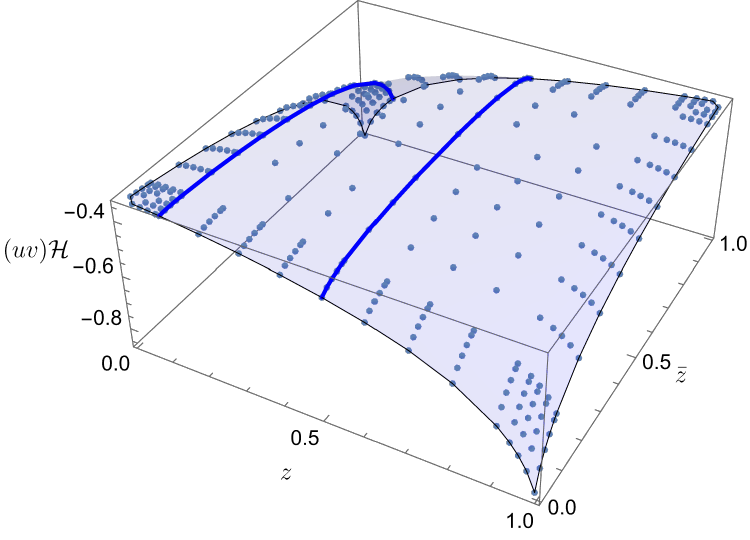}
\caption{Plot of the correlator $uv{\cal H}(z,\bar{z})$ at $g=0.4$ for different independent real $z$ and $\bar{z}$. The blue curves show the $z=0.1$ and $z=0.5$ cases in fig.~\ref{fig:result-Gzzb-0.4}.
The discrete points show the data used to produce the curves and surface.
The shape qualitatively follows that of ${\cal H}^{\rm sugra}$.
}
\label{fig:result-Gzzb-0.43d}
\end{figure}

Next, even though convergence of eq.~\eqref{PR expansion uv} outside the Euclidean regime has not been analyzed in full generality \cite{Caron-Huot:2020adz}, we expect convergence for real and independent $z$ and $\bar{z}$ at least when $0<z,\bar{z}<1$.
We have explored this region at $g=0.4$ and found non-trivial upper and lower bounds. We again start by analyzing the resulting bounds for fixed values of $z$ to get a better understanding of different features. In fig.~\ref{fig:result-Gzzb-0.4}, we focus on $z=0.1$ and $z=0.5$. In the left plots, we plot our result for the correlator, $uv \mathcal{H}$ as well as the contributions of the supergravity to the sum at various values of $\bar{z}$. In the right plots, we remove the supergravity contributions and divide by the Konishi block to only focus on the contribution of unprotected operators. In these plots, we can see the size of the gap between upper and lower bounds as well as the OPE coefficient of Konishi operator. For $z=0.1$, we see that in the s-channel OPE limit ($z,\bar{z}\rightarrow 0$), the sums are dominated by Konishi operator and as we approach the double light-cone limit ($z\rightarrow0$, $\bar{z}\rightarrow1$) they get further away.

In fact upon exploring other values of $z$, we find that in Lorentzian regime, we get a smooth curve for all the points. The result for various $z, \bar{z} \leq 1$ illustrated in fig.~\ref{fig:result-Gzzb-0.43d}. In this plot, we illustrate the two $z=0.1$ and $z=0.5$ curves of fig.~\ref{fig:result-Gzzb-0.4} by blue curves. Here for all the points we studied the gaps are smaller than $0.1 \%$.

As a final remark, we reiterate that an important (and hard to quantify) source of systematic error in the analysis of the correlator is truncation of the numerical problem to finite $J_{\text{max}}$. This source of error does not always get reflected in the size of the gap between upper and lower bounds and thus the gap is not a complete indicator of the error. In fact, we have seen multiple instances (for example in fig.~\ref{fig:HboundsManyCouplings}) where the gap between upper and lower bound is small but the resulting curve displays noise which is suggestive of a larger error. We observed that this tends to be more pronounced away from the s-channel OPE limit where eq.~\eqref{PR expansion uv} converges more slowly. One resolution would be to use OPE sums in other channels instead of eq.~\eqref{PR expansion uv} in regimes where those have better convergence.  Another approach, which could open up larger Lorentzian regions, would be to include asymptotic large spin approximations into the problem. We leave further studies of convergence for future works.

\section{Bounds at strong coupling and the flat space limit}
\label{sec:flat-space}

In this section, we relate the CFT correlator bootstrap at strong `t Hooft coupling and the flat-space amplitude bootstrap.  This will shed light on the question of whether we expect the gap between lower and upper bounds to shrink more with additional spectral information.

We first explain analytically in subsection~\ref{sec:PRvsFlat} how our CFT bootstrap problem can be expressed in terms of S-matrix ingredients in this limit.
In subsection~\ref{sec:flatspaceNumerics} we formulate an analogous flat-space numerical bootstrap following \cite{Caron-Huot:2020cmc} and including additional spectral information (the linear Regge trajectories of flat space string theory) following \cite{Albert:2024yap,Berman:2023jys,Berman:2024wyt,Eckner:2024ggx}.
Finally, in subsection~\ref{sec:numericalCFTvsFlat} we compare the two numerical approaches and show that they give equivalent bounds (and even equivalent extremal functionals!) up to 
$O(1/\Delta_K^2)$ corrections.

\subsection{Flat space limit of Mack polynomials and Polyakov-Regge blocks}\label{sec:PRvsFlat}

Kinematical relations between CFT OPE coefficients and flat space three-point couplings were discussed in section \ref{sec:OPEandFlat}.  Here we extend this dictionary by considering the limit of Polyakov-Regge blocks.

It is useful to start by writing down the flat space \emph{analogs} of the OPE and Polyakov-Regge expansions. The OPE is simply the partial-wave decomposition \cite{Correia:2020xtr}:
\beq\label{eq:MflatPwave}
\mathcal{M}(s,t) =\frac{1}{s^{4+\frac{D-4}{2}}} \sum_{J=0,2,4,\cdots} n_{J}\,a_{J}(s) \mathcal{P}_{J}\left(1+\frac{2t}{s}\right)
\eeq
where the factor $1/s^4$ is appropriate for the reduced super-amplitude (see section~\ref{sec:OPEandFlat}) and $D=d+1$ is the dimension of the bulk spacetime. The normalization $n_J$ and Gegenbauer polynomial
$\propto \mathcal{P}_J$ are given by:
\beq\label{eq:mathPJ}
n_J = \frac{(4\pi)^{D/2}\,(D+2J-3)\,\Gamma(D+J-3)}{\pi\Gamma(\frac{D-2}{2})\,\Gamma(J+1)}\,,\qquad \mathcal{P}_{J}(x)= \,_{2}F_{1}\left(-J, J+D-3, \frac{D-2}{2},\frac{1-x}{2}\right).
\eeq
Conceptually, the flat-space analog of the Polyakov-Regge expansion  \eqref{PR expansion st} is then simply the dispersive representation: 
\begin{align}
 \cM(s,t) &= \frac{8\pi G_D}{stu} + \sum_{\j,m} C^2_{m,\j} \cP^{\rm flat}_{s,t}[m,\j], \label{PR flat}
 \\
\label{PR flat explicit}
\cP^{\rm flat}_{s,t}[m,\j]&\equiv \frac{n_{J}
}{m^{8+D-4}} \cP_\j\left(1-2\frac{s+t}{m^2}\right)\left(\frac{1}{m^2-s}+\frac{1}{m^2-t}\right).
\end{align}
This reconstructs the amplitude from its $s$- and $t$- channel poles, exploiting its vanishing in the Regge limit: $|s|\to\infty$ at fixed $u=-s-t$.
In \eqref{PR flat} we separated out the $s=0$ pole and the 
sum only runs over massive states,
mimicking \eqref{PR expansion st}.
Here the $C^2_{m,J}$ are the residues of particle poles,
${\rm Im}\,a_J(s)=\sum_m C^2_{m,J}\pi\delta(s-m^2)$.\footnote{We assume here that ${\rm Im}\,a_J(s)$ is a discrete sum of $\delta$-functions since we are focusing on the planar limit. In general, \eqref{PR flat} would involve a continuous integral over $m$.}

We would like to show that \eqref{PR flat}
actually arises as the flat space limit of the
CFT Polyakov-Regge expansion.  From now on, we specialize to $d=4$ (so $D=5$).

In the flat space limit, the exchange states have a large dimension $\Delta$ but fixed spin $\j$.  It is easy to check that the coefficients in the Mack polynomials \eqref{eq:Qcal_rep}
go like $\Qmat_{q,k}\sim \Delta^{\j-q-k}$.
On the other hand, the sum over descendants is dominated by $n\sim \Delta^2$,
and it is thus natural to scale $\mT\sim \Delta^2$ to make contact with fixed-angle scattering, as will be clear shortly.
Thus the typical contribution of a given coefficient goes like
$n^q\Qmat_{q,k}\mT^k\sim \Delta^{\j+q+k}$,
and we see that the limit is dominated by the terms with the largest value of $q+k$, namely those along the anti-diagonal $q+k=\j$. Comparing the explicit expressions (which are relatively concise at small $J$) we find that
these precisely reproduce the Gegenbauer polynomials:
\be \label{Q flat}
 \lim_{\Delta^2\sim n\sim \mT\to\infty} Q^{n,a,b}_{\Delta,\j}(\mT) =\ {-}\left(\frac{n}{4}\right)^\j \frac{(d-2)_\j}{\big(\tfrac{d-2}{2}\big)_\j} \cP_{\j}\left(1+\tfrac{\mT}{n}\right),
\ee
where $n$ represents the descendant order. The proportionality constant can be determined from the normalization $Q\sim -(\mT/2)^\j$ at large $\mT$.

Now consider the Polyakov-Regge block in eq.~\eqref{PR Mellin explicit}, and set $\mS,\mT\sim 1\ll \Delta^2$. 
This is the region which is used for our CFT bootstrap functionals with finite $\mS$, $\mT$, $u$, or $v$. This is also the region which dominates the integrated constraints
\eqref{cI2}-\eqref{cI4} thanks to the exponentially decaying kernels.
It is easy to see that, assuming that $n\sim \Delta^2$ dominates the descendant sums,
in this region, we can series-expand the Polyakov-Regge block \eqref{PR Mellin explicit} in $\mS$, $\mT$ or $\mU$ such that each additional power is suppressed by $\Delta^{-2}$.

Comparing the CFT Polyakov-Regge block \eqref{PR Mellin explicit}-\eqref{eq:Qcal_rep} with its flat space counterpart \eqref{PR flat explicit}, we see now thanks to \eqref{Q flat} that their low-energy expansions are simply related to each other, with the coefficients of each monomial related by a specific sum over $n$:
\be\begin{aligned}
\widehat{\cP}^{\Nfour}_{\mS,\mT}[\Delta,\j]\Big|_{\mS^a \mT^b} &\approx
\cP^{\rm flat}_{s,t}[\Delta,\j]\Big|_{s^a t^b} \times \frac{\Delta^9}{128\pi(\j+1)} \sum_n K_{\Delta+4,\j}^{n} \left(\frac{\Delta^2}{2n}\right)^{a+b+1}\left(\frac{n}{4}\right)^\j \,.
\label{PR flat limit}
\end{aligned}\ee
Here we assumed $n\gg 1$ to obtain the powers in parentheses, anticipating that $n\sim \Delta^2$ will dominate. (The preceding approximation is not good for $n=0$, but the coefficient $K$ is suppressed there.)
The sum can be done analytically by replacing it with suitable Pochhammers:
\begin{align}
\sum\limits_{n=0}^\infty K_{\Delta+4,\j}^{n} \left(\frac{\Delta^2}{2n}\right)^{a+b+1}\left(\frac{n}{4}\right)^\j &
=K_{\Delta+4,\j}^{0} \sum_n  \frac{\big(\tfrac{\Delta-\j-2}{2}\big)_n^2}{n!(\Delta+3)_n}  \left(\frac{\Delta^2}{2n}\right)^{a+b+1}\left(\frac{n}{4}\right)^\j
\nonumber\\ &
\approx  \frac{2\sin(\tfrac{\pi(\Delta-\j)}{2})^2}{\OPEfree_{\Delta,\j}}  \frac{\j+1}{\pi^2\Delta^8}2^{a+b+14}\Gamma(a+b+6)
\end{align}
where we used that the sum evaluates exactly to $\frac{\Gamma(a+b+6)\Gamma(\Delta+a+b+4)}{\Gamma\big(\tfrac{\Delta}{2}+a+b+5\big)^2}$,
and expanded at large $\Delta$ to find the second line. In summary:
\be\label{eq:Pstrongvsflat}
\frac{\OPEfree_{\Delta,\j}}{2\sin\big(\tfrac{\pi(\Delta-\j)}{2}\big)^2}
\widehat{\cP}^{\Nfour}_{\mS,\mT}[\Delta,\j]\Big|_{\mS^a \mT^b}
= \cP^{\rm flat}_{s,t}[m,\j]\Big|_{s^a t^b}
\times  c_{a+b}\frac{4m}{\pi\RAdS} \left(1+O(\Delta^{-1})\right)\,,
\ee
with $m=\Delta/\RAdS$ and $c_{n}\equiv\frac{2^{n+5}\Gamma(n+6)}{\pi^2\RAdS^{2n+9}}$.
The factor $\frac{4m}{\pi\RAdS}$ is precisely the conversion between flat-space residues and OPE coefficients in \eqref{flat OPE}!
This ensures that the flat-space limit of the Polyakov-Regge expansion \eqref{PR expansion st} is exactly \eqref{PR flat}. 

The factor $c_{a+b}$ also has a natural interpretation and is related to the normalization of contact interactions in Mellin space.
Let us recall from \cite{Penedones:2010ue,Fitzpatrick:2011hu,Fitzpatrick:2011ia} that the flat-space limit of the Mellin amplitude is a certain Borel transform of the flat-space amplitude; for the super-amplitude we find:\footnote{
The overall normalization factor is in principle determined from \cite{Penedones:2010ue,Fitzpatrick:2011hu,Fitzpatrick:2011ia} but here we have fixed it simply by comparing the supergravity limit.
}
\be
 \widehat{\mathcal{H}}_{\mS,\mT}\approx \frac{32\RAdS^3c}{\pi^2} \int_0^\infty d\beta \beta^5 e^{-\beta\RAdS^2} \cM(2\beta\mS,2\beta\mT).
\ee
A contact interaction in the amplitude then maps to a polynomial in Mellin space, with each monomial rescaled by $c_{a+b}$:
\be \label{contact scaling}
 \cM^{\rm flat}_{s,t}\supset s^a t^b\  \Longleftrightarrow \  \widehat{\mathcal{H}}_{\mS,\mT}\supset c_{a+b}\ \mS^a \mT^b, \qquad c_{n}\equiv\frac{2^{n+5}\Gamma(n+6)}{\pi^2\RAdS^{2n+9}}.
\ee
The full normalization in \eqref{eq:Pstrongvsflat} is thus seen to be correct.

The same logic can be applied to crossing equations $\widehat{X}_{\mS,\mT}\equiv \widehat{\cP}_{\mS,\mT}^{\Nfour}-\widehat{\cP}_{16-\mS-\mT,\mT}^{\Nfour}$, which will thus reduce to crossing equations of the flat-space amplitude \eqref{PR flat}.  Thus our CFT bootstrap problem, when all exchanged operators have a large dimension $\Delta$, reduces to an S-matrix bootstrap problem.
This phenomenon is familiar from other studies  (see for example \cite{Paulos:2016fap}); here we have extended it to massless external particles and to dispersive functionals.

\subsection{Numerical bounds for flat space closed string amplitude}\label{sec:flatspaceNumerics}

We now consider the following flat-space question: given the known spectrum of string theory (linear Regge trajectories), general constraints of unitarity and analyticity, and possibly some ``integrated constraints'', what bounds can we obtain on Wilson coefficients and on-shell three-point couplings?

Closely related questions have been discussed by many authors.  We will follow closely \cite{Albert:2024yap,Berman:2024wyt}, after reviewing 
the general constraints on weakly coupled low-energy effective theories from \cite{Caron-Huot:2020cmc}.
This subsection can be read independently from the rest of this paper.

The superamplitude ${\cal M}$ we are interested in admits a massless pole from (super)graviton exchange, made explicit in \eqref{PR flat}.
We will focus on the lightest non-protected particle,
a massive scalar ($J=0$) of mass $m_K$,
and try to bound its three-point coupling: 
\beq\label{eq:ResFlat}
\tilde{\lambda}_{K}^{2,\text{flat}}\,=\, \frac{m_{K}^4}{8 \pi G_{D}}\times\Res_{s=m_{K}^2}\left[-\mathcal{M}(s,t)\right] \equiv \frac{m_{K}^4}{8\pi G_{D}}\, 
\times \frac{n_{J=0}}{m_{K}^{4+D}}
\Res_{s=m_{K}^2}\left[-a_{J=0}(s)\right].
\eeq
As reference, for the Virasoro-Shapiro amplitude in \eqref{eq:MVSflat}, the lightest mass and the correspondent normalized residue are
\beq
\left[m_K^2\right]^{VS}=\frac{4}{\alpha'}\quad\text{and}\quad 
\left[\tilde{\lambda}_{K}^{2,\text{flat}}\right]^{VS} = 1\,.
\eeq
Our goal is to find upper and lower bounds on the coupling $\tilde{\lambda}^{2,\text{flat}}_K$, defined by \eqref{eq:ResFlat}, by making various assumptions on the high-energy spectrum 
as well as on the low energy expansion of the amplitude. These assumptions resemble the input used in the analog CFT problem in subsection~\ref{sec:numericalCFTvsFlat}.

In the flat-space limit, we know, of course, the
complete spectrum of (tree-level) string theory.
The states organize into linear Regge trajectories,
\be
 m_{n,J}^2 = m_K^2\left(\tfrac{J}{2}+1+n\right) \qquad (n=0,1,2\ldots).
\ee
\begin{figure}
    \centering
    {\centering
\includegraphics[width=1\textwidth,angle=0]{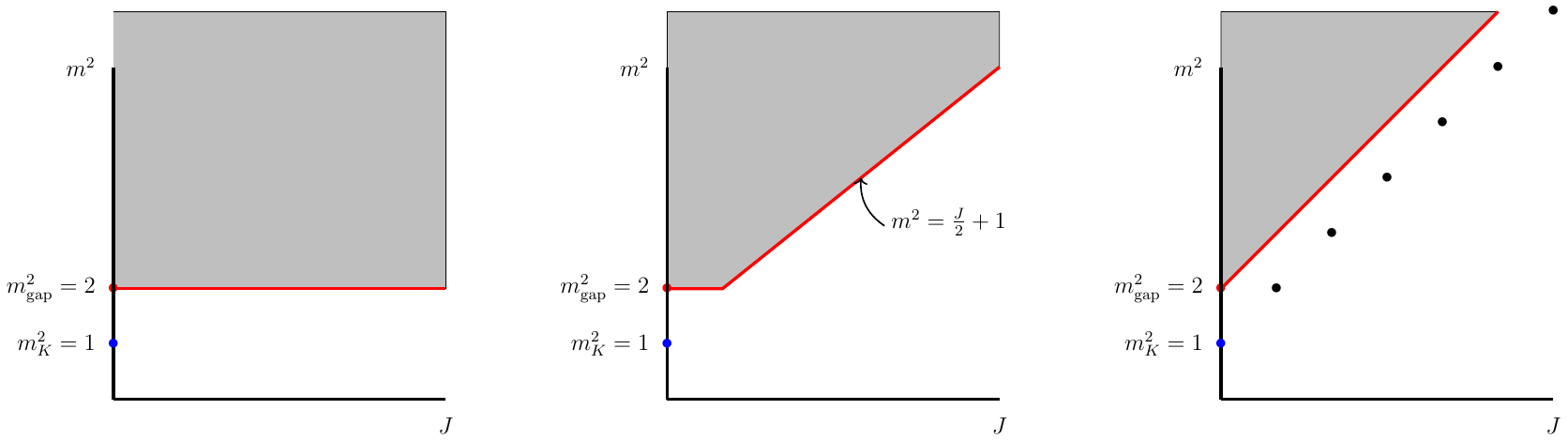}
}
\caption{Three different assumptions on the spectrum in the flat-space amplitude.
     On the first panel, a homogeneous gap for all states of spin $J$. On the second panel, we introduce the first Regge trajectory of Virasoro-Shapiro amplitude. On the third panel, we introduce the second Regge trajectory.}
\label{fig:GridFlatoptions}
\end{figure}
We will implement numerically different levels of information, as illustrated in figure \ref{fig:GridFlatoptions}.
In the simplest case, we assume that the lightest state above $m_K$ can have the same mass for all spin $J$:
\beq\label{eq:flatspec1}
\text{I}:\quad m^{2}_{J} \geq 2 m_{K}^2\,.
\eeq
In the second case we consider the lightest mass for each spin is given by the first Regge trajectory: 
\beq\label{eq:flatspec2}
\text{II}:\quad m^{2}_{J}\geq m_K^2\times {\rm max}\left(\tfrac{J}{2}+1,2\right),
\eeq 
and in a third scenario we include discrete states along the full leading trajectory and a second gap going to the second Regge trajectory: 
\beq\label{eq:flatspec3}
\text{III}:\quad m^{2}_{J} = m_{K}^2\,\frac{(J+2)}{2} \quad \text{or}\quad m^{2}_{J} \geq m_{K}^2\,\frac{(J+4)}{2}\,.
\eeq 
Finally, we specify additional assumptions we make on the low-energy Wilson coefficients, which are meant to account for the supersymmetric localization constraints in subsection~\ref{sec:Integrated constraints}.

The low-energy EFT expansion of the amplitude is: 
\beq\label{eq:Mlow}
\mathcal{M}_{EFT} \overset{s,t\to0}{=}\frac{8\pi G_D}{s\, t\, u}+g_0 + g_2\,(s^2+t^2+u^2) + g_3\,s t u + g_4 (s^2+t^2+u^2)^2 +\cdots 
\eeq
parametrized by the low-energy coefficients $g_k$ where $k$ stands for the degree of the corresponding Mandelstam polynomial. (The absence of a $g_{1}$ term is due to the momentum conservation condition $s+t+u=0$.)
In the flat-space limit, the two integrated constraints amount to assigning the first two Wilson coefficients ($R^4$ and $D^4R^4$ terms in the string effective action)
to the values they take on the Virasoro-Shapiro amplitude in \eqref{eq:MVSflat} (see eq.~\eqref{limits I24}):
\beq\label{eq:g02VS}
g_0\,\frac{m_K^6}{8\pi G_D}\to \left[g_0\,\frac{m_K^6}{8\pi G_D}\right]^{VS}=2\zeta_3 \quad \text{and}\quad g_2\,\frac{m_K^{10}}{8\pi G_D} \to \left[g_2\,\frac{m_K^{10}}{8\pi G_D}\right]^{VS} = \zeta_5 \,.
\eeq

\paragraph{Optimization from dispersive relations and null constraints}
Our main tool to find bounds will be the dispersion relations that relate the low energy limit of the amplitude (EFT coefficients) with high energy sums rules of the massive spectrum 
\cite{Caron-Huot:2020cmc}: 
\beq\label{eq:Bkflat}
B_{k}:\quad\Res_{s=0}\left[\frac{2s+t}{s(s+t)}\,\frac{\mathcal{M}_{EFT}(s,t)}{\left[s(s+t)\right]^{k/2}}\right]\,=\, \left\langle \frac{2m^2+t}{m^2+t}\,\frac{\mathcal{P}_{J}\left(1+\frac{2t}{m^2}\right)}{[m^2(m^2+t)]^{k/2}} \right\rangle\qquad (t<0, k=-2,0,2,\cdots)
\eeq
where the average $\langle\cdots \rangle$ is a sum over high energy states: 
\beq\label{eq:flataverage}
\big{\langle} f(m^2,J) \big{\rangle}  \,=\,  \sum_{J=0,2,\cdots}n_{J}\,\int_{m^{2,\text{min}}_J}^{\infty} \frac{dm^2}{\pi}\,\frac{m^{(4-D)-8}}{m^2}\,\rho_{J}(m^2)\, f(m^2,J) .
\eeq
The density of states is given by the imaginary part of the partial-wave coefficient: $\rho_{J}(m^2) \equiv \text{Im}\,a_{J}(m^2)$.
These sum rules are equivalent to imposing the Polyakov-Regge expansion \eqref{PR flat} order-by-order in an expansion in $s$. (More precisely, $B_k$ is a linear combination of the coefficients of $s^0$, $s^1$, \ldots $s^k$ in \eqref{PR flat} \cite{Caron-Huot:2020cmc}.)
These sum rules are therefore related, by virtue of \eqref{eq:Pstrongvsflat}, to the flat space limit of the CFT functionals.

More precisely, the sum rules with $k\geq 0$ relate to low-energy expansion of \eqref{eq:Pstrongvsflat}, whereas the sum rules with $k=-2$, which probe the graviton pole, are related to the antisubtracted CFT $B$ functionals in \eqref{eq:BtMellin}.
For most of our analysis we will only use the $k\geq 0$ sum rules, which admit smooth forward limits, although we briefly comment on the effect of $k=-2$ sum rules below.

On the left-hand side of \eqref{eq:Bkflat} we use the EFT approximation of \eqref{eq:Mlow}. On the right-hand side, for each spin $J$, we include the spectrum described in figure \ref{fig:GridFlatoptions}: we include a few discrete states followed possibly by a continuum.

In order to introduce the three-point coupling  $\tilde{\lambda}_{K}^{2}$ we isolate the contribution of the lightest scalar $m_{K}$ to the high-energy sum rules: 
\beq
\langle f(m^2,J) \rangle  \,=\, \tilde{\lambda}^{2,\text{flat}}_{K}\times\, \frac{8 \pi G_D}{m_K^6} f(m^{2}_{K},0) +\langle f(m^2,J) \rangle'
\eeq
where the primed average (similar to section \ref{sec:OptimizationSetUp}) excludes the lightest state:
\beq\label{eq:averageprime}
\langle f(m^2,J) \rangle' \equiv \frac{n_{0}}{\pi} \int_{m^2_{\text{gap}}}^{\infty} \frac{dm^2}{m^{D+6}}\rho_{0}(m^2)\, f(m^2,0)  + \sum_{J=2,4\cdots}\frac{n_{J}}{\pi}\int_{m^{2,\text{min}}_J}^{\infty} \frac{dm^{2}}{m^{D+6}}\rho_{J}(m^2)\, f(m^2,J) 
\eeq
here $m_{\text{gap}}$ stands for the mass of the second lightest state at $J=0$ and $m_{J}^{2,\text{min}}$ stands for the lightest mass at spin $J\geq 2$. These latter will be subject to the different spectral assumptions in eqs.~\eqref{eq:flatspec1} to \eqref{eq:flatspec3}.

By using the sum rules in \eqref{eq:Bkflat} with  $k=0$ (un-substracted) and $k=2$ (twice-substracted) and expanding in the forward limit ($t=0$) we can obtain sum rules that equate to the low energy coefficients $g_{0}$ and $g_{2}$ in \eqref{eq:Mlow} as:
\beq\label{eq:g02sumrules}
g_{0} = 2\,\tilde{\lambda}_{K}^{2,\text{flat}}\,\frac{8 \pi G_D}{m_K^6}  \,+\, \big{\langle} 2 \big{\rangle}' \quad \text{and}\quad g_{2} = \frac{\tilde{\lambda}_{K}^{2,\text{flat}}}{m_{K}^4}\,\frac{8 \pi G_D}{m_K^6}  \,+\, \left\langle \frac{1}{m^4}\right\rangle'\,.
\eeq
Using higher $k$-substracted sum rules we can obtain similar results for the other low energy coefficients but we do not make use of them here. Instead, we use only combinations of other sum rules that equate to zero, known as null constraints. The first null constraint can be obtained as the coefficient of $t$ in the $B_{0}$ sum rule:
\beq\label{eq:firstnull}
0 =  \left\langle \frac{\frac{4 \mathcal{J}^2}{D-2}-1}{m^2} \right\rangle = \partial_{t}B_{0}\big{|}_{t=0} 
\eeq
where we denote the conformal Casimir  as $\mathcal{J}\equiv\sqrt{J(J+D-3)}$. Similarly, other null constraints can be obtained by enforcing crossing symmetry on the high energy sum rule on the right-hand side of \eqref{eq:Bkflat}, to match the crossing symmetric left-hand side. In order to organize these crossing constraints, in analogy to the CFT, we define the flat-space crossing functional:
\beq\label{eq:XstFlat}
X^{\rm flat}_{s,t} \,=\,  \mathcal{P}_{J}\left(1+\frac{2u}{m^2}\right)\left(\frac{m^2}{m^2-s}+\frac{m^2}{m^2-t}\right) -\left(s\leftrightarrow u\right)\quad\text{with }\quad u=-s-t
\eeq
and consider its Taylor coefficients at small $s,t$:
\beq\label{eq:Xnullab}
\mathcal{X}_{a,b}(m^2,J) =m_K^{2(a+b)}\,
X^{\rm flat}_{s,t}\big{|}_{s^a u^b}
\eeq
which like \eqref{eq:firstnull} are functions of mass and spin.
Then the null constraints are
\beq\label{eq:nullflat}
\langle \mathcal{X}_{a,b}\rangle =0 \qquad \mbox{for }a,b=0,1,2\ldots
\eeq
For fixed weight $n=a+b$ in $m^2$, it suffices to consider $a$ starting from 0 up to $\lfloor \frac{n-1}{3}\rfloor$ (integer part of the fraction)  to form a complete basis. This is an infinite set of conditions but in our numerics, we consider only a finite number of them up to a certain order in $m$. For instance, up to order $\mathcal{O}(m^{-12})$ we have a set of nine null constraints that we can collect into a vector:
\bba\label{eq:Xcrossflat}
\vec{\mathcal{X}}^{\rm cross} \,&=\, \left(\begin{matrix} \mathcal{X}_{0,1} &  \mathcal{X}_{0,2} &  \mathcal{X}_{0,3} &  \mathcal{X}_{0,4} & 
\mathcal{X}_{1,3} &  \mathcal{X}_{0,5} &  \mathcal{X}_{1,4} &
\mathcal{X}_{0,6} &  
\mathcal{X}_{1,5}   
\end{matrix}\right) \\ 
\,&=\, \left(\begin{matrix}
\frac{(2-D)+4\mathcal{J}^2}{(D-2)\left(m/m_K\right)^2}\,\,
 & \frac{D(2-D)+2(4-3D)\mathcal{J}^2+4\mathcal{J}^4}{D(D-2)\left(m/m_K\right)^4}&\cdots 
 \end{matrix}\right)
\end{align}

In order to place bounds on the coupling $\tilde{\lambda}_{K}^{2,\rm flat}$ we use the same algorithm as described in section~\ref{sec:OptimizationSetUp}. We look for a vector  of dimension $N_{\rm cross}+2$ with components $\alpha_{k}$ which we use to construct a positive functional using as basis: 
\beq
\vec{W}^{\text{flat}}=\{g_{0}(m^2,J)\equiv 2,\,g_{2}(m^2,J)\equiv\frac{m_K^4}{m^4},\,\vec{\mathcal{X}}^{\text{cross}}\}
\eeq
For this new functional we demand the positivity property:
\beq\label{eq:positivityflat}
\vec{\alpha}.\vec{W}^{\rm flat}(m^2,J) = \alpha_{1}\,g_{0}(m^2,J) + \alpha_{2}\, g_{2}(m^2,J) +\sum_{i=1}^{N_{\text{cross}}}\alpha_{i+2}\,\mathcal{X}^{\text{cross}}_{i}(m^2,J) \,\geq\, 0 
\eeq
for all high-energy states that enter in the average $\langle \cdots \rangle'$ in \eqref{eq:averageprime}, excluding the lightest scalar $m_{K}$. Instead, we use this latter to set the normalization as:
\beq\label{eq:normflat}
\vec{\alpha}.\vec{W}^{\text{flat}}(m^2_{K},0)
\,=\,\pm 1\,.
\eeq
By combining the sum rules for the low energy coefficients \eqref{eq:g02sumrules}, null constraints \eqref{eq:nullflat},  the positivity and norm conditions in \eqref{eq:positivityflat} and \eqref{eq:normflat}, we can obtain bounds on the coupling:
\beq
\alpha^{(-)}_{1}\,g_0\,\frac{m_K^6}{8\pi G_D} + \alpha^{(-)}_{2}\,g_2\,\frac{m_K^{10}}{8\pi G_D} \leq \tilde{\lambda}^{2, \text{flat}}_{K} \leq -\alpha^{(+)}_{1}\,g_0\,\frac{m_K^6}{8\pi G_D} - \alpha^{(+)}_{2}\,g_2\,\frac{m_K^{10}}{8\pi G_D}
\eeq
where labels $(\pm)$ correspond to the choice of norm in \eqref{eq:normflat}. The optimized bounds are found by extremizing the corresponding linear combination of low-energy coefficients $g_0$ and $g_2$.  
\begin{table}[t]
\def\same{$''$}
\centering
\resizebox{0.75\textwidth}{!}{
{\small 
\begin{tabular}{|l|ll|ll|ll|}
\hline
 & \multicolumn{2}{c|}{I} & \multicolumn{2}{c|}{II}& \multicolumn{2}{c|}{III}
\\
$D=5$& lower & upper & lower & upper& lower & upper
\\
\hline
$g_0$& $0$ 
& 1.04491 
& 0 
& 1.04488 
& 0 
&  1.04482 
\\
$g_{0},\,B_{-2}$ & \same & \same & \same & \same  & \same & \same \\
$g_0$,\,$g_2$&
0.98938 
& 1.0195 
& 0.98944 
& 1.0189 
& 0.98952 
& 1.0179 
\\
$g_{0},\,g_{2},\,B_{-2}$ & 
 \same & \same & \same & \same  & \same & \same \\
\hline
\hline
$D=10$& lower & upper & lower & upper& lower & upper
\\
\hline
$g_0$&
 0
 & 1.01861
& 0
& 1.01858 
& 0
& 1.01843
\\
$g_{0},\,B_{-2}$ & 0.970
  &\same  & 0.970 & \same & 0.983 
  & \same \\ 
$g_0$\,, $g_2$&
0.99511 
& 1.0086 
& 0.99514
& 1.0082 
& 0.99531 
& 1.0072
\\
$g_{0},\,g_{2},\,B_{-2}$ & 
 0.9966 & \same & 0.9967 & \same  & 0.9968 & \same \\
\hline
\end{tabular}
}
}
\caption{Table with bounds on the residue $\tilde{\lambda}_{K}^{2,\text{flat}}$ at the lightest mass $m_{K}$, defined in \eqref{eq:ResFlat}, as a function of the types of sum rules included (left column); quotes indicate bounds that are unchanged from the line above.
We consider three different assumptions on the spectrum as depicted in figure \ref{fig:GridFlatoptions}. 
Note that the significant figures are not meant to suggest the degree of convergence of the bounds with respect to $N_{\rm cross}$ (described in the text), we kept them to show the differences between different options with comparable parameters.
}
\label{tab:flatbounds}
\end{table}

In table \ref{tab:flatbounds} we show the bounds obtained in spacetime dimensions $D=5$ and $D=10$, with the three different spectral assumptions in figure \ref{fig:GridFlatoptions} and using the Virasoro-Shapiro low-energy coefficients in \eqref{eq:g02VS}. We also report on the bounds obtained by using only the $g_{0}$ sum rule (setting $\alpha_{2}=0$ in \eqref{eq:positivityflat}).
All the bounds that do not include $B_{-2}$ (see below) used $N_{\rm cross}=26$ (up to order $\mathcal{O}(m^{-22})$ in the null constraints of \eqref{eq:Xnullab}).

We use the flat-space bootstrap of this section in $D=5$ (and without $B_{-2}$) to obtain the bounds that we label as ``flat" or $g=\infty$ in the section below, see table \ref{tab:strongXupbounds}. There we compare them against the CFT bounds at large values of the coupling $g$. 

\paragraph{Effect of antisubtracted sum rules}  We also explored the effect of including the ``antisubtracted'' $B_{-2}$ sum rules corresponding to $k=-2$ in \eqref{eq:Bkflat}.
The left-hand-side in this case is $\frac{8\pi G_D}{-t}$.
This is the only place where Newton's constant appear: these sum rules imposes the nontrivial fact that the amplitude must UV-complete gravity.
Technically, the $t=0$ pole precludes a Taylor expansion around the forward limit and instead one should integrate $B_{-2}(-p^2)$ against a wavepacket $\psi(p^2)$ with $0\leq p\leq p_{\rm max}$, as explained in \cite{Caron-Huot:2021rmr}.
This makes the analysis more technically involved: at large spin, the $m$ dependence of the right-hand-side of \eqref{eq:Bkflat} is a polynomial of high degree; in practice we numerically sample the mass instead of treating the polynomial exactly, as described in \cite{Caron-Huot:2021rmr,Albert:2024yap}.
In addition, it is necessary to impose positivity in the Regge limit $m\to\infty$ with fixed impact parameter $b=2J/m$, as explained in these references; this is also similar to our CFT approach described in section \ref{sec:numericalbootstrap}.
To make the table we used $p_{\rm max}=\sqrt{2}$ and $\psi(p)$ equal to
$p^{3/2}(1-p)^2$ (in $D=5$) or $p^3$ (in $D=10$) times an arbitrary polynomial of degree 8, together with $N_{\rm cross}=12$ null constraints.

The main observation from our explorations is that in $D=5$, including or not the antisubtracted functionals does not seem to affect the bounds, within numerical error.
This is consistent with the observation we made in section \ref{sec:numericsKonishi}
that the CFT antisubtracted functionals appear with negligible coefficients in the optimized functionals at strong coupling.

However, we would like to report an intriguing surprise: in $D=10$, inclusion of $B_{-2}$ has a nontrivial impact on the lower bounds when only the $g_0$ integrated constraint is used.
The improvement becomes less clearly significant when $g_2$ is used ($\lesssim 10^{-3}$), but it could still be significant and worth exploring further. It seems conceivable that with a larger basis of functionals, larger $p_{\rm max}$, or more aggressive spectral assumptions, one could obtain tighter bounds around the Virasoro-Shapiro result $\tilde{\lambda}_K^{2,\rm flat}=1$.
Why $B_{-2}$ seems to have a more significant effect in higher dimensions should also be understood better.

\subsection{Comparison with flat space limit at strong coupling}
\label{sec:numericalCFTvsFlat}

We return to the main focus of this paper and consider the CFT optimization problem of section~\ref{sec:numericsKonishi}, now at larger values of the coupling $g$. We also restrict the menu of functionals to the two integrated constraints and a few number of crossing functionals, in order to perform direct comparisons with the flat-space optimization problem just described , where the role of the two integrated constraints is played by the input of the first two EFT coefficients, and the crossing symmetry is encoded in null constraints. Our goal is to compare the results when using the same number of functionals in both the CFT and flat-space numerics. We show that the results for numerical bounds and extremal functionals in flat space approximate those of the CFT context (especially for $g=100$).

We begin by  considering a basis of functionals that includes the integrated constraints $I_{2}$ and $I_{4}$, supplemented with a number $N_{\text{cross}}$ of crossing functionals ($\widehat{X}_{\mS,\mT}$ or $X_{u,v}$):
\beq\label{eq:Wvector}
\vec{W}[\Delta, J] = \left(I_{2}[\Delta,J]\;\; I_{4}[\Delta,J]\;\;\vec{X}^{\text{cross}}[\Delta, J] \right).
\eeq
For instance for $N_{\text{cross}}=9$ we choose the following basis of Mellin crossing functionals $\widehat{X}_{\mS,\mT}$ with $\mS,\mT$ close to the crossing symmetric point:
\beq\label{eq:XcrossCFT}
\vec{X}^{\rm cross} \equiv \left(
\widehat{X}_{\frac{24}{5},\frac{28}{5}}\;\;
\widehat{X}_{\frac{24}{5},\frac{29}{5}}\;\;
\widehat{X}_{\frac{24}{5},\frac{30}{5}}\;\;
\widehat{X}_{\frac{25}{5},\frac{28}{5}}\;\;
\widehat{X}_{\frac{26}{5},\frac{24}{5}}\;\;
\widehat{X}_{\frac{26}{5},\frac{27}{5}}\;\;
\widehat{X}_{\frac{27}{5}\frac{25}{5}}\;\;
\widehat{X}_{\frac{22}{5},\frac{29}{5}}\;\;
\widehat{X}_{\frac{28}{5},\frac{22}{5}}
\right)
\eeq
and for lower $N_{\text{cross}}$ we choose a subset of this latter vector. This basis is the CFT counterpart of the flat-space basis of null constraints. The map between them is explained by the limit in eq.~\eqref{eq:Pstrongvsflat} and the definitions in eqs.~\eqref{eq:XstMellin}
 and \eqref{eq:XstFlat}. We will let the optimization algorithm discover the change of basis from discrete $\mS,\mT$ in \eqref{eq:XcrossCFT} to the flat-space basis of $s,t$ derivatives in \eqref{eq:Xcrossflat}.

Following section \ref{sec:OptimizationSetUp}, we search for the extremal functional that is positive: $\vec{\alpha}.\vec{W}[\Delta,J] \geq 0$, on a grid of high-energy states with scaling dimension at and above the first Regge trajectory (excluding Konishi operator). We use the strong-coupling approximation \eqref{eq:strong1stRegge} for the leading trajectory and introduce the second lightest scalar state at $\Delta_{\rm gap} \approx 4\sqrt{2 \pi g} -2 + \frac{1}{\sqrt{2 \pi g}} +\mathcal{O}(1/g)$. In summary, the grid of high-energy states where we impose positivity of the functional is set by the boundaries:
\beq\label{eq:GridHighEnergy}
\text{High-energy states}: \begin{cases}
    J  = 0 \quad &:\quad 2	\lesssim \frac{(\Delta+2)^2}{(\Delta_{K}+2)^2} \lesssim 5, \\ 
    J  = 2,4,\cdots,J_{\rm max} \quad &:\quad  1 \lesssim \frac{(\Delta+2)^2/(J+2)}{(\Delta_{K}+2)^2/2} \lesssim 5.
\end{cases}
\eeq
We explore the changes of lower and upper bounds on the Konishi OPE coefficients as we vary $N_{\rm cross}$ from $4$ to $9$ in the strong-coupling regime at $g= 3.7,\,10$ and $100$. We report the resulting CFT bounds in table \ref{tab:strongXupbounds}, together with the flat-space bounds of section \ref{sec:flatspaceNumerics} where we set $N_{\rm cross}$ as the number of null constraints.

\begin{table}[t]
\setlength\arrayrulewidth{1pt}
\renewcommand{\arraystretch}{1.25}
\centering
\resizebox{0.495\textwidth}{!}{
{\small 
\begin{tabular}{|c|l|l|l|l|}
\hline
\multicolumn{5}{|c|}{Lower bounds on $\tilde{\lambda}_{K}^2$ at large $g$} \\
\hline
 $N_{\text{cross}}$& \multicolumn{1}{c|}{g=3.7} & \multicolumn{1}{c|}{g=10}& \multicolumn{1}{c|}{g=100}& \multicolumn{1}{c|}{$\text{Flat}_{D=5}$}
\\
\hline
 4 & 1.0188 & 1.0160 & 0.98923 & 0.98636 \\
 5 & 1.0626 & 1.0160 & 0.98983 & 0.98783 \\
 6 & 1.0627 & 1.0172 & 0.99106 & 0.98820 \\
 7 & 1.0629 & 1.0173 & 0.99140 & 0.98856 \\
 8 & 1.0636 & 1.0175 & 0.99141 & 0.98857 \\
 9 & 1.0643 & 1.0178 & 0.99178 & 0.98881 \\
\hline
\end{tabular}
}
}
\resizebox{0.48\textwidth}{!}{
{\small 
\begin{tabular}{|c|l|l|l|l|}
\hline
\multicolumn{5}{|c|}{Upper bounds on $\tilde{\lambda}_{K}^2$ at large $g$} \\
\hline
 $N_{\text{cross}}$ & \multicolumn{1}{c|}{g=3.7} & \multicolumn{1}{c|}{g=10}& \multicolumn{1}{c|}{g=100}& \multicolumn{1}{c|}{$\text{Flat}_{D=5}$
 }
\\
\hline
 4 & 1.1078 & 1.0536 & 1.0321 & 1.0280 \\
 5 & 1.1065 & 1.0528 & 1.0296 & 1.0271 \\
 6 & 1.0965 & 1.0519 & 1.0293 & 1.0269 \\
 7 & 1.0959 & 1.0518 & 1.0293 & 1.0269 \\
 8 & 1.0899 & 1.0475 & 1.0243 & 1.0219 \\
 9 & 1.0895 & 1.0470 & 1.0241 & 1.0218 \\
\hline
\end{tabular}
}
}
\caption{Table with numerical bounds on  $\tilde{\lambda}^2_K$ in the strong coupling regime with $g = 3.7,\,10$ and $100$. For comparison, we include the flat-space bounds from sec.~\ref{sec:flatspaceNumerics} in spacetime dimension $D=5$. The columns show how the bounds change as we increase the number $N_{\text{cross}}$ of  CFT crossing functionals or null constraints in the flat-space context.  For all these bounds we use a grid given by \eqref{eq:GridHighEnergy} with $J_{\text{max}}=60$ and supplemented with few extra points at much larger spin.
}
\label{tab:strongXupbounds}
\end{table}
All these bounds were obtained with a relatively small choice for the maximum spin on the grid: $J_{\rm max}=60$. This choice is sufficient to give bounds convergent with spin for some fixed number of crossing functionals, notably $N_{\rm cross}=6\text{ and } 8$. However for other values such as $N_{\rm cross}= 7\text{ and }9$, convergence with spin requires supplementing the grid with extra few points at much larger values of spin in both  CFT and flat-space numerics. After doing this large-spin fix we compare the numerical bounds. 

\begin{figure}[t]
\centering
\includegraphics[width=.54\textwidth]{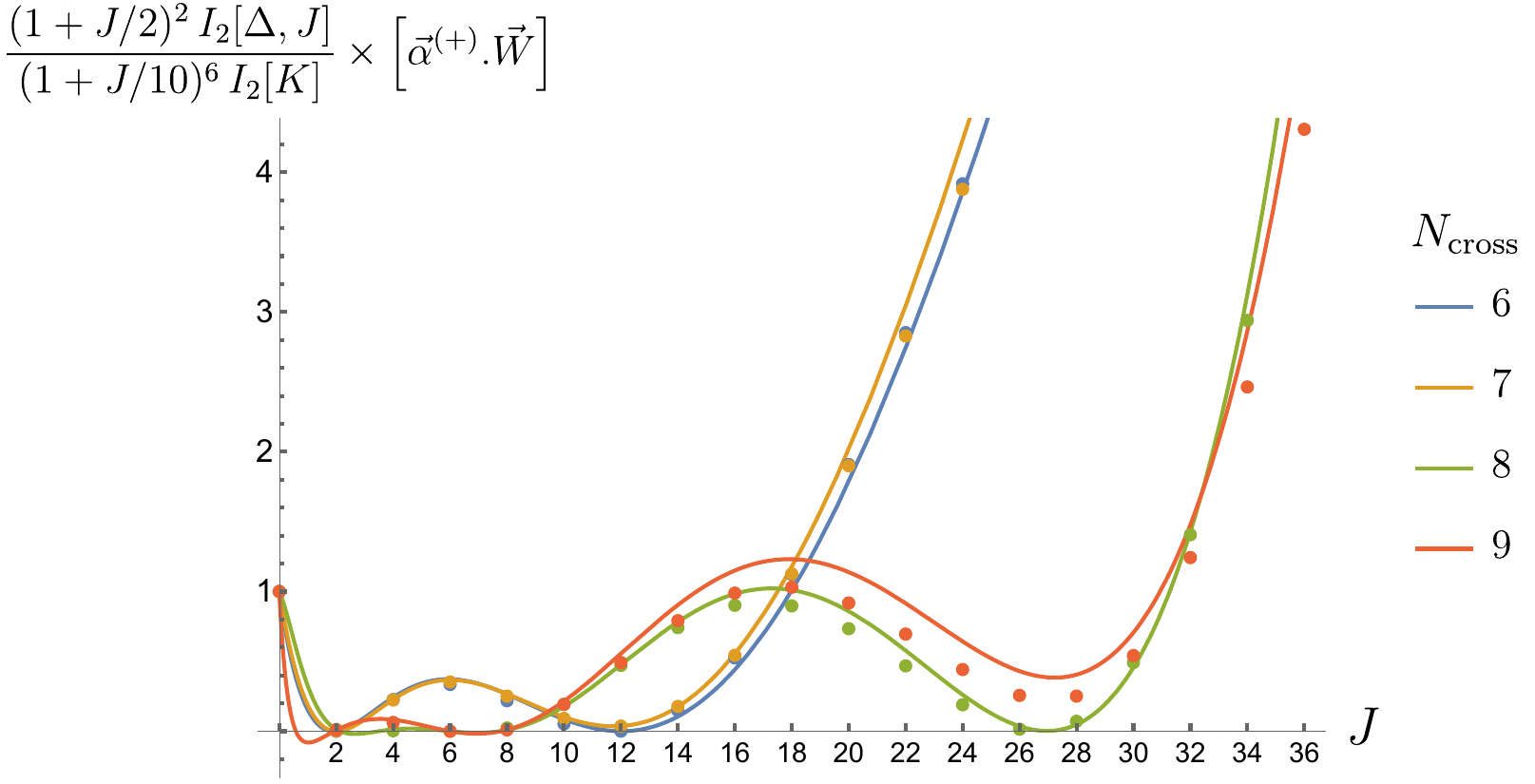}
\includegraphics[width=.45\textwidth]{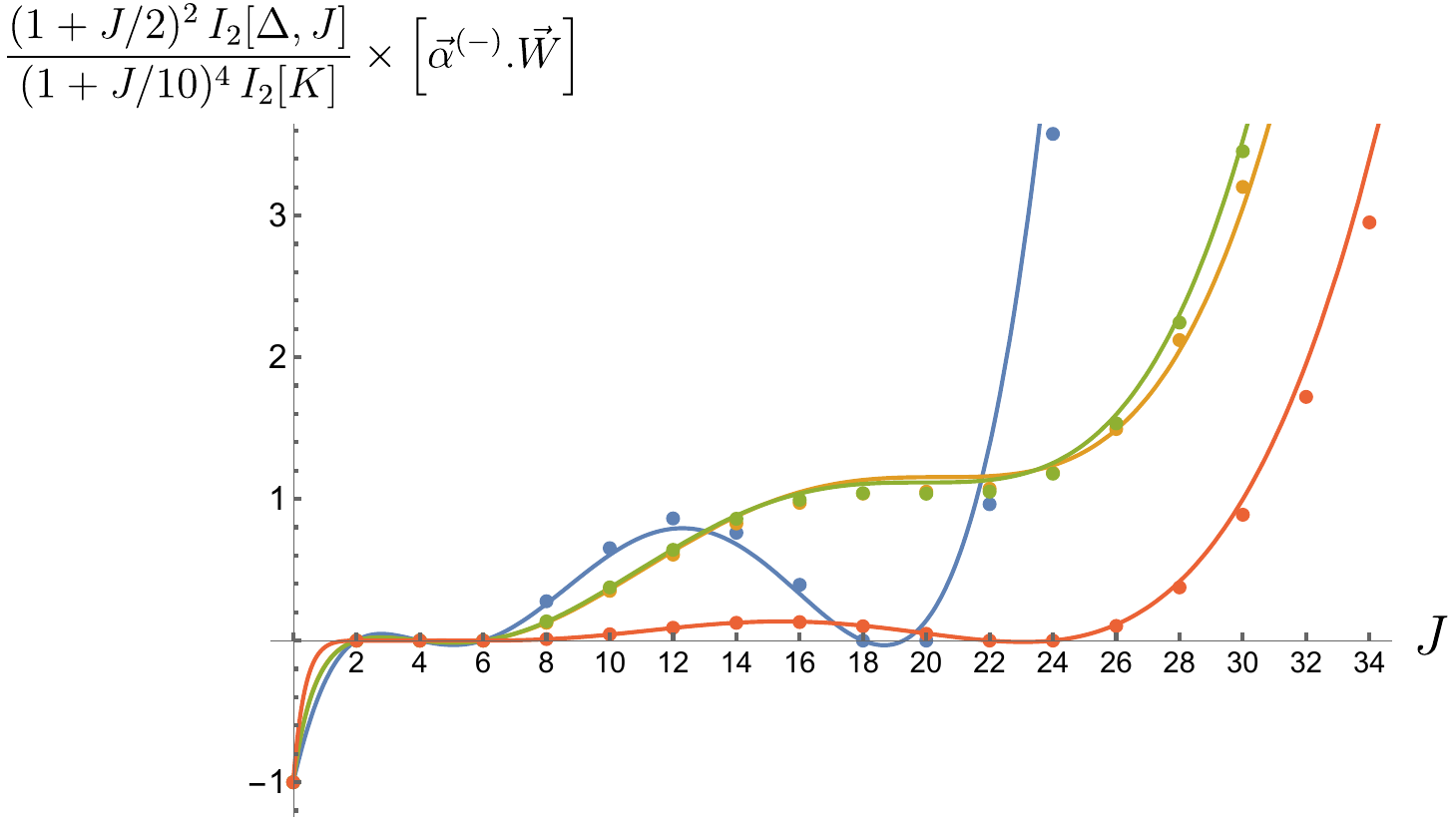}
\caption{Action of the extremal functional $\vec{\alpha}.\vec{W}$ on the first Regge trajectory. We show the CFT extremal functional (points), with coupling $g=100$, as well as the flat-space functional (solid lines) from subsection~\ref{sec:flatspaceNumerics}, as we change the number of crossing functionals $N_{\rm cross}$. On the left, the upper-bound extremal functionals and on the right the lower-bound case. We choose to normalize the vertical axis by $J$-dependent factors for better visualization. 
}
\label{fig:Fun2DplotNcross}
\end{figure}

\begin{figure}[t]
\includegraphics[width=1\textwidth]{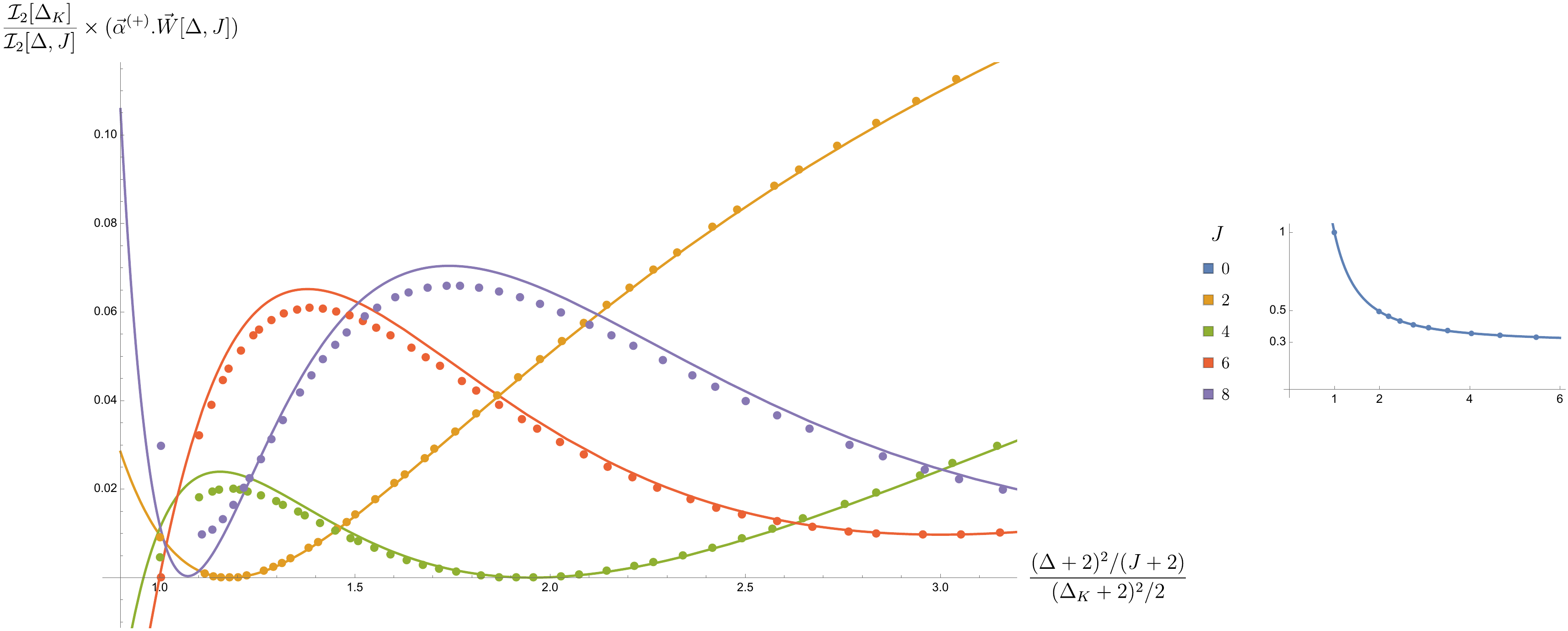}
\caption{Comparison of extremal functionals in CFT, with $g=100$,  and flat space for the upper-bound case. The solid color lines represent the flat-space functional and the dots correspond to the CFT functional. We use $N_{\text{cross}}=8,\, J_{\text{max}}=60$. On the right, we isolate the spin $J=0$ curve which lies high above the higher-spin curves. In the flat-space context the horizontal axis is given by $\frac{m^{2}/(J+2)}{m^{2}_{K}/2}$.}
\label{fig:upperCFTvsFlatg100}
\end{figure}

\begin{figure}[t]
\includegraphics[width=.9\textwidth]{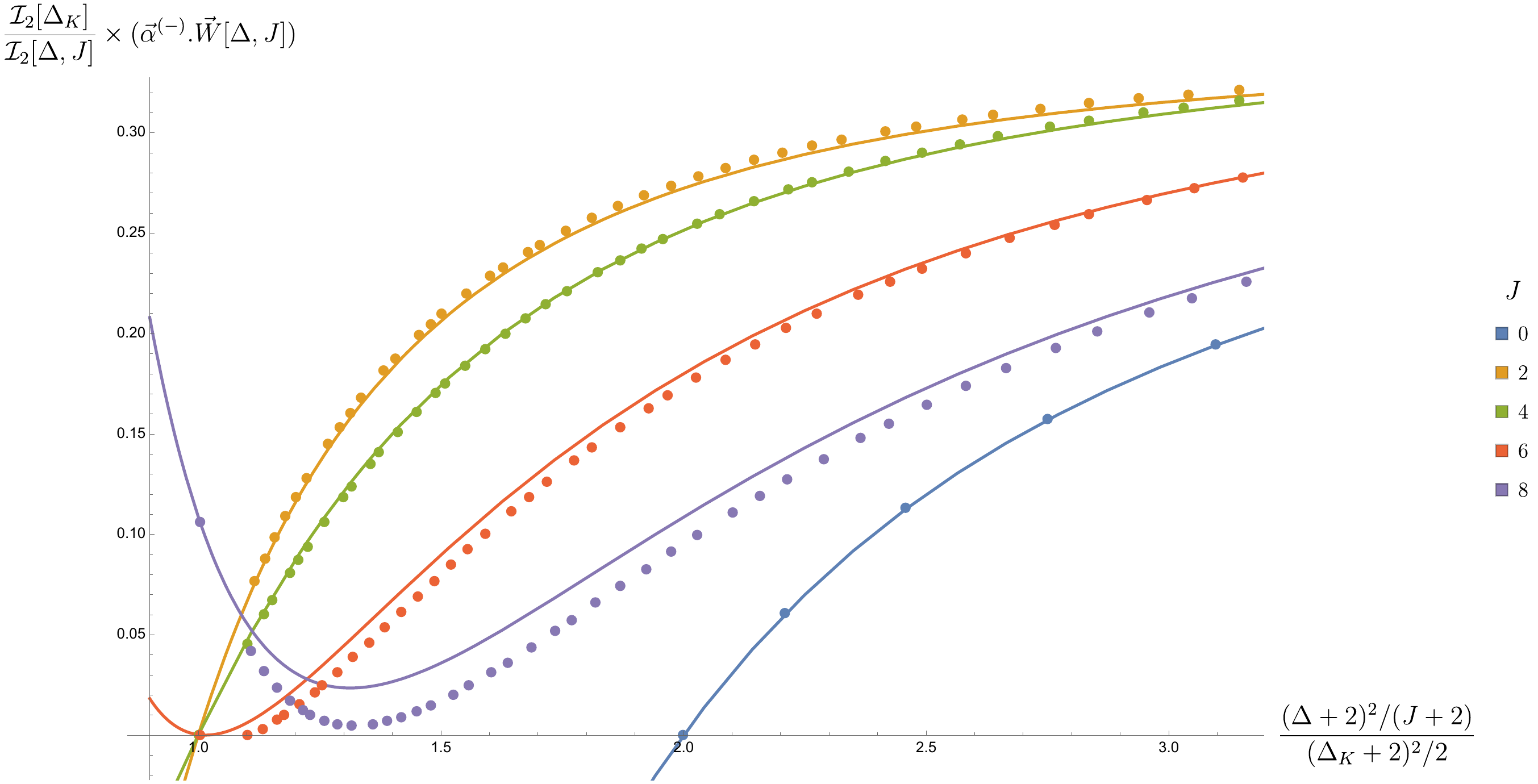}
\caption{Comparison of extremal functionals in CFT, with $g=100$,  and flat space in the lower-bound case. The solid color lines represent the flat-space functional and the dots correspond to the CFT functional. We use $N_{\rm cross}=6,\,J_{\rm max}=60$. In the flat space context, the horizontal axis is given by $\frac{m^{2}/(J+2)}{m^{2}_{K}/2}$.}
\label{fig:lowerCFTvsFlatg100}
\end{figure}

For the upper-bound case, we find that the difference between CFT and flat space is of order $\approx \frac{0.25}{g}$, which is of the same magnitude as the first $1/g$ correction in the analytic series of $\tilde{\lambda}_{K}^2$, see \eqref{eq:KtildeStrongSeries}. Similarly for the lower bound, we find the difference is of order $\approx \frac{0.30}{g}$. For instance, with $N_{\text{cross}}=9$, we obtain  the upper-bound difference: 
\beq
(\tilde{\lambda}^{2,\,\rm CFT}_{K})_{\text{upper}} - (\tilde{\lambda}^{2,\,\rm flat}_{K})_{\text{upper}} = \begin{cases}
    0.232\times 10^{-2}\,,\quad &g=100 \\
    0.252\times 10^{-1}\,,\quad &g=10 \\
    0.0677\,,\quad &g=37/10 \\
\end{cases}
\eeq
and for the lower bound:
\beq
(\tilde{\lambda}^{2,\,\rm CFT}_{K})_{\text{lower}} - (\tilde{\lambda}^{2,\,\rm flat}_{K})_{\text{lower}}  = \begin{cases}
    0.297\times 10^{-2}\,,\quad &g=100 \\
    0.290\times 10^{-1}\,,\quad &g=10 \\
    0.0755\,,\quad &g=37/10 \\
\end{cases}
\eeq
The comparison of CFT and flat-space numerics can also be made at the level of the extremal functional itself: $\vec{\alpha}.\vec{W}$.  For this purpose, after appropriate normalization,  we plot the action of the extremal functionals on the leading Regge trajectory in figure \ref{fig:Fun2DplotNcross}, for various values of $N_{\rm cross}$.
There, we see how the CFT extremal functional, with coupling $g=100$, lies close to the flat-space functional, including the position of their simple zeroes. Furthermore, in figures \ref{fig:upperCFTvsFlatg100} (upper bound) and \ref{fig:lowerCFTvsFlatg100} (lower bound), we show plots of the functionals acting on states with fixed spin and energy above the leading trajectory. These plots also show a good match between CFT and flat-space extremal functionals, specially for low values of spin. 

It is also interesting to analyze the relative weight of the contributions of functionals: $I_{2}[\Delta,J]$ and $I_{4}[\Delta,J]$, to the extremal functional. This corresponds to the ratio of the first two components in the SDPB vector $\vec{\alpha}$, given by:
\beq
\frac{\alpha^{(+)}_{1}}{\alpha^{(+)}_{2}} = \begin{cases}
    -127.99997\,,\quad &g=100, \\
    -127.997\,,\quad &g=10, \\
\end{cases} \quad\text{and}\quad 
\frac{\alpha^{(-)}_{1}}{\alpha^{(-)}_{2}} = \begin{cases}
    -128.00002\,,\quad &g=100, \\
    -128.001\,,\quad &g=10. \\
\end{cases}
\eeq
This result means that, at strong coupling, the optimization algorithm favors the linear combination of functionals: $I_4-128 I_2$. As anticipated below eq.~\eqref{limits I24},  at leading order in the strong coupling limit, the ``protected" part of this combination isolates the flat-space low-energy coefficient: $g_2\propto \zeta_5$. 

This observation motives a change of basis of functionals. So, we replace the second component of the vector $\vec{W}$, defined in \eqref{eq:Wvector}, by a linear combination of the integrated constraints $I_{2}$ and $I_{4}$:
\beq\label{eq:W2new}
W_{2} \equiv I_{4} \quad\longrightarrow\quad  W^{\text{new}}_{2} \,\equiv\,  \left(I_{4} -128\,I_{2}\right)\times \frac{\left(\Delta_{K}+2\right)^4}{3\times 2^{10}}\,.
\eeq
The rescaling factor serves to bring $W_{2}^{\text{new}}$ to the same order as the first component $W_{1}\equiv I_{2}$ in the strong coupling limit. From \eqref{limits I24}, the protected part of this new component is: 
\beq
\frac{\left[W_{2}^{\text{new}}\right]^{\text{protected}}}{\left[W_{1}\right]^{\text{protected}}} = \frac{\zeta_5}{2\zeta_3} -\frac{\zeta_3^2-2 \zeta_5}{\zeta_3\,(4\pi g)} +O(g^{-2})\,.
\eeq
At leading order this gives the same normalization we use in section~\ref{sec:flatspaceNumerics} for the flat-space vector, see eq. \ref{eq:g02VS}. It corresponds to the ratio between second and first flat-space Wilson coefficients (when setting string length to 1).

Under the change of basis \eqref{eq:W2new} the optimized bounds stay the same (e.g. $\vec{\alpha}.\vec{W} = \vec{\alpha}^{\text{new}}.\vec{W}^{\text{\text{new}}}$) but now we can compare the first two components of the new SDPB vector with the flat-space case:
\beq
\frac{\alpha^{\text{new}(+)}_{2}}{\alpha^{\text{new}(+)}_{1}} = \begin{cases}
 4.49\,,\quad &  g\to\infty\;(\text{flat space})  \\
    4.45\,,\quad &g=100 \\
    4.71\,,\quad &g=10 \\
     5.33\,,\quad &g=3.7 
\end{cases} \quad\text{and}\quad 
\frac{\alpha^{\text{new}(-)}_{2}}{\alpha^{\text{new}(-)}_{1}} = \begin{cases}
-8.01\,,\quad & g \to \infty \\
    -7.99\,,\quad &g=100 \\
    -7.88\,,\quad &g=10 \\
     -7.63\,,\quad &g=3.7 
\end{cases}
\eeq
This neat comparison gives more evidence of how good our CFT numerical bootstrap approximates the flat-space numerics at strong coupling.

\section{Discussion}
\label{sec:discussion}

We explored nonperturbative constraints on the correlator of four stress tensor multiplets in planar four-dimensional $\mathcal{N}=4$ sYM, 
extending the previous analysis from \cite{Caron-Huot:2022sdy} by adding ``integrated constraints'' from supersymmetric localization, detailed in 
section~\ref{sec:Spectral assumptions}.
We then combined this information with
a number of single-trace scaling dimensions known from integrability (section~\ref{sec:Spectral assumptions}), and general constraints from crossing symmetry (section~\ref{sec:Dispersive Constraints}).

Our main result is a set of rigorous lower and upper bounds on the OPE coefficient of the lightest unprotected single-trace scalar (the Konishi operator) at finite values of the `t Hooft coupling, shown in fig.~\ref{fig:ope-results} and recorded
in table~\ref{tab:bounds}.
The lower and upper bounds are relatively close to each other, thus providing a rigorous and nonperturbative error estimate on the value of these quantities at
finite coupling, which cannot be reached via any perturbative technique. 

We consider a range of couplings between $g=0.1$ to $g=3.7$.  The lower end is well within the regime of validity of the weak coupling expansion and fully agrees with it.  In this regime, certain ``antisubtracted'' functionals which were designed in \cite{Caron-Huot:2022sdy} to analytically solve the one-loop problem, namely $\Psi_0$ and $\Phi_{0,2}$, are found to be very important. Physically, antisubtraction means that these functionals 
(see \eqref{Bt}) are sensitive to the ultraviolet completion of the graviton exchange.
At intermediate couplings $g\sim 0.4$, the numerics transition to using a different set of functionals, and at strong coupling the unsubtracted crossing relations seem preferred.

We also showed how to obtain similar bounds for the correlator itself at various values of the cross-ratios (see figures in section \ref{sec:corr}).
Here we found it more difficult to control in practice the errors coming from the finite-spin truncation of the bootstrap problem, which makes the gap between lower and upper bounds an unreliable error estimate. The difficulty becomes particularly acute 
away from OPE limits. This could potentially be improved by incorporating large spin approximations to the blocks.

\paragraph{Do we expect error bars to firther reduce with more effort?}

We can answer this question in the limit of strong `t Hooft coupling. As discussed in section~\ref{sec:flat-space}, our CFT bootstrap problem then reduces to an analogous S-matrix problem: to bootstrap the Virasoro-Shapiro amplitude given the linear Regge trajectories and general principles.  In this context, supersymmetry localization fixes the coefficients of the leading two corrections to supergravity ($R^4$ and $D^4R^4$ terms).

Unfortunately, the solution of this toy problem appears to converge to bounds with finite error bars, albeit small (see table \ref{tab:flatbounds}).
This resonates with the observation in \cite{Berman:2023jys,Albert:2024yap} that the coefficients of $R^4$ and $D^4R^4$ in string theory are ``close'' to the boundary of the allowed region of theory space allowed by causality and unitarity, although
string theory is not exactly on the boundary.
Reference \cite{Eckner:2024ggx} provided a somewhat different approach to the problem and it would be interesting to see whether it
has already converged or can be improved further.

In addition,  as we mentioned in section~\ref{sec:numericalbootstrap},  the bounds obtained on the Konishi OPE coefficient depend most strongly on its scaling dimension, and to some extent on the gap to the next spin 0 single-trace; the bounds depend continuously on the latter, as shown in figure \ref{fig:boundspin0gap}. Further explorations show that the addition of the exact higher-spin spectrum has a subleading effect on our error bars and this becomes marginal at stronger coupling. 

This suggests that the study of the single-correlator bootstrap problem for the Konishi OPE coefficient is, in some sense, concluded: we only expect marginal improvements to fig.~\ref{fig:ope-results}.

\paragraph{How to further reduce the error bars?}  A common observation in the numerical CFT bootstrap is that mixed correlators help \cite{Poland:2022qrs}.
This means to consider a matrix of correlators and imposing positivity on the matrix itself.
In sYM theory it would be natural indeed to add other protected single-trace operators with higher R-charges, corresponding to graviton Kaluza-Klein modes in the dual $AdS_5\times S^5$. In the CFT this would be technically hard but feasible.

In the flat space limit it is relatively straightforward to estimate the effect of adding KK modes: the most optimistic outcome is that this will convert the 5d bootstrap problem to a 10d one.  As seen in table~\ref{tab:flatbounds}, imposing 10d symmetry reduces the error bars by a factor two or so, but does not eliminate them.  Curiously, we find that while adding the antisubtracted sum rules (sensitive to the ultraviolet completion of the graviton pole) has a negligible effect on the bounds in $D=5$, it can have a meaningful effect in $D=10$.
It could be worth exploring more aggressive assumptions in this context.

A more novel and potentially fruitful approach would be to consider mixed correlators involving massive states like the Konishi operator, either from the CFT or S-matrix perspective.

\paragraph{Other directions}  In \cite{Chester:2023ehi} upper bounds on the OPE data of the lightest unprotected operator
in the same theory where obtained at finite rank $N_c$ using a combination of 
integrated constraints and standard bootstrap techniques (crossing symmetry and unitarity).
For large values of $N_c$ and weak `t Hooft coupling, their lightest unprotected operator is mostly single-trace and can be identified with the Konishi operator, and in this regime we expect our bounds to be compatible. However at finite `t Hooft coupling there is a transition (level repulsion) above which the lightest unprotected operator becomes mostly double-trace. On the other hand, at all couplings we follow the single-trace thanks to our use of dispersive sum rules. Nevertheless, as noted in section~\ref{sec:OptimizationSetUp}, it is possible to design dispersive functionals
that probe the anomalous dimension and OPE coefficient of individual double-traces. 
This could allow for a more direct comparison with \cite{Chester:2023ehi}. 

As discussed, the exact knowledge of the higher-spin spectrum seems to have a subleading effect on our numerical bounds. Therefore, it could be interesting to study analytically some variant of the problem, where for example one would input some spin-0 spectral information but only impose the unitarity bound for other spins.

A technical improvement, which could also be important for applications of dispersive functionals beyond sYM theory,  would be to derive polynomial approximations for the $\Delta$-dependence of dispersive functionals.
These should make it possible to include a vastly larger number of functionals.
For standard crossing functionals this was historically a very important step \cite{Poland:2011ey}.

Lastly, one can wonder about the generalization of our setup to other theories such as ABJM theory \cite{Aharony:2008ug} for which many similar ingredients are available.

\acknowledgments 
We gratefully acknowledge Julius Julius and Nika Sokolova for discussions and sharing integrability data with us. We also acknowledge the participants of the workshop ``Integrability meets bootstrap'' at ICTP-SAIFR, Sao-Paulo,  where this work was presented. We also thank Andrea Guerrieri, Shota Komatsu and Sasha Zhiboedov for useful discussions.
S.C.H.'s work is supported in parts by the National Science and Engineering Council of Canada (NSERC) and by the Canada Research Chair program, reference number CRC-2022-00421.  S.C.H.'s work is additionally supported by a Simons Fellowships in Theoretical Physics and by the Simons Collaboration on the non-perturbative Bootstrap.  Z.Z. is funded by the European Research Council (ERC) under the European Union's Horizon 2020 research and innovation programme (grant agreement number 949077). The work of F.C. is supported in part by the Simons Foundation
grant 994306 (Simons Collaboration on Confinement and QCD Strings), as well the NCCR
SwissMAP that is also funded by the Swiss National Science Foundation. 

\appendix
\section{Regge limit of dispersive functionals}
\label{app:regge-limit}

In this section we derive analytic expressions for the Regge limits of the functionals used in this paper.

\subsection{Regge limit of position functional \texorpdfstring{$B_v$}{}}

Let us begin with the ``antisubtracted'' $B_v$ functional:
\be
B_v[\Delta,\j] = \int\limits_v^\infty dv'\!\! \int\limits_0^{(\sqrt{v'}-\sqrt{v})^2} du' \frac{v'-u'}{\pi^2v\sqrt{v^2-2(u'+v')v+(u'-v')^2}}
\dDisc_s[ G^{\Nfour}_{\Delta,\j}(u',v')],
\ee
with $G^{\Nfour}_{\Delta,\j}(u',v')=u'^{-4}G_{\Delta+4,\j}(u',v')$. At large twist, the integral will be pushed to the Regge limit $u',v'\to\infty$ with $\sqrt{v'}-\sqrt{u'}$ fixed. We follow the method of \cite{Caron-Huot:2021enk}. Switching to the radial variables of \cite{Hogervorst:2013sma},
\be
 r=\sqrt{\r\rb},\quad \eta=\frac{\r+\rb}{2\sqrt{r\rb}},\quad \r=\frac{1}{(\sqrt{1-w}+\sqrt{-w})^2},\quad
 \rb=\frac{1}{(\sqrt{1-\wb}+\sqrt{-\wb})^2},
\ee
the integral can be rewritten exactly as
\be \label{Bv r}
B_v[\Delta,\j] = \frac{1}{16\pi^2v}\int_{\sqrt{v}}^\infty \frac{\eta d\eta}{\sqrt{\eta^2-v}}
\int_0^{r_{\rm max}(\eta)} \frac{dr}{r^4}(1-r^4) \frac{(1+r^2)^2-4\eta^2r^2}{\sqrt{(1+r^2)^2-4vr^2}} \dDisc_s[G^{\Nfour}_{\Delta,\j}(u',v')]
\ee
with $r_{\rm max}(\eta)=\eta-\sqrt{\eta^2-1}$ determined from the constraint $\r<1$.
This matches with $B_{2,v}$ in eq.~(2.45) of \cite{Caron-Huot:2021enk} up to multiplication by the factor
$\frac{u'v'}{v}=\frac{((1+r^2)^2-4\eta^2r^2)^2}{(4r)^4v}$
which implements the antisubtraction.
From here on we'll plug in the explicit formula for $G^{\Nfour}$ to write things in terms of normal blocks. Defining Regge moments in the standard way
\be
 \Pi_{k,\eta}^{\Df}[\Delta,\j] = \int_0^{\rm r_{\rm max}(\eta)} dr r^{k-2} \dDisc_s[u'^{-\Df}G_{\Delta,\j}(u',v')],
\ee
we can expand \eqref{Bv r} as $r\to 0$ into Regge moments
\be \label{Bv Regge}
B_v[\Delta,\j] \!=\! \frac{1}{16\pi^2}\int_{\sqrt{v}}^\infty \frac{\eta d\eta}{v\sqrt{\eta^2-v}}
\left( \Pi_{-2,\eta}^{\Df=4}[\Delta{+}4,\j] \ + \ (2v{+}1{-}4\eta^2)\Pi_{0,\eta}^{\Df=4}[\Delta{+}4,\j] \ + \ O(\Pi_{2,\eta}) \right).
\ee
The idea is that when acting on heavy operators, the omitted higher moments are suppressed by powers of the twist.
The dependence on $(\Delta,\j)$ will be best captured by the following Casimir-friendly combinations,
related to those of \cite{Caron-Huot:2021enk} by $\Delta\mapsto\Delta{+}4$:
\be
 m^2\equiv (\Delta-\j+1)(\Delta+\j+3),\qquad
 \eta_{\rm AdS} \equiv  1+ \frac{2(\j+1)^2}{m^2}.
\ee

Explicit expressions for these Regge moments,
or more precisely their harmonic decomposition,
are given in eqs.~(3.19) and (2.52) of \cite{Caron-Huot:2021enk} (omitting $(-1)^\j=1$ since $\j$ is even throughout the present paper):
\begin{align}
 \frac{\Pi_{k,\eta}^{\Df}[\Delta{+}4,\j]}{2\sin^2(\frac{\Delta-\j-2\Df}{2})} &=
 \frac{4^{2\Df-1}\pi^{d-2}}{(m^2)^{2\Df+k-1}b_{\Delta+4,\j}}
 \int_0^\infty \frac{d\nu}{2\pi} \gamma_{2\Df+k-1}(\nu)^2 \rho(\nu)
 \cP_{\frac{2-d}{2}+i\nu}(\eta)\cP_{\frac{2-d}{2}+i\nu}(\eta_{\rm AdS})
 \nonumber\\ &\qquad\qquad \times
 (1+O(m^{-2})),
\end{align}
where $b_{\Delta+4,\j}$ is a product of $\Gamma$-functions (independent of $\Df$) in (2.32) of \cite{Caron-Huot:2021enk}, 
$\rho(\nu)=\nu^2$ in $d=4$, and
\begin{align}
 \gamma_{a}(\nu) &= \Gamma\!\left(\tfrac{1+a-d/2-i\nu}{2}\right)\Gamma\!\left(\tfrac{1+a-d/2+i\nu}{2}\right),
\\
 \cP_\j(\eta) &= {}_2F_1\big({-}\j,\j+d-2,\tfrac{d-1}{2},\tfrac{1-\eta}{2}\big).
\end{align}
To deduce the limit of $B_v$ in \eqref{Bv Regge}, we integrate over $\eta$ using the following integral:
\be
 \int_{\sqrt{v}}^\infty \frac{\eta d\eta}{v\sqrt{\eta^2-v}} \cP_{-1+i\nu}(\eta)\big|_{d=4} =  \frac{\pi}{2\nu} \coth(\tfrac{\pi\nu}{2})
 B_v(\nu),
 \qquad B_v(\nu)\equiv \frac{{}_2F_1\big(\frac{-i\nu}{2},\frac{i\nu}{2},1,1{-}v\big)}{v}, \label{Bv transform}
\ee
which gives finally
\be
 \frac{B_v[\Delta,\j]}{2\sin^2(\frac{\Delta-\j}{2})} \to \frac{2^9\pi}{m^{10} b_{\Delta+4,\j}}
\int_0^\infty \frac{d\nu}{2\pi} \coth(\tfrac{\pi\nu}{2}) \gamma_5(\nu)^2 \nu \cP_{-1+i\nu}(\eta_{\rm AdS})
B_v(\nu) (1+O(m^{-2})) \label{Bv asympt 0}.
\ee
This is the key step in this calculation.
The dependence on $m$ is explicit, and the integral depends only on the functional parameter $v$ and bulk impact parameter $\eta_{\rm AdS}$.

For our applications, it is useful rewrite the normalization in terms of the mean free coefficients of  $\mathcal{N}=4$ sYM in \eqref{OPEfree N4}, which we repeat here for convenience
(see ie. \cite{Alday:2017vkk,Caron-Huot:2018kta} for analogous expressions for arbitrary Kaluza-Klein modes)
\be
 \OPEfree_{\Delta,\j} = 2(\Delta+2)(\j+1)
 \frac{\Gamma\big(\tfrac{\Delta-\j}{2}+1\big)^2\Gamma\big(\tfrac{\Delta+\j}{2}+2\big)^2}{\Gamma(\Delta-\j+1)\Gamma(\Delta+\j+3)}.
\ee
We then find (exactly, not only approximately):
\be
\frac{2^9\pi}{m^6 b_{\Delta+4,\j}} =\frac{2^{9}(\eta_{\rm AdS}^2-1)}{\pi \lambda^{2,\rm free}_{\Delta,\j}}\,.
\ee
Using further the explicit form of $\cP$ in $d=4$, with $\beta_{\rm AdS}=\cosh^{-1}(\eta_{\rm AdS}) \equiv \log\frac{\Delta+\j+3}{\Delta-\j+1}$,
\be
 \nu \cP_{-1+i\nu}(\eta_{\rm AdS}) = \frac{\sin(\nu\beta_{\rm AdS})}{\sinh(\beta_{\rm AdS})},
\ee
we finally rewrite \eqref{Bv asympt 0} explicitly as
\be\begin{aligned}
 \frac{B_v[\Delta,\j]\ m^4\OPEfree_{\Delta,\j}}{2\sin^2(\frac{\Delta-\j}{2})} & \to B_v^{\rm Regge}[\etaAdS] \times (1+O(m^{-2})),
 \\
B_v^{\rm Regge}[\etaAdS]&\equiv 4\sinh(\beta_{\rm AdS}) \int_0^\infty d\nu \frac{\cosh\big(\tfrac{\pi \nu}{2}\big)}{\sinh\big(\tfrac{\pi \nu}{2}\big)^3}
 \nu^2(\nu^2+4)^2 \sin\left(\nu\beta_{\rm AdS}\right) B_v(\nu).
\label{Bv asympt}
\end{aligned}\ee
This is the main result of this subsection.
For $v>1$, $B_v(\nu)$ is a bounded oscillatory function and therefore the integral converges exponentially in $\nu$.
For $v<1$, $B_v(\nu)$ grows exponentially with $\nu$ 
although the integral \eqref{Bv asympt} remains convergent.
However, the rapid oscillations (especially at large impact parameter $\beta_{\rm AdS}$) make the integral numerically nontrivial; below we describe a different, more stable, formula.

\subsection{Regge limit of \texorpdfstring{$\widehat{B}_\mT$}{} and other antisubtracted functionals}

We can deduce asymptotics of $\widehat{B}_\mT$ by using a Mellin representation of \eqref{Bv transform}:
\be\label{Bv Mtransform}
B_v(\nu) = \frac12\int \frac{d\mT\ v^{\frac{\mT}{2}-4}}{4\pi i} \Gamma(\tfrac{\mT-2}{2})^2 \Gamma(\tfrac{8-\mT}{2})^2
\widehat{B}_{\mT}(\nu), \quad
\widehat{B}_{\mT}(\nu)\equiv \frac{4}{\pi^2}\sinh(\tfrac{\pi\nu}{2})^2\frac{\Gamma(\tfrac{\mT-i\nu-6}{2})\Gamma(\tfrac{\mT+i\nu-6}{2})}{(6-\mT)\Gamma(\tfrac{\mT-2}{2})^2}.
\ee
Since the $B_v$ and $\widehat{B}_{\mT}$ functionals are related to each other by a similar Mellin transform (see (3.13) of \cite{Caron-Huot:2022sdy}), one can imagine that
the asymptotics of $\widehat{B}_{\mT}[\Delta,\j]$ are simply obtained by substituting $B_v(\nu)$ in \eqref{Bv asympt} for $\widehat{B}_{\mT}(\nu)$.
This is essentially correct up to one subtlety: the contours don't match since the usual Mellin contour is to the left of ${\rm Re}\,\mT=6$, but \eqref{Bv Mtransform} holds with ${\rm Re}\,\mT> 6\pm i\nu$.
We can correct for the discrepancy by adding a residue
and we find
It is actually slightly more subtle: contours don't match, so one needs to add to \eqref{Bv asympt} its residue at $\nu=i(6-\mT)$.
The exact expression, normalized with the factors on the left of \eqref{Bv asympt}, is
\begin{align}
\widehat{B}^{\rm Regge}_{\mT}[\etaAdS]
=&\ 4\sinh(\beta_{\rm AdS}) \Bigg[\int_0^\infty d\nu \frac{\cosh\big(\tfrac{\pi \nu}{2}\big)}{\sinh\big(\tfrac{\pi \nu}{2}\big)^3}
 \nu^2(\nu^2+4)^2 \sin\left(\nu\beta_{\rm AdS}\right) \widehat{B}_{\mT}(\nu)
\nonumber\\ &\hspace{25mm} + \frac{16 (\mT-4)^2(\mT-8)^2\Gamma(\mT-5)}{\pi\tan\big(\pi\tfrac{\mT-6}{2}\big)\Gamma\big(\tfrac{\mT-2}{2}\big)^2}
\sinh((\mT{-}6)\beta_{\rm AdS})\Bigg].
\label{Bt asympt}
\end{align}
This can be simplified further by observing that the $\nu$ integral is essentially a representation of a hypergeometric function, and so it can be done exactly.
Omitting steps, we thus obtain a compact expression for the Regge limit of $\widehat{B}_\mT$:
\be
\widehat{B}^{\rm Regge}_{\mT}[\etaAdS] =
 \frac{2^{12}\times 3}{\pi^2(6-\mT)\eta_{\rm AdS}^3}
\Bigg[
x\ {}_2F_1\big(4,\tfrac{5-\mT}{2},\tfrac{3}{2},x\big)+x^2\ {}_2F_1\big(4,\tfrac{7-\mT}{2},\tfrac{5}{2},x\big) \Bigg]_{x=1-\eta_{\rm AdS}^{-2}}\,.
\label{Bhat nice}
\ee
Possibly, this could be derived directly by taking a suitably limit of the Mack polynomials.

As a sanity check, we verified numerically that performing the Mellin transforms recovers \eqref{Bv asympt}. In fact, this approach provides a stable way to evaluate the Regge asymptotics of $B_v$ and other derived antisubtracted functionals,
which we can all write in the form:
\be \label{mellin X regge}
 X^{\rm Regge}[\etaAdS] = \int_{-\infty}^{+\infty} \frac{dy}{4\pi}\ \tilde{X}(y)\ \widehat{B}_{5+iy}^{\rm Regge}[\etaAdS],
\ee
where $X\in \{ B_v,\Phi_{\ell_1,\ell_2},\Psi_\ell\}$.
For example, the kernel corresponding to $B_v$ is simply (from \eqref{Bv Mtransform})
\be \label{Bv Regge numerics}
 \tilde{B}_v(y) = \frac{\pi^2 (1+y^2)^2}{32\cosh\big(\tfrac{\pi y}{2}\big)^2} v^{\frac{iy-3}{2}}\,.
\ee
To evaluate a large number of functionals, a very efficient method 
is to change variable to $y=\sinh(Y)$ and simply perform a Riemann sum  from $Y\sim -5$ to $Y\sim 5$ in steps of 0.1 or smaller,
adjusting the parameters to achieve the desired accuracy (smaller step sizes are required for large $\etaAdS$).
As the integral is exponentially convergent, this simple method can produce hundreds of digits of precision with very little effort.
In addition, the same values
$\widehat{B}^{\rm Regge}_{\mT}[\etaAdS]$ can be recycled to compute many different functionals.


\paragraph{Other derived functionals}
The $\Phi_{\ell_1,\ell_2}$ kernel is expressed as a difference between $\Phi_\ell$ kernels in \cite{Caron-Huot:2022sdy}:\footnote{
The $\Phi_\ell(y)$ kernel is defined in that reference with a different endpoint, however this integration constant cancels out in all relevant combinations $\Phi_{\ell_1,\ell_2}$ so we can ignore it here.}
\begin{align} \label{Phi ker}
\Phi_{\ell_1,\ell_2}(y) &\equiv \Phi_{\ell_1}(y)-\frac{\Phi_{\ell_1}^\infty}{\Phi_{\ell_2}^\infty}\Phi_{\ell_2}(y),\\
\Phi_\ell(y) &= \frac{i\pi^2}{16}(1+y^2)^2\int_{\infty}^y \frac{dy'\ A_\ell(y')}{\cosh\big(\tfrac{\pi y'}{2}\big)^2},
\end{align}
where $\Phi_\ell^\infty$ is the leading term in the $y\to\infty$ limit and $A_\ell(y)\equiv c_\ell a_\ell(y)$ is a Mack polynomial satisfying the following recursion
\be
 A_\j(y) = \frac{(\j+1)^2(\j+3)}{4\j(2\j+1)(2\j+3)} A_{\j-2}(y) - \frac{i y}{2\j} A_{\j-1}(y),
\ee
with seeds $A_0(y)=1$, $A_1(y) = \frac{-iy}{2}$.
The integral for $\Phi_\ell$ can then be computed as a finite sum using the following primitive for a power law:
\be
 \frac{\pi}{4} \int_\infty^y \frac{dy'\ y'^n}{\cosh\big(\tfrac{\pi y'}{2}\big)^2} = \sum_{a=0}^{n} \frac{n! y^{n-a}}{(n-a)!\pi^a} {\rm Li}_a(-e^{-\pi y})\,.
\ee
Evaluating $\Phi_{\ell_1,\ell_2}^{\rm Regge}[\eta_{\rm AdS}]$ using
\eqref{mellin X regge} numerically boils down to evaluating $\Phi_\ell(y)$ on a number of sampling points;
the same kernel can then be used for any value of $\etaAdS$.

Finally, we evaluated the Regge limit of the $\Psi_\ell$ kernel
by writing it in the following way:
\be
\Psi_\ell(y) = \frac{\pi^2 (1+y^2)^2}{32\cosh\big(\tfrac{\pi y}{2}\big)^2} A_\ell(y)
-\int_y^\infty \frac{dy'\ \tilde{\Psi}_\ell(y')}{\cosh\big(\tfrac{\pi y'}{2}\big)^2}\,, \label{Psi kernel exact}
\ee
where $\tilde{\Psi}_\ell(y)$ turns out to be a combination of $A_\ell(y)$ times harmonic numbers,
plus a polynomial part, given by the following closed-form expression:
\be
\tilde{\Psi}_\ell(y) = A_\ell(y)\left( \frac12 H_{\frac{-1+iy}{2}} + \frac12 H_{\frac{-1-iy}{2}} + \Psi_\beta \right)
+\sum_{j=0\,\rm even}^{\ell-2}
\frac{(\ell+2)!^2(2j+5)!}{(j+2)!^2(2\ell+4)!}\frac{A_j(y)}{(\ell-j)(\ell+j+5)}\,.
\ee
The kernel $\Psi_\ell(y)$ can thus be evaluated to high accuracy at the sampling points mentioned above by numerically integrating \eqref{Psi kernel exact}.

\subsection{Regge limit of integrated constraints and Polyakov-Regge blocks}

At strong coupling, the integrated constraints vanish like a power of $1/\lambda$ (see \eqref{limits I2}-\eqref{limits I4}) and so they project out gravity and probe only contact interactions.
They are thus expected to behave like sum rules of spin 4 or faster convergence.  Indeed, we observe numerically that $\frac{I_k[\Delta,\j]\ m^4\OPEfree_{\Delta,\j}}{2\sin^2(\frac{\Delta-\j}{2})}\sim \frac{1}{m^4}$.

Similarly, the unsubtracted Polyakov-Regge block $\widehat{\cal P}^{\Nfour}(\mS,\mT)$  in \eqref{PR Mellin explicit} vanishes like $1/m^4$ (this is more or less explicit from its behavior $\sim \mS^{-2}$ at large $\mS$ with fixed $\mU$).
The same is true for the Polyakov-Regge block and $\widehat{\cal P}^{\Nfour}(u,v)$ for generic fixed $u,v$ since the Mellin transform is dominated by $\mS,\mT=O(1)$. We can thus neglect both integrated constraints and unsubtracted blocks in the Regge limit, compared with antisubtracted functionals.

\section{Integrated constraints in Mellin space} \label{app:Integrated Mellin}

\subsection{Derivation and checks}

As discussed in section~\ref{sec:Integrated constraints}, supersymmetric localization constraints the integrals \eqref{eq:I-prot}
over the four-point stress-tensor correlator.
By substituting the Mellin representation \eqref{eq:mellinRep}, we obtain the Mellin form of the integraed constraints
\be \label{Ip Mellin appendix}
 I_p =
\iint\!\!\frac{d\mS\,d\mT}{(4\pi i)^2}\,
\Gamma\!\left(\Df-\tfrac{\mS}{2}\right)^2\Gamma\!\left(\Df-\tfrac{\mT}{2}\right)^2\Gamma\!\left(\Df-\tfrac{\mU}{2}\right)^2
I_p[\mS,\mT] \widehat{\cH}(\mS,\mT)
\ee
where $\mS+\mT+\mU=16$ as in the main text and the kernels are given as the Mellin transforms:
\begin{subequations}
\label{I24 kernels def}
\begin{align}
 I_2[\mS,\mT]&\equiv -\frac{1}{2\pi^2} \int d^4x\ |x|^{\mS-8}|e-x|^{\mT-8}\,, \label{I2 kernel def} \\
 I_4[\mS,\mT]&\equiv -\frac{16}{2\pi^2} \int d^4x\ |x|^{\mS-8}|e-x|^{\mT-8} (1+x^2+(e-x)^2) \bar{D}_{1,1,1,1}(x^2,(e-x)^2)\,.
\label{I4 kernel def}
\end{align}
\end{subequations}
The Mellin transform $I_2[\mS,\mT]$ was given in \cite{Alday:2021vfb} (with a shift $\mS^{\rm here}=\mS^{\rm there}+4$), but not that for $I_4$.
Let us thus rederive the former here, using a method which will generalize.

Our strategy is based on the fact that the Mellin transform is a meromorphic function. By expanding the integral \eqref{I2 kernel def} around $x\to 0$ we can predict its $\mS$-channel poles.  Since the kernels are crossing symmetric, this predicts all poles, and hopefully we can spot a pattern and resum.
Explicitly, passing to radial coordinates, the first few $\mS$-poles are
\begin{align} I_2[\mS,\mT]\Big|_{\rm \mS-poles} &= -\frac{2}{\pi} \int_0^\pi \sin^2\theta d\theta \int_0^1
\frac{dr}{r} r^{\mS-4} (1+r^2-2r\cos\theta)^{\frac{\mT}{2}-4}
\\
 &= \frac{-1}{\mS-4} -\frac{(3-\tfrac12\mT)(4-\tfrac12\mT)}{2(\mS-2)} - \frac{(3-\tfrac12\mT)(4-\tfrac12\mT)^2(5-\tfrac12\mT)}{12 \mS}
\nonumber\\ &\quad -\frac{(3-\tfrac12\mT)(4-\tfrac12\mT)^2(5-\tfrac12\mT)^2(5-\tfrac12\mT)}{144 (\mS+2)} + \ldots
\label{I2 s-poles}
\end{align}
The numerators can be recognized as Pochhammer functions, ie. ratios of Gamma functions. This suggests that $I_2$ itself is a product of Gamma functions, and indeed we find that all the residues precisely match a simple ansatz:
\be \label{I2 kernel}
 I_2[\mS,\mT] = -\frac12\frac{\Gamma\big(\tfrac{\mS}{2}-2\big)\Gamma\big(\tfrac{\mT}{2}-2\big)\Gamma\big(\tfrac{\mU}{2}-2\big)}
 {\Gamma\big(4-\tfrac{\mS}{2}\big)\Gamma\big(4-\tfrac{\mT}{2}\big)\Gamma\big(4-\tfrac{\mU}{2}\big)}\,.
\ee
We confirm this guess by comparing with direct numerical integration of \eqref{I2 kernel def}, finding precise agreement.

For $I_4$ we proceed with the same method.
The novel feature is the presence of $\log(r)$ in $\bar{D}_{1,1,1,1}$, which produces double poles at the locations
in \eqref{I2 s-poles}, instead of single poles.
The residues again take a simple form, which suggests that $I_4[\mS,\mT]$
contains the same overall product of Gamma functions.
From here, we can create double poles simply by multiplying by the harmonic function $H_{\frac{\mS}{2}-3}$, and summing over channels.
Another salient feature is that the residues are simple polynomials in $\mT$, owing to the nature of the $r\to0$ limit of the integral.
However, the single-poles found after multiplying \eqref{I2 kernel} by $H_{\frac{\mS}{2}-3}$ generally have harmonic numbers in them;
we observe that these can be canceled by adding $H_{3-\frac{\mS}{2}}$. This allows us to make a credible ansatz, and we then find indeed that the remainder is a simple rational function for each pole.  In this way, we find that
\begin{align} \label{I4 app}
 I_4[\mS,\mT] =&\ {-}8\frac{\Gamma\big(\tfrac{\mS}{2}-2\big)\Gamma\big(\tfrac{\mT}{2}-2\big)\Gamma\big(\tfrac{\mU}{2}-2\big)}
 {\Gamma\big(4-\tfrac{\mS}{2}\big)\Gamma\big(4-\tfrac{\mT}{2}\big)\Gamma\big(4-\tfrac{\mU}{2}\big)}
 \nonumber\\ & \times\left[  \frac{4(\mS - 5)}{(\mT - 6) (\mU - 6)}+
 \left( \frac{\mT-\mS}{\mU-6}+\frac{\mU-\mS}{\mT-6}\right)\left( H_{\frac{\mS}{2}-3}+H_{3-\frac{\mS}{2}}\right)+  \mbox{(2 cyclic)}\right]
\end{align}
where the whole expression inside the square bracket is to be cyclically symmetrized.
Again, this formula, which has all the correct poles, can be confirmed by numerical integration of \eqref{I4 kernel def}.
Equation \eqref{cI24} in the main text is equivalent to \eqref{I4 app} upon using crossing symmetry.

Supersymetric localization predicts the right-hand-side of the integral \eqref{Ip Mellin appendix} as a function of $g$.
While this is in principle the ``easy'' part of the calculation (it only needs to be evaluated once for each $g$),
we find that the representation of $I_4(g)$ in \eqref{cI24} is not very stable numerically.  Instead, we found useful the following identity (from (A.8) of \cite{Beisert:2006ez}):
\be
 \frac{t J_0(2gt)J_1(2gt')-(t{\leftrightarrow}t')}{t'^2-t^2}
=\frac{1}{g t t'} \sum_{n=1}^\infty 2n J_{2n}(2gt)J_{2n}(2gt')\,,
\ee
which decouples the $t$ and $t'$ integrals.
Using the relation $2m/t J_{m}(t)=J_{m-1}(t)+J_{m+1}(t)$
and defining the Bessel integrals $J_{i,j}(g)=\int_0^\infty \frac{t\,dt\,e^{-t}}{(1-e^{-t})^2}J_i(2gt)J_j(2gt)$,
we then have
\begin{subequations} \label{I24g}
\begin{align}
    I_2(g) &= J_{1,1}(g)-J_{2,2}(g)\,, \\
    I_4(g) &= 48\zeta_3 -8g^{-2}J_{1,1}(g) -
    \sum_{n=1}^\infty
    \frac{96}{n}(J_{1,2n-1}(g)+J_{1,2n+1}(g))^2\,.
\end{align}
\end{subequations}
These formulas are straightforward to expand at weak coupling, where $J_{i,j}\propto g^{i+j}$, and we also find that the $n$ sum converges exponentially at any coupling provided that more than $O(\sqrt{g})$
terms are kept.

\paragraph{Cross-check in supergravity limit} As a cross-check, we computed the Mellin integral \eqref{cI24} for a simple model amplitude which includes supergravity and two contact interactions:
\begin{align}\label{Mellin model}
\widehat{\cH}^{\rm model} &= \frac{1}{(\tfrac{\mS}{2}-3)(\tfrac{\mT}{2}-3)(\tfrac{\mU}{2}-3)}+ \tilde{g}_0 + \tilde{g}_2 \big((\mS{-}4)^2+(\mT{-}4)^2+(\mU{-}4)^2\big)\,.
\end{align}
By performing the $\mU$ integral analytically using Barne's lemma and then the $\mS$ integral numerically, we find the following simple rational numbers:
\begin{align}
I_2^{\rm model} &= \frac14-\frac{1}{40}\tilde{g}_0-\frac{2}{35}\tilde{g}_2\,,\\
I_4^{\rm model} &= 48\zeta_3-24- \frac{16}{5}\tilde{g}_0-\frac{272}{35}\tilde{g}_2\,.
\end{align}
All these numbers agree precisely with eq.~(2.17) of \cite{Chester:2020dja}, including the sign flip of $I_4$ 
mentioned in the main text.
These numbers are relevant to the supergravity limit $g\to \infty$, where terms not included in $\widehat{\cH}^{\rm model}$ can be neglected to order $g^{-7}$ and comparison with \eqref{limits I24} yields:
\begin{align}
 \tilde{g}_0 = \frac{120\zeta_3}{(4\pi g)^3} - \frac{1890\zeta_5}{(4\pi g)^5} +O(g^{-7}),
\qquad
 \tilde{g}_2 = \frac{630\zeta_5}{(4\pi g)^5} + O(g^{-7})\,.
\end{align}
Using \eqref{contact scaling} to extract a flat space amplitude, we see that the low-energy amplitude which satisfies the integrated constraints is
\be
 \mathcal{M}^{\rm model,flat}(s,t)\propto \frac{1}{stu} + 
 g_0
+ g_2
(s^2+t^2+u^2) + O(m_K^{-12}),
\ee
where (using $m_K^2 \RAdS^2\approx 16\pi g$) we find:
\be
g_0=\frac{c_{-3}}{8c_0}\tilde{g}_0=\frac{2\zeta_3}{m_K^6},\qquad
g_2=\frac{c_{-3}}{8c_{2}}\tilde{g}_2=\frac{\zeta_5}{m_K^{10}}\,.
\ee
This is precisely the low-energy expansion of the Virasoro-Shapiro amplitude \eqref{eq:MVSflat}.  Of course, the same comparison was performed in \cite{Chester:2020dja} (who also interpreted the $1/\RAdS^2$ part of $\tilde{g}_0$) and here we are only replicating this result while verifying the Mellin transform \eqref{I4 app}.

\subsection{Evaluation on Polyakov-Regge blocks}


We now detail how we numerically evaluate the integrated constraint 
$I_p[\Delta,\j]$ in \eqref{integrated constraints}.
As explained in the main text, we evaluate it interms of single-trace data in the planar limit by inserting the Polyakov-Regge block \eqref{PR Mellin explicit}.  We deal with the $\mT$-channel poles by symmetrizing in $\mS{\leftrightarrow}\mT$ the kernel \eqref{cI24}.
The dependence on descendant index $n$ is then solely through the following sums, for $q=0\ldots \j$:
\be
\widehat{\cP}^{\mathcal{N}=4}_{\mS}[\Delta,\j]_{q} \equiv
\sum_{n=0}^\infty \frac{K_{\Delta,\j}^{n,\{\Delta_i\}}}{\tilde{K}_{\Delta,\j}^{\Nfour}}
\frac{(-n)_q}{\mS-(\Delta{-}\j{+}2n{+}4)}\,. \label{Ps vector}
\ee
The same sums enter any evaluation of Polyakov-Regge blocks and has been optimized in \cite{Caron-Huot:2022sdy}; $\tilde{K}$ is defined below eq.~(C.14) there.
The only ingredient that is specific to the particular sum rules at hands are the fixed-$\mS$ integrals:
\begin{align}
[I_{p,\mS}]_k&= \int \frac{d\mT}{4\pi i}
\frac{\Gamma\big(\tfrac{\mT}{2}-2\big)\Gamma\!\left(4-\tfrac{\mT}{2}\right)
\Gamma\big(\tfrac{\mU}{2}-2\big)\Gamma\!\left(4-\tfrac{\mU}{2}+k\right)}
{\Gamma\big(\tfrac{\mS}{2}\big)\Gamma\big(4-\tfrac{\mS}{2}\big)}
\\ &\phantom{=} \times \left\{
\begin{array}{l@{{\qquad}}l}
{-}1, & \mbox{for $p=2$},\\
{-}48\left[
\frac{4(\mU- 5)}{(\mS- 6) (\mT - 6)}+
\left(\frac{\mT-\mS}{\mU-6}\left( H_{\frac{\mS}{2}-3}+H_{3-\frac{\mS}{2}}\right)+(\mS{\leftrightarrow}\mT)\right)\right],
& \mbox{for $p=4$}.
\end{array}\right.
\end{align}
The first integral is elementary,
whereas the second is considerably more involved, in particular the parts with harmonic functions of $\mT$.
We were able to perform it analytically by writing $H_{x} = \sum_{n=1}^\infty \left(\frac{1}{n}-\frac{1}{n+x}\right)$ and integrating term-by-term.
We omit steps and record our final results:
\begin{align}
[I_{2,\mS}]_k&=\frac{-\big(\tfrac{\mS}{2}\big)_k}{(k+2)(k+3)}, \\
[I_{4,\mS}]_k&=
\frac{-768(\mS-5)^2k!}{(\mS-6)^2(\mS-4)(\mS-2)}+
\frac{48\big(\tfrac{\mS}{2}\big)_k}{(k+1)_3(\mS-6)}
\Bigg[2\frac{(k+3)(\mS-5)(\mS-4)}{2k+\mS-2} + 8(k+\mS-4)
\nonumber\\ & \hspace{20mm} + ((k + 5) (\mS - 4) - 4)\left(H_{\frac{\mS}{2}+k-1} - H_{\frac{\mS}{2}-4} -2 H_{k+3}\right)\Bigg]\,.
\end{align}
We tried other permutations of \eqref{cI4} but we found that they did not lead to simpler expressions.
Having performed one of the Mellin integrals analytically, we are left with a single integral:
\be \label{Ip app}
I_p[\Delta,\j]= \tilde{K}_{\Delta,\j}^{\Nfour} \int \frac{d\mS}{4\pi i}
\Gamma\big(\tfrac{\mS}{2}-2\big)\Gamma\big(\tfrac{\mS}{2}\big)\Gamma\big(4-\tfrac{\mS}{2}\big)^2 \sum_{q,k=0}^\j
\widehat{\cP}_{\mS}^{\Nfour}[\Delta,\j]_{q} \left[Q_{\Delta+4,\j}\right]_{q,k}[I_{p,\mS}]_k 
\,.
\ee
Like the other functionals discussed in \cite{Caron-Huot:2022sdy} and
below \eqref{Bv Regge numerics}, this can be evaluated as an exponentially convergent Riemann sum after the simple change of variable
$\mS=5+i\sinh x$, allowing for maximal recycling of the ingredients involved.

\section{OPE coefficients for first Regge trajectory}
Here we record the strong coupling series for the OPE data in the first Regge trajectory. There is a single operator for each spin $J$ with scaling dimension given by: 
\begin{align}\label{eq:strong1stRegge}
\frac{(\Delta+2)^2}{J+2} &= 2 \lambda^{1/2}+ \left(\frac{4}{J+2}-1+\frac{3}{2}(J+2)\right)+\frac{\frac{15}{4}+\left(\frac{3}{8}-3\zeta_3\right)(J+2)-\frac{3}{8}(J+2)^2} {\lambda^{1/2}} \nonumber\\
&\quad + \frac{\frac{15}{4}-\frac{9(3+8\zeta_3)}{16}(J+2)+\frac{-9+60(\zeta_3+\zeta_5)}{16}(J+2)^2+\frac{31}{64}(J+2)^3}{\lambda} + \mathcal{O}(\lambda^{-3/2})
\end{align}
At leading order the OPE coefficient $\lambda^2_{\Delta,J}$ can be obtained by using the relation to the flat-space coefficient $a^{\rm flat}_{J}$ given in section \ref{sec:OPEandFlat}. This latter coefficient can be obtained by using the flat-space projection formula:
\be \label{C2 integral flat}
 a_\j(s) = \frac{s^{\frac{D-4}{2}}}{2^{2D-3}\pi^{\frac{D-2}{2}}\,\Gamma\left(\frac{D-2}{2}\right)} 
 \int_{-1}^1 dx (1-x^2)^{\frac{d-4}{2}} \cP_\j(x) s^4\cM(s,t)_{t=-\frac12s(1-x)} 
\ee
on the Virasoro-Shapiro amplitude in \eqref{eq:MVSflat}, specializing to $D=5$ (so $d=4$). Then, applying the relation in eq.~\eqref{flat OPE}, we generalized the result 
 of \eqref{Konishi strong prediction} for Konishi operator to all spinning states: 
\be
 C^2_{2\sqrt{(1+\j/2)/\alpha'},\j} = \frac{2G_5}{(\alpha')^{\frac52}} \frac{2^{-2\j}(1+\frac{\j}{2})^{\j+\frac52}}{(\j{+}1) \Gamma(1+\frac{\j}{2})^2}
\quad \Rightarrow \quad
\frac{2\sin^2(\tfrac{\pi\Delta}{2}\big) \OPE_{\Delta,\j}}{\OPEfree_{\Delta,\j}}  = \frac{\pi^2\lambda}{32c} \frac{2^{-2\j}(1+\frac{\j}{2})^{\j+2}}{(\j{+}1) \Gamma(1+\frac{\j}{2})^2},
\ee
This result is in perfect agreement with
(E.10) of \cite{Alday:2022uxp} and confirms that the simple dictionary \eqref{flat OPE} works also for spinning operators.

\begin{table}[t]
\setlength\arrayrulewidth{1pt}
\renewcommand{\arraystretch}{1.2}
\centering\begin{tabular}{|l|c|c|c|}
\hline
$g$ & $\tilde{\lambda}_K^{2,\,\text{lower}}$  & $\tilde{\lambda}_K^{2,\,\text{upper}}$ & 
\text{diff.}\\ \hline
 0.1 & 1.911 & 1.912 & 0.00067 \\
 0.2 & 1.730 & 1.735 & 0.0052 \\
 0.25 & 1.637 & 1.653 & 0.016 \\
 0.275 & 1.596 & 1.618 & 0.022 \\
 0.3 & 1.558 & 1.585 & 0.027 \\
 0.325 & 1.522 & 1.555 & 0.033 \\
 0.35 & 1.491 & 1.529 & 0.038 \\
 0.375 & 1.461 & 1.505 & 0.044 \\
 0.4 & 1.441 
 & 1.484 & 0.043 \\
 0.425 & 1.417 & 1.465 & 0.048 \\
 0.45 & 1.403 & 1.448 & 0.045 \\
 0.475 & 1.391 & 1.432 & 0.041 \\
 0.5 & 1.379 & 1.418 & 0.039 \\
 0.525 & 1.368 & 1.405 & 0.036 \\
 0.55 & 1.359 & 1.393 & 0.034 \\
 0.6 & 1.341 & 1.371 & 0.030 \\
 0.65 & 1.323 & 1.352 & 0.029 \\
 0.7 & 1.307 & 1.332 & 0.026 \\
 0.75 & 1.292 & 1.314 & 0.023 \\
\hline
\end{tabular}
\quad\quad
\begin{tabular}{|l|c|c|c|}
\hline
$g$ & $\tilde{\lambda}_K^{2,\,\text{lower}}$  & $\tilde{\lambda}_K^{2,\,\text{upper}}$ & 
\text{diff.}\\ \hline
  0.8 & 1.278 & 1.297 & 0.019 \\
 0.85 & 1.265 & 1.283 & 0.018 \\
 0.9 & 1.253 & 1.270 & 0.017 \\
 0.95 & 1.242 & 1.259 & 0.017 \\
 1. & 1.232 & 1.249 & 0.017 \\
 1.05 & 1.223 & 1.240 & 0.017 \\
 1.1 & 1.214 & 1.231 & 0.017 \\
 1.15 & 1.206 & 1.223 & 0.018 \\
 1.2 & 1.198 & 1.216 & 0.018 \\
 1.25 & 1.191 & 1.209 & 0.018 \\
 1.3 & 1.184 & 1.203 & 0.019 \\
 1.35 & 1.178 & 1.197 & 0.019 \\
 1.5 & 1.161 & 1.181 & 0.020 \\
 1.8 & 1.136 & 1.157 & 0.021 \\
 2.1 & 1.117 & 1.138 & 0.022 \\
 2.3 & 1.107 & 1.129 & 0.022 \\
 2.7 & 1.090 & 1.113 & 0.023 \\
 3.2 & 1.075 & 1.089 & 0.014 \\
 3.7 & 1.066 & 1.081 & 0.014 \\
 \hline
\end{tabular}
\caption{List of lower and upper bounds for the OPE coefficient of Konishi operator at various values of the `t Hooft coupling ($\lambda= 16\pi^2 g^2$). We use the normalization introduced in eq.~\eqref{eq:Ktilde}. 
}
\label{tab:bounds}
\end{table}
The latter reference also provides sub-leading corrections and we recast them by first generalizing our tilde normalization in \eqref{eq:Ktilde} to spinning operators: 
\begin{align}
\tilde{\lambda}^{2}_{\Delta,J} = \frac{\lambda_{\Delta,J}^{2}}{\lambda^{2,\text{free}}_{\Delta,J}}\times\frac{2^{8}}{\left(\Delta+J+6\right)^2}\frac{\sin^2\left(\frac{\pi}{2}(\Delta-J)\right)}{\left[\frac{\pi}{2}(\Delta-J-2)\right]^2}\times\frac{2^{3J}(J+1)\Gamma\left(\frac{J}{2}+1\right)^2}{(J+2)^J}
\end{align}
with the free double-trace coefficient given in eq.~\eqref{OPEfree N4}. In this normalization we recast the results of \cite{Alday:2022uxp,Alday:2023mvu} for the OPE coefficient as:
\begin{align}\label{eq:logOPEhigherSpin}
\frac{\log\tilde{\lambda}^{2}_{\Delta,J}}{J+2} \,&=\,  \frac{\frac{17}{6}+(J+2)+\left(-\frac{7}{12}+\zeta_3\right)(J+2)^2}{(\Delta+2)^2} \nonumber\\
&\quad + \frac{\frac{511}{60}+6(J+2)+\left(\frac{1}{12}-2\zeta_3\right)(J+2)^2 -\left(\frac{13}{8}+6\zeta_3\right)(J+2)^3+\left(\frac{31}{40}-\frac{3}{2}\zeta_5\right)\,(J+2)^4}{(\Delta+2)^{4}}\nonumber\\
&\quad  +\mathcal{O}(\Delta^{-6})
\end{align}

\section{Lower and upper bounds on Konishi OPE coefficient}\label{app:tablebounds}
In table \ref{tab:bounds} we record the bounds on  Konishi OPE coefficient for various values of the coupling obtained from our numerical analysis in section \ref{sec:numericsKonishi}. This data was plotted in fig.~\ref{fig:ope-results} in the main text.

\bibliographystyle{JHEP}
\bibliography{references}

\providecommand{\href}[2]{#2}\begingroup\raggedright\begin{thebibliography}{10}

\bibitem{Beisert:2010jr}
N.~Beisert et~al., \emph{{Review of AdS/CFT Integrability: An Overview}}, \href{https://doi.org/10.1007/s11005-011-0529-2}{\emph{Lett. Math. Phys.} {\bfseries 99} (2012) 3--32}, [\href{https://arxiv.org/abs/1012.3982}{{\ttfamily 1012.3982}}].

\bibitem{Gromov:2013pga}
N.~Gromov, V.~Kazakov, S.~Leurent and D.~Volin, \emph{{Quantum Spectral Curve for Planar $\mathcal{N} = 4$ Super-Yang-Mills Theory}}, \href{https://doi.org/10.1103/PhysRevLett.112.011602}{\emph{Phys. Rev. Lett.} {\bfseries 112} (2014) 011602}, [\href{https://arxiv.org/abs/1305.1939}{{\ttfamily 1305.1939}}].

\bibitem{Gromov:2014caa}
N.~Gromov, V.~Kazakov, S.~Leurent and D.~Volin, \emph{{Quantum spectral curve for arbitrary state/operator in AdS$_{5}$/CFT$_{4}$}}, \href{https://doi.org/10.1007/JHEP09(2015)187}{\emph{JHEP} {\bfseries 09} (2015) 187}, [\href{https://arxiv.org/abs/1405.4857}{{\ttfamily 1405.4857}}].

\bibitem{Gromov:2009zb}
N.~Gromov, V.~Kazakov and P.~Vieira, \emph{{Exact Spectrum of Planar ${\cal N}=4$ Supersymmetric Yang-Mills Theory: Konishi Dimension at Any Coupling}}, \href{https://doi.org/10.1103/PhysRevLett.104.211601}{\emph{Phys. Rev. Lett.} {\bfseries 104} (2010) 211601}, [\href{https://arxiv.org/abs/0906.4240}{{\ttfamily 0906.4240}}].

\bibitem{Gromov:2023hzc}
N.~Gromov, A.~Hegedus, J.~Julius and N.~Sokolova, \emph{{Fast QSC solver: tool for systematic study of $ \mathcal{N} $ = 4 Super-Yang-Mills spectrum}}, \href{https://doi.org/10.1007/JHEP05(2024)185}{\emph{JHEP} {\bfseries 05} (2024) 185}, [\href{https://arxiv.org/abs/2306.12379}{{\ttfamily 2306.12379}}].

\bibitem{Ekhammar:2024rfj}
S.~Ekhammar, N.~Gromov and P.~Ryan, \emph{{New Approach to Strongly Coupled N = 4 SYM via Integrability}},  \href{https://arxiv.org/abs/2406.02698}{{\ttfamily 2406.02698}}.

\bibitem{Basso:2015zoa}
B.~Basso, S.~Komatsu and P.~Vieira, \emph{{Structure Constants and Integrable Bootstrap in Planar N=4 SYM Theory}},  \href{https://arxiv.org/abs/1505.06745}{{\ttfamily 1505.06745}}.

\bibitem{Fleury:2016ykk}
T.~Fleury and S.~Komatsu, \emph{{Hexagonalization of Correlation Functions}}, \href{https://doi.org/10.1007/JHEP01(2017)130}{\emph{JHEP} {\bfseries 01} (2017) 130}, [\href{https://arxiv.org/abs/1611.05577}{{\ttfamily 1611.05577}}].

\bibitem{Basso:2022nny}
B.~Basso, A.~Georgoudis and A.~K. Sueiro, \emph{{Structure Constants of Short Operators in Planar N=4 Supersymmetric Yang-Mills Theory}}, \href{https://doi.org/10.1103/PhysRevLett.130.131603}{\emph{Phys. Rev. Lett.} {\bfseries 130} (2023) 131603}, [\href{https://arxiv.org/abs/2207.01315}{{\ttfamily 2207.01315}}].

\bibitem{Bercini:2022jxo}
C.~Bercini, A.~Homrich and P.~Vieira, \emph{{Structure Constants in $\mathcal{N} = 4$ SYM and Separation of Variables}},  \href{https://arxiv.org/abs/2210.04923}{{\ttfamily 2210.04923}}.

\bibitem{Basso:2019diw}
B.~Basso and D.-L. Zhong, \emph{{Three-point functions at strong coupling in the BMN limit}}, \href{https://doi.org/10.1007/JHEP04(2020)076}{\emph{JHEP} {\bfseries 04} (2020) 076}, [\href{https://arxiv.org/abs/1907.01534}{{\ttfamily 1907.01534}}].

\bibitem{Jiang:2019zig}
Y.~Jiang, S.~Komatsu and E.~Vescovi, \emph{{Exact Three-Point Functions of Determinant Operators in Planar $N=4$ Supersymmetric Yang-Mills Theory}}, \href{https://doi.org/10.1103/PhysRevLett.123.191601}{\emph{Phys. Rev. Lett.} {\bfseries 123} (2019) 191601}, [\href{https://arxiv.org/abs/1907.11242}{{\ttfamily 1907.11242}}].

\bibitem{Coronado:2018cxj}
F.~Coronado, \emph{{Bootstrapping the Simplest Correlator in Planar $\mathcal N = 4$ Supersymmetric Yang-Mills Theory to All Loops}}, \href{https://doi.org/10.1103/PhysRevLett.124.171601}{\emph{Phys. Rev. Lett.} {\bfseries 124} (2020) 171601}, [\href{https://arxiv.org/abs/1811.03282}{{\ttfamily 1811.03282}}].

\bibitem{Kostov:2019stn}
I.~Kostov, V.~B. Petkova and D.~Serban, \emph{{Determinant Formula for the Octagon Form Factor in $N$=4 Supersymmetric Yang-Mills Theory}}, \href{https://doi.org/10.1103/PhysRevLett.122.231601}{\emph{Phys. Rev. Lett.} {\bfseries 122} (2019) 231601}, [\href{https://arxiv.org/abs/1903.05038}{{\ttfamily 1903.05038}}].

\bibitem{Belitsky:2020qrm}
A.~V. Belitsky and G.~P. Korchemsky, \emph{{Octagon at finite coupling}}, \href{https://doi.org/10.1007/JHEP07(2020)219}{\emph{JHEP} {\bfseries 07} (2020) 219}, [\href{https://arxiv.org/abs/2003.01121}{{\ttfamily 2003.01121}}].

\bibitem{Alday:2022uxp}
L.~F. Alday, T.~Hansen and J.~A. Silva, \emph{{AdS Virasoro-Shapiro from dispersive sum rules}},  \href{https://arxiv.org/abs/2204.07542}{{\ttfamily 2204.07542}}.

\bibitem{Alday:2023mvu}
L.~F. Alday and T.~Hansen, \emph{{The AdS Virasoro-Shapiro amplitude}}, \href{https://doi.org/10.1007/JHEP10(2023)023}{\emph{JHEP} {\bfseries 10} (2023) 023}, [\href{https://arxiv.org/abs/2306.12786}{{\ttfamily 2306.12786}}].

\bibitem{Rattazzi:2008pe}
R.~Rattazzi, V.~S. Rychkov, E.~Tonni and A.~Vichi, \emph{{Bounding scalar operator dimensions in 4D CFT}}, \href{https://doi.org/10.1088/1126-6708/2008/12/031}{\emph{JHEP} {\bfseries 12} (2008) 031}, [\href{https://arxiv.org/abs/0807.0004}{{\ttfamily 0807.0004}}].

\bibitem{El-Showk:2012cjh}
S.~El-Showk, M.~F. Paulos, D.~Poland, S.~Rychkov, D.~Simmons-Duffin and A.~Vichi, \emph{{Solving the 3D Ising Model with the Conformal Bootstrap}}, \href{https://doi.org/10.1103/PhysRevD.86.025022}{\emph{Phys. Rev. D} {\bfseries 86} (2012) 025022}, [\href{https://arxiv.org/abs/1203.6064}{{\ttfamily 1203.6064}}].

\bibitem{Poland:2018epd}
D.~Poland, S.~Rychkov and A.~Vichi, \emph{{The Conformal Bootstrap: Theory, Numerical Techniques, and Applications}}, \href{https://doi.org/10.1103/RevModPhys.91.015002}{\emph{Rev. Mod. Phys.} {\bfseries 91} (2019) 015002}, [\href{https://arxiv.org/abs/1805.04405}{{\ttfamily 1805.04405}}].

\bibitem{Poland:2022qrs}
D.~Poland and D.~Simmons-Duffin, \emph{{Snowmass White Paper: The Numerical Conformal Bootstrap}},  in \emph{{2022 Snowmass Summer Study}}, 3, 2022, \href{https://arxiv.org/abs/2203.08117}{{\ttfamily 2203.08117}}.

\bibitem{Beem:2016wfs}
C.~Beem, L.~Rastelli and B.~C. van Rees, \emph{{More ${\mathcal N}=4$ superconformal bootstrap}}, \href{https://doi.org/10.1103/PhysRevD.96.046014}{\emph{Phys. Rev. D} {\bfseries 96} (2017) 046014}, [\href{https://arxiv.org/abs/1612.02363}{{\ttfamily 1612.02363}}].

\bibitem{Chester:2021aun}
S.~M. Chester, R.~Dempsey and S.~S. Pufu, \emph{{Bootstrapping $ \mathcal{N} $ = 4 super-Yang-Mills on the conformal manifold}}, \href{https://doi.org/10.1007/JHEP01(2023)038}{\emph{JHEP} {\bfseries 01} (2023) 038}, [\href{https://arxiv.org/abs/2111.07989}{{\ttfamily 2111.07989}}].

\bibitem{Chester:2023ehi}
S.~M. Chester, R.~Dempsey and S.~S. Pufu, \emph{{Level repulsion in $ \mathcal{N} $ = 4 super-Yang-Mills via integrability, holography, and the bootstrap}}, \href{https://doi.org/10.1007/JHEP07(2024)059}{\emph{JHEP} {\bfseries 07} (2024) 059}, [\href{https://arxiv.org/abs/2312.12576}{{\ttfamily 2312.12576}}].

\bibitem{Cavaglia:2021bnz}
A.~Cavagli\`a, N.~Gromov, J.~Julius and M.~Preti, \emph{{Integrability and conformal bootstrap: One dimensional defect conformal field theory}}, \href{https://doi.org/10.1103/PhysRevD.105.L021902}{\emph{Phys. Rev. D} {\bfseries 105} (2022) L021902}, [\href{https://arxiv.org/abs/2107.08510}{{\ttfamily 2107.08510}}].

\bibitem{Cavaglia:2022qpg}
A.~Cavagli\`a, N.~Gromov, J.~Julius and M.~Preti, \emph{{Bootstrability in defect CFT: integrated correlators and sharper bounds}}, \href{https://doi.org/10.1007/JHEP05(2022)164}{\emph{JHEP} {\bfseries 05} (2022) 164}, [\href{https://arxiv.org/abs/2203.09556}{{\ttfamily 2203.09556}}].

\bibitem{Cavaglia:2023mmu}
A.~Cavagli\`a, N.~Gromov and M.~Preti, \emph{{Computing Four-Point Functions with Integrability, Bootstrap and Parity Symmetry}},  \href{https://arxiv.org/abs/2312.11604}{{\ttfamily 2312.11604}}.

\bibitem{Caron-Huot:2022sdy}
S.~Caron-Huot, F.~Coronado, A.-K. Trinh and Z.~Zahraee, \emph{{Bootstrapping $\mathcal{N}=4$ sYM correlators using integrability}},  \href{https://arxiv.org/abs/2207.01615}{{\ttfamily 2207.01615}}.

\bibitem{Chester:2020dja}
S.~M. Chester and S.~S. Pufu, \emph{{Far beyond the planar limit in strongly-coupled $ \mathcal{N} $ = 4 SYM}}, \href{https://doi.org/10.1007/JHEP01(2021)103}{\emph{JHEP} {\bfseries 01} (2021) 103}, [\href{https://arxiv.org/abs/2003.08412}{{\ttfamily 2003.08412}}].

\bibitem{Eden:2000bk}
B.~Eden, A.~C. Petkou, C.~Schubert and E.~Sokatchev, \emph{{Partial nonrenormalization of the stress tensor four point function in N=4 SYM and AdS / CFT}}, \href{https://doi.org/10.1016/S0550-3213(01)00151-1}{\emph{Nucl. Phys. B} {\bfseries 607} (2001) 191--212}, [\href{https://arxiv.org/abs/hep-th/0009106}{{\ttfamily hep-th/0009106}}].

\bibitem{Dolan:2004mu}
F.~A. Dolan, L.~Gallot and E.~Sokatchev, \emph{{On four-point functions of 1/2-BPS operators in general dimensions}}, \href{https://doi.org/10.1088/1126-6708/2004/09/056}{\emph{JHEP} {\bfseries 09} (2004) 056}, [\href{https://arxiv.org/abs/hep-th/0405180}{{\ttfamily hep-th/0405180}}].

\bibitem{Nirschl:2004pa}
M.~Nirschl and H.~Osborn, \emph{{Superconformal Ward identities and their solution}}, \href{https://doi.org/10.1016/j.nuclphysb.2005.01.013}{\emph{Nucl. Phys. B} {\bfseries 711} (2005) 409--479}, [\href{https://arxiv.org/abs/hep-th/0407060}{{\ttfamily hep-th/0407060}}].

\bibitem{Caron-Huot:2020adz}
S.~Caron-Huot, D.~Mazac, L.~Rastelli and D.~Simmons-Duffin, \emph{{Dispersive CFT Sum Rules}}, \href{https://doi.org/10.1007/JHEP05(2021)243}{\emph{JHEP} {\bfseries 05} (2021) 243}, [\href{https://arxiv.org/abs/2008.04931}{{\ttfamily 2008.04931}}].

\bibitem{Costa:2012cb}
M.~S. Costa, V.~Goncalves and J.~Penedones, \emph{{Conformal Regge theory}}, \href{https://doi.org/10.1007/JHEP12(2012)091}{\emph{JHEP} {\bfseries 12} (2012) 091}, [\href{https://arxiv.org/abs/1209.4355}{{\ttfamily 1209.4355}}].

\bibitem{Fitzpatrick:2011ia}
A.~L. Fitzpatrick, J.~Kaplan, J.~Penedones, S.~Raju and B.~C. van Rees, \emph{{A Natural Language for AdS/CFT Correlators}}, \href{https://doi.org/10.1007/JHEP11(2011)095}{\emph{JHEP} {\bfseries 11} (2011) 095}, [\href{https://arxiv.org/abs/1107.1499}{{\ttfamily 1107.1499}}].

\bibitem{Alday:2021vfb}
L.~F. Alday, S.~M. Chester and T.~Hansen, \emph{{Modular invariant holographic correlators for $ \mathcal{N} $ = 4 SYM with general gauge group}}, \href{https://doi.org/10.1007/JHEP12(2021)159}{\emph{JHEP} {\bfseries 12} (2021) 159}, [\href{https://arxiv.org/abs/2110.13106}{{\ttfamily 2110.13106}}].

\bibitem{Beisert:2006ez}
N.~Beisert, B.~Eden and M.~Staudacher, \emph{{Transcendentality and Crossing}}, \href{https://doi.org/10.1088/1742-5468/2007/01/P01021}{\emph{J. Stat. Mech.} {\bfseries 0701} (2007) P01021}, [\href{https://arxiv.org/abs/hep-th/0610251}{{\ttfamily hep-th/0610251}}].

\bibitem{Gary:2009ae}
M.~Gary, S.~B. Giddings and J.~Penedones, \emph{{Local bulk S-matrix elements and CFT singularities}}, \href{https://doi.org/10.1103/PhysRevD.80.085005}{\emph{Phys. Rev. D} {\bfseries 80} (2009) 085005}, [\href{https://arxiv.org/abs/0903.4437}{{\ttfamily 0903.4437}}].

\bibitem{Paulos:2016fap}
M.~F. Paulos, J.~Penedones, J.~Toledo, B.~C. van Rees and P.~Vieira, \emph{{The S-matrix bootstrap. Part I: QFT in AdS}}, \href{https://doi.org/10.1007/JHEP11(2017)133}{\emph{JHEP} {\bfseries 11} (2017) 133}, [\href{https://arxiv.org/abs/1607.06109}{{\ttfamily 1607.06109}}].

\bibitem{Li:2021snj}
Y.-Z. Li, \emph{{Notes on flat-space limit of AdS/CFT}}, \href{https://doi.org/10.1007/JHEP09(2021)027}{\emph{JHEP} {\bfseries 09} (2021) 027}, [\href{https://arxiv.org/abs/2106.04606}{{\ttfamily 2106.04606}}].

\bibitem{Mukhametzhanov:2018zja}
B.~Mukhametzhanov and A.~Zhiboedov, \emph{{Analytic Euclidean Bootstrap}}, \href{https://doi.org/10.1007/JHEP10(2019)270}{\emph{JHEP} {\bfseries 10} (2019) 270}, [\href{https://arxiv.org/abs/1808.03212}{{\ttfamily 1808.03212}}].

\bibitem{vanRees:2023fcf}
B.~C. van Rees and X.~Zhao, \emph{{Flat-space Partial Waves From Conformal OPE Densities}},  \href{https://arxiv.org/abs/2312.02273}{{\ttfamily 2312.02273}}.

\bibitem{Carmi:2019cub}
D.~Carmi and S.~Caron-Huot, \emph{{A Conformal Dispersion Relation: Correlations from Absorption}}, \href{https://doi.org/10.1007/JHEP09(2020)009}{\emph{JHEP} {\bfseries 09} (2020) 009}, [\href{https://arxiv.org/abs/1910.12123}{{\ttfamily 1910.12123}}].

\bibitem{Caron-Huot:2017vep}
S.~Caron-Huot, \emph{{Analyticity in Spin in Conformal Theories}}, \href{https://doi.org/10.1007/JHEP09(2017)078}{\emph{JHEP} {\bfseries 09} (2017) 078}, [\href{https://arxiv.org/abs/1703.00278}{{\ttfamily 1703.00278}}].

\bibitem{Minahan:2014usa}
J.~A. Minahan and R.~Pereira, \emph{{Three-point correlators from string amplitudes: Mixing and Regge spins}}, \href{https://doi.org/10.1007/JHEP04(2015)134}{\emph{JHEP} {\bfseries 04} (2015) 134}, [\href{https://arxiv.org/abs/1410.4746}{{\ttfamily 1410.4746}}].

\bibitem{Goncalves:2014ffa}
V.~Gon\c{c}alves, \emph{{Four point function of $\mathcal{N}=4$ stress-tensor multiplet at strong coupling}}, \href{https://doi.org/10.1007/JHEP04(2015)150}{\emph{JHEP} {\bfseries 04} (2015) 150}, [\href{https://arxiv.org/abs/1411.1675}{{\ttfamily 1411.1675}}].

\bibitem{Collier:2017shs}
S.~Collier, P.~Kravchuk, Y.-H. Lin and X.~Yin, \emph{{Bootstrapping the Spectral Function: On the Uniqueness of Liouville and the Universality of BTZ}}, \href{https://doi.org/10.1007/JHEP09(2018)150}{\emph{JHEP} {\bfseries 09} (2018) 150}, [\href{https://arxiv.org/abs/1702.00423}{{\ttfamily 1702.00423}}].

\bibitem{Antunes:2021abs}
A.~Antunes, M.~S. Costa, J.~a. Penedones, A.~Salgarkar and B.~C. van Rees, \emph{{Towards bootstrapping RG flows: sine-Gordon in AdS}}, \href{https://doi.org/10.1007/JHEP12(2021)094}{\emph{JHEP} {\bfseries 12} (2021) 094}, [\href{https://arxiv.org/abs/2109.13261}{{\ttfamily 2109.13261}}].

\bibitem{Paulos:2020zxx}
M.~F. Paulos, \emph{{Dispersion relations and exact bounds on CFT correlators}}, \href{https://doi.org/10.1007/JHEP08(2021)166}{\emph{JHEP} {\bfseries 08} (2021) 166}, [\href{https://arxiv.org/abs/2012.10454}{{\ttfamily 2012.10454}}].

\bibitem{Simmons-Duffin:2015qma}
D.~Simmons-Duffin, \emph{{A Semidefinite Program Solver for the Conformal Bootstrap}}, \href{https://doi.org/10.1007/JHEP06(2015)174}{\emph{JHEP} {\bfseries 06} (2015) 174}, [\href{https://arxiv.org/abs/1502.02033}{{\ttfamily 1502.02033}}].

\bibitem{Mazac:2019shk}
D.~Maz\'a\v{c}, L.~Rastelli and X.~Zhou, \emph{{A basis of analytic functionals for CFTs in general dimension}}, \href{https://doi.org/10.1007/JHEP08(2021)140}{\emph{JHEP} {\bfseries 08} (2021) 140}, [\href{https://arxiv.org/abs/1910.12855}{{\ttfamily 1910.12855}}].

\bibitem{Hegedus:2016eop}
A.~Heged\'{u}s and J.~Konczer, \emph{{Strong coupling results in the AdS$_{5}$ /CF T$_{4}$ correspondence from the numerical solution of the quantum spectral curve}}, \href{https://doi.org/10.1007/JHEP08(2016)061}{\emph{JHEP} {\bfseries 08} (2016) 061}, [\href{https://arxiv.org/abs/1604.02346}{{\ttfamily 1604.02346}}].

\bibitem{Basso:2010in}
B.~Basso, \emph{{Exciting the GKP string at any coupling}}, \href{https://doi.org/10.1016/j.nuclphysb.2011.12.010}{\emph{Nucl. Phys. B} {\bfseries 857} (2012) 254--334}, [\href{https://arxiv.org/abs/1010.5237}{{\ttfamily 1010.5237}}].

\bibitem{Basso:2013aha}
B.~Basso, A.~Sever and P.~Vieira, \emph{{Space-time S-matrix and Flux tube S-matrix II. Extracting and Matching Data}}, \href{https://doi.org/10.1007/JHEP01(2014)008}{\emph{JHEP} {\bfseries 01} (2014) 008}, [\href{https://arxiv.org/abs/1306.2058}{{\ttfamily 1306.2058}}].

\bibitem{Gubser:2002tv}
S.~S. Gubser, I.~R. Klebanov and A.~M. Polyakov, \emph{{A Semiclassical limit of the gauge / string correspondence}}, \href{https://doi.org/10.1016/S0550-3213(02)00373-5}{\emph{Nucl. Phys. B} {\bfseries 636} (2002) 99--114}, [\href{https://arxiv.org/abs/hep-th/0204051}{{\ttfamily hep-th/0204051}}].

\bibitem{Su:2022xnj}
N.~Su, \emph{{The Hybrid Bootstrap}},  \href{https://arxiv.org/abs/2202.07607}{{\ttfamily 2202.07607}}.

\bibitem{Caron-Huot:2021enk}
S.~Caron-Huot, D.~Mazac, L.~Rastelli and D.~Simmons-Duffin, \emph{{AdS bulk locality from sharp CFT bounds}}, \href{https://doi.org/10.1007/JHEP11(2021)164}{\emph{JHEP} {\bfseries 11} (2021) 164}, [\href{https://arxiv.org/abs/2106.10274}{{\ttfamily 2106.10274}}].

\bibitem{Georgoudis:2017meq}
A.~Georgoudis, V.~Goncalves and R.~Pereira, \emph{{Konishi OPE coefficient at the five loop order}}, \href{https://doi.org/10.1007/JHEP11(2018)184}{\emph{JHEP} {\bfseries 11} (2018) 184}, [\href{https://arxiv.org/abs/1710.06419}{{\ttfamily 1710.06419}}].

\bibitem{Drummond:2013nda}
J.~Drummond, C.~Duhr, B.~Eden, P.~Heslop, J.~Pennington and V.~A. Smirnov, \emph{{Leading singularities and off-shell conformal integrals}}, \href{https://doi.org/10.1007/JHEP08(2013)133}{\emph{JHEP} {\bfseries 08} (2013) 133}, [\href{https://arxiv.org/abs/1303.6909}{{\ttfamily 1303.6909}}].

\bibitem{Eden:2011we}
B.~Eden, P.~Heslop, G.~P. Korchemsky and E.~Sokatchev, \emph{{Hidden symmetry of four-point correlation functions and amplitudes in N=4 SYM}}, \href{https://doi.org/10.1016/j.nuclphysb.2012.04.007}{\emph{Nucl. Phys. B} {\bfseries 862} (2012) 193--231}, [\href{https://arxiv.org/abs/1108.3557}{{\ttfamily 1108.3557}}].

\bibitem{Caron-Huot:2020cmc}
S.~Caron-Huot and V.~Van~Duong, \emph{{Extremal Effective Field Theories}}, \href{https://doi.org/10.1007/JHEP05(2021)280}{\emph{JHEP} {\bfseries 05} (2021) 280}, [\href{https://arxiv.org/abs/2011.02957}{{\ttfamily 2011.02957}}].

\bibitem{Albert:2024yap}
J.~Albert, W.~Knop and L.~Rastelli, \emph{{Where is tree-level string theory?}},  \href{https://arxiv.org/abs/2406.12959}{{\ttfamily 2406.12959}}.

\bibitem{Berman:2023jys}
J.~Berman, H.~Elvang and A.~Herderschee, \emph{{Flattening of the EFT-hedron: supersymmetric positivity bounds and the search for string theory}}, \href{https://doi.org/10.1007/JHEP03(2024)021}{\emph{JHEP} {\bfseries 03} (2024) 021}, [\href{https://arxiv.org/abs/2310.10729}{{\ttfamily 2310.10729}}].

\bibitem{Berman:2024wyt}
J.~Berman and H.~Elvang, \emph{{Corners and islands in the S-matrix bootstrap of the open superstring}}, \href{https://doi.org/10.1007/JHEP09(2024)076}{\emph{JHEP} {\bfseries 09} (2024) 076}, [\href{https://arxiv.org/abs/2406.03543}{{\ttfamily 2406.03543}}].

\bibitem{Eckner:2024ggx}
C.~Eckner, F.~Figueroa and P.~Tourkine, \emph{{The Regge bootstrap, from linear to non-linear trajectories}},  \href{https://arxiv.org/abs/2401.08736}{{\ttfamily 2401.08736}}.

\bibitem{Correia:2020xtr}
M.~Correia, A.~Sever and A.~Zhiboedov, \emph{{An analytical toolkit for the S-matrix bootstrap}}, \href{https://doi.org/10.1007/JHEP03(2021)013}{\emph{JHEP} {\bfseries 03} (2021) 013}, [\href{https://arxiv.org/abs/2006.08221}{{\ttfamily 2006.08221}}].

\bibitem{Penedones:2010ue}
J.~Penedones, \emph{{Writing CFT correlation functions as AdS scattering amplitudes}}, \href{https://doi.org/10.1007/JHEP03(2011)025}{\emph{JHEP} {\bfseries 03} (2011) 025}, [\href{https://arxiv.org/abs/1011.1485}{{\ttfamily 1011.1485}}].

\bibitem{Fitzpatrick:2011hu}
A.~L. Fitzpatrick and J.~Kaplan, \emph{{Analyticity and the Holographic S-Matrix}}, \href{https://doi.org/10.1007/JHEP10(2012)127}{\emph{JHEP} {\bfseries 10} (2012) 127}, [\href{https://arxiv.org/abs/1111.6972}{{\ttfamily 1111.6972}}].

\bibitem{Caron-Huot:2021rmr}
S.~Caron-Huot, D.~Mazac, L.~Rastelli and D.~Simmons-Duffin, \emph{{Sharp boundaries for the swampland}}, \href{https://doi.org/10.1007/JHEP07(2021)110}{\emph{JHEP} {\bfseries 07} (2021) 110}, [\href{https://arxiv.org/abs/2102.08951}{{\ttfamily 2102.08951}}].

\bibitem{Poland:2011ey}
D.~Poland, D.~Simmons-Duffin and A.~Vichi, \emph{{Carving Out the Space of 4D CFTs}}, \href{https://doi.org/10.1007/JHEP05(2012)110}{\emph{JHEP} {\bfseries 05} (2012) 110}, [\href{https://arxiv.org/abs/1109.5176}{{\ttfamily 1109.5176}}].

\bibitem{Aharony:2008ug}
O.~Aharony, O.~Bergman, D.~L. Jafferis and J.~Maldacena, \emph{{N=6 superconformal Chern-Simons-matter theories, M2-branes and their gravity duals}}, \href{https://doi.org/10.1088/1126-6708/2008/10/091}{\emph{JHEP} {\bfseries 10} (2008) 091}, [\href{https://arxiv.org/abs/0806.1218}{{\ttfamily 0806.1218}}].

\bibitem{Hogervorst:2013sma}
M.~Hogervorst and S.~Rychkov, \emph{{Radial Coordinates for Conformal Blocks}}, \href{https://doi.org/10.1103/PhysRevD.87.106004}{\emph{Phys. Rev. D} {\bfseries 87} (2013) 106004}, [\href{https://arxiv.org/abs/1303.1111}{{\ttfamily 1303.1111}}].

\bibitem{Alday:2017vkk}
L.~F. Alday and S.~Caron-Huot, \emph{{Gravitational S-matrix from CFT dispersion relations}}, \href{https://doi.org/10.1007/JHEP12(2018)017}{\emph{JHEP} {\bfseries 12} (2018) 017}, [\href{https://arxiv.org/abs/1711.02031}{{\ttfamily 1711.02031}}].

\bibitem{Caron-Huot:2018kta}
S.~Caron-Huot and A.-K. Trinh, \emph{{All tree-level correlators in AdS$_{5}$\texttimes{}S$_{5}$ supergravity: hidden ten-dimensional conformal symmetry}}, \href{https://doi.org/10.1007/JHEP01(2019)196}{\emph{JHEP} {\bfseries 01} (2019) 196}, [\href{https://arxiv.org/abs/1809.09173}{{\ttfamily 1809.09173}}].

\end{thebibliography}\endgroup

\end{document}